\documentclass{emulateapj} 
\pdfoutput=1
\usepackage{natbib}
\usepackage{amsmath}
\def\be{\begin{equation}}
\def\ee{\end{equation}}
\def\bea{\begin{eqnarray}}
\def\eea{\end{eqnarray}}

\def\Var{\mathrm{Var}}

\newcommand{\griz}{\protect\hbox{$griz$} }

\def\lsim{\hbox{\rlap{\raise 0.425ex\hbox{$<$}}\lower 0.65ex\hbox{$\sim$}}}
\def\gsim{\hbox{\rlap{\raise 0.425ex\hbox{$>$}}\lower 0.65ex\hbox{$\sim$}}}

\def\arcdeg{\mbox{$^\circ$}}

\begin{document}

% We may want the title to be more precise One way could be to say how
% many clusters we are talking about, but since the sample includes
% stuff from literature as well, it gets tricky.  \title{Optical
% Spectroscopy and Velocity Dispersions of \\ Galaxy Clusters from the
% South Pole Telescope Survey}
\title{Optical Spectroscopy and Velocity Dispersions \\ of Galaxy
  Clusters from the SPT-SZ Survey}

%auto-ignore
\altaffiltext{\Harvard}{Department of Physics, Harvard University, 17 Oxford Street, Cambridge, MA 02138}
\altaffiltext{\Munich}{Department of Physics,
Ludwig-Maximilians-Universit\"{a}t,
Scheinerstr.\ 1, 81679 M\"{u}nchen, Germany}
\altaffiltext{\ExcellenceCluster}{Excellence Cluster Universe,
Boltzmannstr.\ 2, 85748 Garching, Germany}
\altaffiltext{\CfA}{Harvard-Smithsonian Center for Astrophysics,
60 Garden Street, Cambridge, MA 02138}
\altaffiltext{\Miss}{Department of Physics and Astronomy, University of Missouri, 5110 Rockhill Road, Kansas City, MO 64110}
\altaffiltext{\illast}{
Astronomy Department,
University of Illinois at Urbana-Champaign,
1002 W.\ Green Street,
Urbana, IL 61801 USA
}
\altaffiltext{\illphy}{
Department of Physics,
University of Illinois Urbana-Champaign,
1110 W.\ Green Street,
Urbana, IL 61801 USA
}
\altaffiltext{\UChicago}{University of Chicago,
5640 South Ellis Avenue, Chicago, IL 60637}
\altaffiltext{\UPenn}{Department of Physics and Astronomy,University of Pennsylvania, Philadelphia, PA 19104}
\altaffiltext{\MIT}{Kavli Institute for Astrophysics and Space
Research, Massachusetts Institute of Technology, 77 Massachusetts Avenue,
Cambridge, MA 02139}
\altaffiltext{\KICPChicago}{Kavli Institute for Cosmological Physics,
University of Chicago,
5640 South Ellis Avenue, Chicago, IL 60637}
\altaffiltext{\EFIChicago}{Enrico Fermi Institute,
University of Chicago,
5640 South Ellis Avenue, Chicago, IL 60637}
%\altaffiltext{\IAPFrance}{Institut d'Astrophysique de Paris, UMR 7095
%CNRS, Universit\'e Pierre et Marie Curie, 98 bis boulevard Arago, F-75014
%Paris, France}
\altaffiltext{\PhysicsUChicago}{Department of Physics,
University of Chicago,
5640 South Ellis Avenue, Chicago, IL 60637}
\altaffiltext{\ANL}{Argonne National Laboratory, 9700 S. Cass Avenue, Argonne, IL, USA 60439}
\altaffiltext{\AAUChicago}{Department of Astronomy and Astrophysics,
University of Chicago,
5640 South Ellis Avenue, Chicago, IL 60637}
\altaffiltext{\Cambridge}{Institute of Astronomy, University of Cambridge, Madingley Road,
Cambridge CB3 0HA, UK}
\altaffiltext{\NIST}{NIST Quantum Devices Group, 325 Broadway Mailcode 817.03, Boulder, CO, USA 80305}
\altaffiltext{\PUC}{Instituto de Astrofisica, Pontificia Universidad Catolica,
Chile}
\altaffiltext{\McGill}{Department of Physics,
McGill University,
3600 Rue University, Montreal, Quebec H3A 2T8, Canada}
\altaffiltext{\Berkeley}{Department of Physics,
University of California, Berkeley, CA 94720}
\altaffiltext{\UFlorida}{Department of Astronomy, University of Florida, Gainesville, FL 32611}
\altaffiltext{\Colorado}{Department of Astrophysical and Planetary Sciences and Department of Physics,
University of Colorado,
Boulder, CO 80309}
\altaffiltext{\NASA}{Department of Space Science, VP62,
NASA Marshall Space Flight Center,
Huntsville, AL 35812}
\altaffiltext{\Davis}{Department of Physics, 
University of California, One Shields Avenue, Davis, CA 95616}
\altaffiltext{\LBNL}{Physics Division,
Lawrence Berkeley National Laboratory,
Berkeley, CA 94720}
\altaffiltext{\Caltech}{California Institute of Technology, 1200 E. California Blvd., Pasadena, CA 91125}
\altaffiltext{\Arizona}{Steward Observatory, University of Arizona, 933 North Cherry Avenue, Tucson, AZ 85721}
\altaffiltext{\Michigan}{Department of Physics, University of Michigan, 450 Church Street, Ann  
Arbor, MI, 48109}
\altaffiltext{\MPE}{Max-Planck-Institut f\"{u}r extraterrestrische Physik,
Giessenbachstr.\ 85748 Garching, Germany}
\altaffiltext{\CaseWestern}{Physics Department, Center for Education and Research in Cosmology 
and Astrophysics, 
Case Western Reserve University,
Cleveland, OH 44106}
\altaffiltext{\Minnesota}{Physics Department, University of Minnesota, 116 Church Street S.E., Minneapolis, MN 55455}
\altaffiltext{\STScI}{Space Telescope Science Institute, 3700 San Martin
Dr., Baltimore, MD 21218}
\altaffiltext{\SAIC}{Liberal Arts Department, 
School of the Art Institute of Chicago, 
112 S Michigan Ave, Chicago, IL 60603}
\altaffiltext{\Yale}{Department of Physics, Yale University, P.O. Box 208210, New Haven,
CT 06520-8120}
\altaffiltext{\LLNL}{Institute of Geophysics and Planetary Physics, Lawrence
Livermore National Laboratory, Livermore, CA 94551}
\altaffiltext{\Dunlap}{Dunlap Institute for Astronomy \& Astrophysics, University of Toronto, 50 St George St, Toronto, ON, M5S 3H4, Canada}
\altaffiltext{\Toronto}{Department of Astronomy \& Astrophysics, University of Toronto, 50 St George St, Toronto, ON, M5S 3H4, Canada}
\altaffiltext{\BCCP}{Berkeley Center for Cosmological Physics,
Department of Physics, University of California, and Lawrence Berkeley
National Labs, Berkeley, CA 94720}

\def\Harvard{1}
\def\Munich{2}
\def\ExcellenceCluster{3}
\def\CfA{4}
\def\Miss{5}
\def\illast{6}
\def\illphy{7}
\def\UChicago{8}
\def\UPenn{9}
\def\MIT{10}
\def\KICPChicago{11}
\def\EFIChicago{12}
\def\PhysicsUChicago{13}
\def\ANL{14}
\def\AAUChicago{15}
\def\Cambridge{16}
\def\NIST{17}
\def\PUC{18}
\def\McGill{19}
\def\Berkeley{20}
\def\UFlorida{21}
\def\Colorado{22}
\def\NASA{23}
\def\Davis{24}
\def\LBNL{25}
\def\Caltech{26}
\def\Arizona{27}
\def\Michigan{28}
\def\MPE{29}
\def\CaseWestern{30}
\def\Minnesota{31}
\def\STScI{32}
\def\SAIC{33}
\def\Yale{34}
\def\LLNL{35}
\def\Dunlap{36}
\def\Toronto{37}
\def\BCCP{38}

\author{J.~Ruel\altaffilmark{\Harvard},
G.~Bazin\altaffilmark{\Munich,\ExcellenceCluster},
M.~Bayliss\altaffilmark{\Harvard,\CfA}, 
M.~Brodwin\altaffilmark{\Miss},
R.~J.~Foley\altaffilmark{\CfA,\illast,\illphy}, 
B.~Stalder\altaffilmark{\CfA},
K.~A.~Aird\altaffilmark{\UChicago},
R.~Armstrong\altaffilmark{\UPenn},
M.~L.~N.~Ashby\altaffilmark{\CfA},
M.~Bautz\altaffilmark{\MIT},
B.~A.~Benson\altaffilmark{\KICPChicago,\EFIChicago},
%E.~Bertin\altaffilmark{\IAPFrance},
L.~E.~Bleem\altaffilmark{\KICPChicago,\PhysicsUChicago,\ANL},
S.~Bocquet\altaffilmark{\Munich,\ExcellenceCluster},
J.~E.~Carlstrom\altaffilmark{\KICPChicago,\EFIChicago,\PhysicsUChicago,\ANL,\AAUChicago}, 
C.~L.~Chang\altaffilmark{\KICPChicago,\EFIChicago,\ANL}, 
S.~C.~Chapman\altaffilmark{\Cambridge}, 
H.~M. Cho\altaffilmark{\NIST}, 
A.~Clocchiatti\altaffilmark{\PUC},
T.~M.~Crawford\altaffilmark{\KICPChicago,\AAUChicago},
A.~T.~Crites\altaffilmark{\KICPChicago,\AAUChicago},
T.~de~Haan\altaffilmark{\McGill},
S.~Desai\altaffilmark{\Munich,\ExcellenceCluster},
M.~A.~Dobbs\altaffilmark{\McGill},
J.~P.~Dudley\altaffilmark{\McGill},
W.~R.~Forman\altaffilmark{\CfA},
E.~M.~George\altaffilmark{\Berkeley},
%D.~Gettings\altaffilmark{\UFlorida},
M.~D.~Gladders\altaffilmark{\KICPChicago,\AAUChicago},
A.~H.~Gonzalez\altaffilmark{\UFlorida},
N.~W.~Halverson\altaffilmark{\Colorado},
N.~L.~Harrington\altaffilmark{\Berkeley},
F.~W.~High\altaffilmark{\KICPChicago,\AAUChicago}, 
G.~P.~Holder\altaffilmark{\McGill},
W.~L.~Holzapfel\altaffilmark{\Berkeley},
%S.~Hoover\altaffilmark{\KICPChicago,\EFIChicago},
J.~D.~Hrubes\altaffilmark{\UChicago},
C.~Jones\altaffilmark{\CfA},
M.~Joy\altaffilmark{\NASA},
R.~Keisler\altaffilmark{\KICPChicago,\PhysicsUChicago},
L.~Knox\altaffilmark{\Davis},
A.~T.~Lee\altaffilmark{\Berkeley,\LBNL},
E.~M.~Leitch\altaffilmark{\KICPChicago,\AAUChicago},
J.~Liu\altaffilmark{\Munich,\ExcellenceCluster},
M.~Lueker\altaffilmark{\Berkeley,\Caltech},
D.~Luong-Van\altaffilmark{\UChicago},
A.~Mantz\altaffilmark{\KICPChicago},
D.~P.~Marrone\altaffilmark{\Arizona},
M.~McDonald\altaffilmark{\MIT},
J.~J.~McMahon\altaffilmark{\Michigan},
J.~Mehl\altaffilmark{\KICPChicago,\AAUChicago},
S.~S.~Meyer\altaffilmark{\KICPChicago,\EFIChicago,\PhysicsUChicago,\AAUChicago},
L.~Mocanu\altaffilmark{\KICPChicago,\AAUChicago},
J.~J.~Mohr\altaffilmark{\Munich,\ExcellenceCluster,\MPE},
T.~E.~Montroy\altaffilmark{\CaseWestern},
S.~S.~Murray\altaffilmark{\CfA},
T.~Natoli\altaffilmark{\KICPChicago,\PhysicsUChicago},
D.~Nurgaliev\altaffilmark{\Harvard}, 
S.~Padin\altaffilmark{\KICPChicago,\AAUChicago,\Caltech},
T.~Plagge\altaffilmark{\KICPChicago,\AAUChicago},
C.~Pryke\altaffilmark{\Minnesota}, 
C.~L.~Reichardt\altaffilmark{\Berkeley},
A.~Rest\altaffilmark{\STScI},
J.~E.~Ruhl\altaffilmark{\CaseWestern}, 
B.~R.~Saliwanchik\altaffilmark{\CaseWestern}, 
A.~Saro\altaffilmark{\Munich},
J.~T.~Sayre\altaffilmark{\CaseWestern}, 
K.~K.~Schaffer\altaffilmark{\KICPChicago,\EFIChicago,\SAIC}, 
L.~Shaw\altaffilmark{\McGill,\Yale},
E.~Shirokoff\altaffilmark{\Berkeley,\Caltech}, 
J.~Song\altaffilmark{\Michigan},
R.~\v{S}uhada\altaffilmark{\Munich},
H.~G.~Spieler\altaffilmark{\LBNL},
S.~A.~Stanford\altaffilmark{\Davis,\LLNL},
Z.~Staniszewski\altaffilmark{\CaseWestern},
A.~A.~Stark\altaffilmark{\CfA}, 
K.~Story\altaffilmark{\KICPChicago,\PhysicsUChicago},
C.~W.~Stubbs\altaffilmark{\Harvard,\CfA}, 
A.~van~Engelen\altaffilmark{\McGill},
K.~Vanderlinde\altaffilmark{\Dunlap,\Toronto},
J.~D.~Vieira\altaffilmark{\illast,\Caltech},
A. Vikhlinin\altaffilmark{\CfA},
R.~Williamson\altaffilmark{\KICPChicago,\AAUChicago}, 
O.~Zahn\altaffilmark{\Berkeley,\BCCP},
A.~Zenteno\altaffilmark{\Munich,\ExcellenceCluster}
}

%%%%%%%%%%%%%%%%%%%%
%% Abstract %%
%%%%%%%%%%%%%%%%%%%%

% Abstract has a maximum of 250 words

\email{Electronic address: mbayliss@cfa.harvard.edu}
 \slugcomment{Submitted to \apj} %\slugcomment{}

\begin{abstract}
  \begin{small}
    We present optical spectroscopy of galaxies in clusters detected
    through the Sunyaev-Zel'dovich (SZ) effect with the South Pole
    Telescope (SPT).  We report our own measurements of $61$
    spectroscopic cluster redshifts, and $48$ velocity dispersions
    each calculated with more than $15$ member galaxies. This catalog
    also includes $19$ dispersions of SPT-observed clusters previously
    reported in the literature.  The majority of the clusters in this
    paper are SPT-discovered; of these, most have been previously
    reported in other SPT cluster catalogs, and five are reported here
    as SPT discoveries for the first time.  By performing a resampling
    analysis of galaxy velocities, we find that unbiased velocity
    dispersions can be obtained from a relatively small number of
    member galaxies ($\lesssim 30$), but with increased systematic scatter. We use
    this analysis to determine statistical confidence intervals that
    include the effect of membership selection.  We fit scaling
    relations between the observed cluster velocity dispersions and
    mass estimates from SZ and X-ray observables.  In both cases, the
    results are consistent with the scaling relation between velocity
    dispersion and mass expected from dark-matter simulations.
    We measure a $\sim$30\% log-normal scatter in dispersion at fixed
    mass, and a $\sim$10\% offset in the normalization of the
    dispersion-mass relation when compared to the expectation from
    simulations, which is within the expected level of systematic
    uncertainty.
  \end{small}
\end{abstract}

\begin{small}
\end{small}

% --- galaxies: formation --- galaxies: evolution --- early universe}

%%%%%%%%%%%%%%%%%%%%
%% Introduction %%
%%%%%%%%%%%%%%%%%%%%

% The intro is meant to be as short as possible and refer to other
% articles for context.  I do not wish to resummarize the history of
% clusters / SZ / SPT / as every paper has been doing.

\section{Introduction}\label{s:introduction}

\setcounter{footnote}{0}

% Cosmic Microwave Background (CMB) surveys such as those conducted by
% the South Pole Telescope (SPT), the Atacama Cosmology Telescope
% (ACT), and {\it Planck} are now reliably finding massive clusters of
% galaxies through their Sunyaev-Zel'dovich \citep[SZ;][]{sunyaev72}
% signature \citep[see, e.g.,][]{staniszewski09, vanderlinde10,
% williamson11, marriage11b, hasselfield13, planck11-5.1a_arxiv,
% planck13}. For a discussion of the astrophysical and cosmological
% contexts, and the most recent results from the SPT-SZ cluster
% survey, see \citet{reichardt12}.

Clusters of galaxies cause a distortion in the cosmic microwave
background (CMB) from the inverse Compton scattering of the CMB
photons with the hot intra-cluster gas, commonly called the
Sunyaev-Zel'dovich (SZ) effect \citep{sunyaev72}. SZ cluster surveys
efficiently find massive, high-redshift clusters, primarily due to the
redshift independence of the brightness of the SZ effect, with
completed SZ surveys by the South Pole Telescope (SPT), Atacama
Cosmology Telescope (ACT), and {\it Planck} having identified over
1000 clusters by their SZ distortion \citep[see,
e.g.,][]{staniszewski09, vanderlinde10, williamson11, reichardt12,
  marriage11b, hasselfield13, planck11-5.1a_arxiv, planck13}.
SZ-selected samples have provided a unique window into high-redshfit
cluster evolution \citep[see, e.g.,][]{mcdonald12, mcdonald13,
  bayliss13}, and have also been used to constrain cosmological
parameters \citep[see, e.g.,][]{benson11, reichardt12}.

In this paper, we report spectroscopic observations of galaxies
associated with 61 galaxy clusters detected in the 2500 deg$^2$ SPT-SZ
survey.  This work is focused on measuring spectroscopic redshifts,
which can inform cosmological studies in two ways. First, we present
spectroscopically determined cosmological redshifts for most clusters.
The measured spectroscopic redshifts are useful as a training set for
photometric redshift measurements \citep{high10, song12}.

Second, we present velocity dispersions, which are a potentially
useful observable for measuring cluster mass \citep{white10, saro12}.
The cosmological constraints from the SPT-SZ cluster survey are
currently limited by the uncertainty in the normalization of the
SZ-mass relation \citep{benson11, reichardt12}.  This motivates using
multiple mass estimation methods, ideally in a joint likelihood
analysis. Our group is pursuing X-ray observations
\citep{andersson10}, weak lensing \citep{high12}, and velocity
dispersions to address the cluster mass calibration challenge.
Currently, the relationship between the SZ observable and mass is
primarily calibrated in a joint fit of SZ and X-ray data to a model
that includes cosmological and scaling relation parameters
\citep{benson11}. Like the SZ effect, X-ray emission is produced by
the hot gas component of the cluster, so velocity dispersions and weak
lensing are important for assessing any systematic biases from
gas-based proxies. Velocity dispersions also have the advantage of
being obtainable from ground-based telescopes up to high redshift.

The velocities of SPT cluster galaxies presented here are primarily
derived from our spectroscopic measurements of 61 massive galaxy
clusters. These data are used to produce $48$ velocity dispersions for
clusters with more than $15$ member galaxies, several of which we have
already presented elsewhere \citep{brodwin10, foley11,
  williamson11,mcdonald12, stalder12, reichardt12, bayliss13}. These
are, for the most part, the data obtained through 2011 in our ongoing
spectroscopy program.
% we take this paper as an opportunity to evaluate our few-member
% observing strategy, and to identify questions that need to be
% answered in view of the mass calibration.
We also list dispersions collected from the literature, including 
observations of 14 clusters that were also detected by ACT and 
targeted for spectroscopic followup by the ACT collaboration \citep{sifon12}.
% alongside a simple fit of the SZ properties to the corresponding
% dynamical masses.

This paper is organized as follows. We describe the observations and
observing strategy in Section \ref{s:observations}. In Section
\ref{s:results}, we present our results, including the individual
galaxy velocities and cluster velocity dispersions, and we investigate
the phase-space galaxy selection using a stacking analysis. In Section
\ref{s:resampling}, we use a resampling analysis to calculate cluster
redshift and dispersion uncertainties that take the effect of the
membership selection into account. We explore the properties of our
sample of velocity dispersions by comparing them with SZ-based SPT
masses, X-ray temperatures, and X-ray-derived masses in Section
\ref{s:comp}. The evaluation of our observing strategy and outstanding
questions are summarized in the conclusion, Section
\ref{s:conclusion}.

Throughout this paper, we define $M_{500c}$ ($M_{200c}$) as the mass
contained within $R_{500c}$ ($R_{200c}$), the radius from the cluster
center within which the average density is 500 (200) times the
critical density at the cluster redshift. Conversion between
$M_{500c}$ and $M_{200c}$ is made assuming an NFW density profile and
the \cite{duffy08} mass-concentration relation. We report
uncertainties at the 68\% confidence level, and we adopt a
WMAP7$+$BAO$+H_0$ flat $\Lambda$CDM cosmology with $\Omega_M = 0.272$,
$\Omega_\Lambda = 0.728$, and $H_0 = 70.2~$km\,s$^{-1}$\,Mpc$^{-1}$
\citep{komatsu11}.

%%%%%%%%%%%%%%%%%%%%
%% Observations %%
%%%%%%%%%%%%%%%%%%%%

\section{Observations}\label{s:observations}
\subsection{South Pole Telescope}\label{ss:obsspt}
Most of the galaxy clusters for which we report spectroscopic
observations were published as SPT cluster detections (and new
discoveries) in \cite{vanderlinde10}, \cite{williamson11}, and
\citet{reichardt12}; we refer the reader to those publications for
details of the SPT observations. In Table \ref{tab:spt}, we give the
SPT identification (ID) of the clusters and their essential SZ
properties. This includes the right ascension and declination of the
SZ center, the cluster redshift, and the SPT detection significance
$\xi$.  We also report the SPT cluster mass estimate,
$M_{500c,\mathrm{SPT}}$, as reported in \citet{reichardt12}, for those
clusters at redshift $z \ge 0.3$, the redshift threshold used in the
SPT cosmological analysis.  As described in \cite{reichardt12}, the
SPT mass estimate is measured from the SPT SZ significance and X-ray
measurements, where available, while accounting for the SPT selection,
and marginalizing over all uncertainties in cosmology and the cluster
observable scaling relations.  The last columns indicate the source of
the spectroscopy, our own measurements for 61 clusters, and a
literature reference for 19 of them. Five clusters have data from both
sources.

There are $11$ clusters that do not appear in prior SPT publications,
and are presented here as SPT detections for the first time. Five of
them are new discoveries (identified with * in Table \ref{tab:spt}),
and the other six were previously published as ACT detections
\citep[][identified with ** in Table \ref{tab:spt}]{marriage11b}.
These SPT detections will be reported in an upcoming cluster catalog
from the full 2500 deg$^2$ SPT-SZ survey.
% \citep[][]{bleem13}.

One cluster, SPT-CL~J0245-5302, is detected by SPT at high
significance, however because of its proximity to a bright point
source ($<8$ arcmin away), it is not included in the official catalog.
SPT-CL~J2347-5158 had a higher SPT significance in early maps of the
survey, but has $\xi < 4.0$ in the 2500-deg$^2$ survey. The SPT
significance and mass are not given for these two clusters.

% \LongTables
\tabletypesize{\scriptsize}
\begin{deluxetable*}{lrcc | rr | ccc}
  \setlength{\tabcolsep}{0.05in} \tablecaption{SPT properties and
    source of spectroscopic data \label{tab:spt}} \tablehead{
    \multicolumn{4}{c}{{\bf ID \& coordinates}} & \multicolumn{2}{c}{}
    &
    \multicolumn{3}{c}{{\bf Source of spectroscopy}} \\
    \colhead{SPT ID} & \colhead{R.A.} & \colhead{Dec.}  &
    \colhead{$z$} & \colhead{\bf $\xi$} & \colhead{ \bf
      $M_{500c,\mathrm{SPT}}$} & \colhead{this work} &
    \colhead{literature} \\
    \colhead{} & \colhead{(J2000 deg.)} & \colhead{(J2000 deg.)} &
    \colhead{} & \colhead{} & \colhead{($10^{14} h_{70}^{-1}
      M_\odot$)} & \colhead{} & \colhead{} } \startdata
  SPT-CL J0000-5748 & $0.2496$ & $-57.8066$ & $0.702$ & $5.48$ & $4.29 \pm 0.71$ & \checkmark &  \\
SPT-CL J0014-4952* & $3.6969$ & $-49.8772$ & $0.752$ & $8.87$ & $5.14 \pm 0.86$ & \checkmark &  \\
SPT-CL J0037-5047* & $9.4441$ & $-50.7971$ & $1.026$ & $6.93$ & $3.64 \pm 0.79$ & \checkmark &  \\
SPT-CL J0040-4407 & $10.2048$ & $-44.1329$ & $0.350$ & $19.34$ & $10.18 \pm 1.32$ & \checkmark &  \\
SPT-CL J0102-4915 & $15.7294$ & $-49.2611$ & $0.870$ & $39.91$ & $15.69 \pm 1.89$ &  & 1 \\
SPT-CL J0118-5156* & $19.5990$ & $-51.9434$ & $0.705$ & $5.97$ & $3.39 \pm 0.82$ & \checkmark &  \\
SPT-CL J0205-5829 & $31.4437$ & $-58.4856$ & $1.322$ & $10.54$ & $4.79 \pm 1.00$ & \checkmark &  \\
SPT-CL J0205-6432 & $31.2786$ & $-64.5461$ & $0.744$ & $6.02$ & $3.29 \pm 0.79$ & \checkmark &  \\
SPT-CL J0232-5257** & $38.1876$ & $-52.9578$ & $0.556$ & $8.65$ & $5.04 \pm 0.89$ &  & 1 \\
SPT-CL J0233-5819 & $38.2561$ & $-58.3269$ & $0.663$ & $6.64$ & $3.71 \pm 0.86$ & \checkmark &  \\
SPT-CL J0234-5831 & $38.6790$ & $-58.5217$ & $0.415$ & $14.65$ & $7.64 \pm 1.50$ & \checkmark &  \\
SPT-CL J0235-5121** & $38.9468$ & $-51.3516$ & $0.278$ & $9.78$ & - &  & 1 \\
SPT-CL J0236-4938** & $39.2477$ & $-49.6356$ & $0.334$ & $5.80$ & $3.39 \pm 0.89$ &  & 1 \\
SPT-CL J0240-5946 & $40.1620$ & $-59.7703$ & $0.400$ & $9.04$ & $5.29 \pm 1.07$ & \checkmark &  \\
SPT-CL J0245-5302 & $41.3780$ & $-53.0360$ & $0.300$ & - & - & \checkmark &  \\
SPT-CL J0254-5857 & $43.5729$ & $-58.9526$ & $0.437$ & $14.42$ & $7.46 \pm 1.46$ & \checkmark &  \\
SPT-CL J0257-5732 & $44.3516$ & $-57.5423$ & $0.434$ & $5.40$ & $3.14 \pm 0.86$ & \checkmark &  \\
SPT-CL J0304-4921** & $46.0619$ & $-49.3612$ & $0.392$ & $12.75$ & $7.32 \pm 1.04$ &  & 1 \\
SPT-CL J0317-5935 & $49.3208$ & $-59.5856$ & $0.469$ & $5.91$ & $3.46 \pm 0.89$ & \checkmark &  \\
SPT-CL J0328-5541 & $52.1663$ & $-55.6975$ & $0.084$ & $7.08$ & - &  & 3 \\
SPT-CL J0330-5228** & $52.7287$ & $-52.4698$ & $0.442$ & $11.57$ & $6.36 \pm 1.00$ &  & 1 \\
SPT-CL J0346-5439** & $56.7247$ & $-54.6505$ & $0.530$ & $9.25$ & $5.07 \pm 0.93$ &  & 1 \\
SPT-CL J0431-6126 & $67.8393$ & $-61.4438$ & $0.059$ & $6.40$ & - &  & 2 \\
SPT-CL J0433-5630 & $68.2522$ & $-56.5038$ & $0.692$ & $5.35$ & $2.89 \pm 0.82$ & \checkmark &  \\
SPT-CL J0438-5419 & $69.5749$ & $-54.3212$ & $0.422$ & $22.88$ & $10.82 \pm 1.39$ & \checkmark & 1 \\
SPT-CL J0449-4901* & $72.2742$ & $-49.0246$ & $0.790$ & $8.91$ & $4.57 \pm 0.86$ & \checkmark &  \\
SPT-CL J0509-5342 & $77.3360$ & $-53.7045$ & $0.462$ & $6.61$ & $5.36 \pm 0.71$ & \checkmark & 1 \\
SPT-CL J0511-5154 & $77.9202$ & $-51.9044$ & $0.645$ & $5.63$ & $3.61 \pm 0.96$ & \checkmark &  \\
SPT-CL J0516-5430 & $79.1480$ & $-54.5062$ & $0.294$ & $9.42$ & - & \checkmark &  \\
SPT-CL J0521-5104 & $80.2983$ & $-51.0812$ & $0.675$ & $5.45$ & $3.46 \pm 0.96$ &  & 1 \\
SPT-CL J0528-5300 & $82.0173$ & $-53.0001$ & $0.769$ & $5.45$ & $3.18 \pm 0.61$ & \checkmark & 1 \\
SPT-CL J0533-5005 & $83.3984$ & $-50.0918$ & $0.881$ & $5.59$ & $2.68 \pm 0.61$ & \checkmark &  \\
SPT-CL J0534-5937 & $83.6018$ & $-59.6289$ & $0.576$ & $4.57$ & $2.71 \pm 1.00$ & \checkmark &  \\
SPT-CL J0546-5345 & $86.6541$ & $-53.7615$ & $1.066$ & $7.69$ & $5.25 \pm 0.75$ & \checkmark & 1 \\
SPT-CL J0551-5709 & $87.9016$ & $-57.1565$ & $0.424$ & $6.13$ & $3.75 \pm 0.54$ & \checkmark &  \\
SPT-CL J0559-5249 & $89.9245$ & $-52.8265$ & $0.609$ & $9.28$ & $6.79 \pm 0.86$ & \checkmark & 1 \\
SPT-CL J0658-5556 & $104.6317$ & $-55.9465$ & $0.296$ & $39.05$ & - &  & 4 \\
SPT-CL J2012-5649 & $303.1132$ & $-56.8308$ & $0.055$ & $5.99$ & - &  & 2 \\
SPT-CL J2022-6323 & $305.5235$ & $-63.3973$ & $0.383$ & $6.58$ & $3.82 \pm 0.89$ & \checkmark &  \\
SPT-CL J2032-5627 & $308.0800$ & $-56.4557$ & $0.284$ & $8.14$ & - & \checkmark &  \\
SPT-CL J2040-4451 & $310.2468$ & $-44.8599$ & $1.478$ & $6.28$ & $3.21 \pm 0.79$ & \checkmark &  \\
SPT-CL J2040-5725 & $310.0631$ & $-57.4287$ & $0.930$ & $6.38$ & $3.25 \pm 0.75$ & \checkmark &  \\
SPT-CL J2043-5035 & $310.8285$ & $-50.5929$ & $0.723$ & $7.81$ & $4.71 \pm 1.00$ & \checkmark &  \\
SPT-CL J2056-5459 & $314.2199$ & $-54.9892$ & $0.718$ & $6.05$ & $3.68 \pm 0.89$ & \checkmark &  \\
SPT-CL J2058-5608 & $314.5893$ & $-56.1454$ & $0.606$ & $5.02$ & $2.64 \pm 0.79$ & \checkmark &  \\
SPT-CL J2100-4548 & $315.0936$ & $-45.8057$ & $0.712$ & $4.84$ & $2.71 \pm 0.93$ & \checkmark &  \\
SPT-CL J2104-5224 & $316.2283$ & $-52.4044$ & $0.799$ & $5.32$ & $3.04 \pm 0.89$ & \checkmark &  \\
SPT-CL J2106-5844 & $316.5210$ & $-58.7448$ & $1.131$ & $22.08$ & $8.36 \pm 1.71$ & \checkmark &  \\
SPT-CL J2118-5055 & $319.7291$ & $-50.9329$ & $0.625$ & $5.62$ & $3.43 \pm 0.93$ & \checkmark &  \\
SPT-CL J2124-6124 & $321.1488$ & $-61.4141$ & $0.435$ & $8.21$ & $4.68 \pm 0.96$ & \checkmark &  \\
SPT-CL J2130-6458 & $322.7285$ & $-64.9764$ & $0.316$ & $7.57$ & $4.46 \pm 0.96$ & \checkmark &  \\
SPT-CL J2135-5726 & $323.9158$ & $-57.4415$ & $0.427$ & $10.43$ & $5.68 \pm 1.11$ & \checkmark &  \\
SPT-CL J2136-4704 & $324.1175$ & $-47.0803$ & $0.425$ & $6.17$ & $4.04 \pm 0.96$ & \checkmark &  \\
SPT-CL J2136-6307 & $324.2334$ & $-63.1233$ & $0.926$ & $6.25$ & $3.18 \pm 0.75$ & \checkmark &  \\
SPT-CL J2138-6007 & $324.5060$ & $-60.1324$ & $0.319$ & $12.64$ & $6.75 \pm 1.32$ & \checkmark &  \\
SPT-CL J2145-5644 & $326.4694$ & $-56.7477$ & $0.480$ & $12.30$ & $6.39 \pm 1.25$ & \checkmark &  \\
SPT-CL J2146-4633 & $326.6473$ & $-46.5505$ & $0.932$ & $9.59$ & $5.36 \pm 1.07$ & \checkmark &  \\
SPT-CL J2146-4846 & $326.5346$ & $-48.7774$ & $0.623$ & $5.88$ & $3.64 \pm 0.93$ & \checkmark &  \\
SPT-CL J2148-6116 & $327.1798$ & $-61.2791$ & $0.571$ & $7.27$ & $4.04 \pm 0.89$ & \checkmark &  \\
SPT-CL J2155-6048 & $328.9851$ & $-60.8072$ & $0.539$ & $5.24$ & $2.82 \pm 0.82$ & \checkmark &  \\
SPT-CL J2201-5956 & $330.4727$ & $-59.9473$ & $0.098$ & $13.99$ & - &  & 5 \\
SPT-CL J2248-4431 & $342.1907$ & $-44.5269$ & $0.351$ & $42.36$ & $17.97 \pm 2.18$ & \checkmark &  \\
SPT-CL J2300-5331 & $345.1765$ & $-53.5170$ & $0.262$ & $5.29$ & - & \checkmark &  \\
SPT-CL J2301-5546 & $345.4688$ & $-55.7758$ & $0.748$ & $5.19$ & $3.11 \pm 0.96$ & \checkmark &  \\
SPT-CL J2325-4111 & $351.3043$ & $-41.1959$ & $0.358$ & $12.50$ & $7.29 \pm 1.07$ & \checkmark &  \\
SPT-CL J2331-5051 & $352.9584$ & $-50.8641$ & $0.575$ & $8.04$ & $5.14 \pm 0.71$ & \checkmark &  \\
SPT-CL J2332-5358 & $353.1040$ & $-53.9733$ & $0.402$ & $7.30$ & $6.50 \pm 0.79$ & \checkmark &  \\
SPT-CL J2337-5942 & $354.3544$ & $-59.7052$ & $0.776$ & $14.94$ & $8.14 \pm 1.14$ & \checkmark &  \\
SPT-CL J2341-5119 & $355.2994$ & $-51.3328$ & $1.002$ & $9.65$ & $5.61 \pm 0.82$ & \checkmark &  \\
SPT-CL J2342-5411 & $355.6903$ & $-54.1887$ & $1.075$ & $6.18$ & $3.00 \pm 0.50$ & \checkmark &  \\
SPT-CL J2344-4243 & $356.1847$ & $-42.7209$ & $0.595$ & $27.44$ & $12.50 \pm 1.57$ & \checkmark &  \\
SPT-CL J2347-5158* & $356.9423$ & $-51.9766$ & $0.869$ & - & - & \checkmark &  \\
SPT-CL J2351-5452 & $357.8877$ & $-54.8753$ & $0.384$ & $4.89$ & $3.18 \pm 1.04$ &  & 6 \\
SPT-CL J2355-5056 & $358.9551$ & $-50.9367$ & $0.320$ & $5.89$ & $4.07 \pm 0.57$ & \checkmark &  \\
SPT-CL J2359-5009 & $359.9208$ & $-50.1600$ & $0.775$ & $6.35$ & $3.54 \pm 0.54$ & \checkmark &  
  \enddata
  % \tablenotetext{A}{Unconfirmed cluster candidate which is either
  % above the quoted redshift limit or a false detection.}
  \tablecomments{SPT ID of each cluster, right ascension and
    declination of its SZ center, and redshift $z$ (from Tables
    \ref{tab:meta} and \ref{tab:lit}, for reference). Also given are
    the SPT significance $\xi$ and the SZ-based SPT mass, marginalized
    over cosmological parameters as in \cite{reichardt12}, for those
    clusters at $z \ge 0.3$, except for two, as described in Section
    \ref{ss:obsspt}. Clusters marked with ** are reported here as SPT
    detections for the first time, and those with * are new
    discoveries.  } \tablerefs{(1) \cite{sifon12}; (2)
    \cite{girardi96}; (3) \citet[][]{struble99}; (4)
    \cite{barrena2002}; (5) \cite{katgert98}; (6)
    \cite{buckleygeer11}.}

\end{deluxetable*}

% \subsection{Optical and Infrared Imaging}\label{ss:obsoir}

\subsection{Optical Spectroscopy}\label{ss:obsopt}

The spectroscopic observations presented in this work are the first of
our ongoing follow-up program. The data were taken from 2008 to 2012
using the Gemini Multi Object Spectrograph \citep[GMOS;][]{hook04} on
Gemini South, the Focal Reducer and low dispersion Spectrograph
\citep[FORS2;][]{appenzeller98} on VLT Antu, the Inamori Magellan
Areal Camera and Spectrograph \citep[IMACS;][]{dressler06} on Magellan
Baade, and the Low Dispersion Survey Spectrograph
\citep[LDSS3\footnote{http://www.lco.cl/telescopes-information/\\magellan/instruments/ldss-3};][]{Allington-Smith94}
on Magellan Clay.

In order to place a large number of slitlets in the central region of
the cluster, most of the IMACS observations were conducted with the
Gladders Image-Slicing Multi-slit Option
(GISMO\footnote{http://www.lco.cl/telescopes-information/\\magellan/instruments/imacs/gismo/gismoquickmanual.pdf}). GISMO
optically remaps the central region of the IMACS field-of-view
(roughly $3.5' \times 3.2'$) to sixteen evenly-spaced regions of the
focal plane, allowing for a large density of slitlets in the cluster
core while minimizing slit collisions on the CCD.

Details about the observations pertaining to each cluster, including
the instrument, optical configuration, number of masks, total exposure
time, and measured spectral resolution are listed in Table
\ref{tab:observations}.

Optical and infrared follow-up imaging observations of SPT clusters
are presented alongside our group's photometric redshift methodology
in \cite{high10}, \cite{song12} and \cite{desai12}. Those photometric
redshifts (and in a few cases, spectroscopic redshifts from the
literature) were used to guide the design of the spectroscopic
observations.  Multislit masks were designed using the best imaging
available to us, usually a combination of ground-based \griz (on
Blanco/MOSAIC~II, Magellan/IMACS, Magellan/LDSS3, or $BVRI$ on Swope)
and {\it Spitzer}/IRAC $3.6 \mathrm{\mu m}$.  In addition,
spectroscopic observations at Gemini and VLT were preceded by at least
one-band ($r$ or $i$) pre-imaging for relative astrometry, and
two-band ($r$ and $i$) pre-imaging for red-sequence target selection
in the cases where the existing imaging was not deep enough. The
exposure times for this pre-imaging were chosen to reach a magnitude
depth for galaxy photometry of $m^\star + 1$ at $10\sigma$ at the
cluster redshift.

In designing the multislit masks, top priority for slit placement was
given to bright red-sequence galaxies \citep[the red sequence of SPT
clusters is discussed in the context of photometric redshifts
in][]{high10, song12}, as defined by their distance to either a
theoretical or an empirically-fit red-sequence model. The details
varied depending on the quality of the available imaging, the program
and the prioritization weighting scheme of the instrument's
mask-making software.  In many of the GISMO observations, blue
galaxies
% of magnitudes comparable to the bright red-sequence galaxies
were given higher priority than faint red galaxies because, especially
at high redshift, they were expected to be more likely to yield a
redshift. The results from the different red-sequence weighting
schemes are very similar, and few emission lines are found, even at $z
\gtrsim 1$ \citep[][these articles also provide more details about the
red-sequence nature of spectroscopic members]{brodwin10, foley11,
  stalder12}. The case of SPT-CL~J2040-4451 at $z=1.478$ is different
and redshifts were only obtained for emission-line galaxies
\citep{bayliss13}. In all cases, non-red-sequence objects were used to
fill out any remaining space in the mask.

The dispersers and filters, listed in Table \ref{tab:observations},
were chosen (within the uncertainty on the photo-$z$) to obtain low-
to medium-resolution spectra covering at least the wavelengths of the
main spectral features that we use to identify the galaxy redshifts:
[\ion{O}{2}] emission, and the \ion{Ca}{2} H\&K absorption lines and
break.

The spectroscopic exposure times (also in Table
\ref{tab:observations}) for GMOS and FORS2 observations were chosen to
reach $S/N = 5$ ($S/N = 3$) per spectral element just below the $4000
\mathrm \AA$ break for a red galaxy of magnitude $m^\star + 1$
($m^\star + 0.5$) at $z < 1$ ($z > 1$). Under the conditions
prevailing at the telescope during classical observing, the exposure
times for the Magellan observations were determined by a combination
of experience, real-time quick-look reductions, and airmass
limitations.
% and the vagaries of weather and instrumentation.

\begin{deluxetable*}{lclcccrrr}
  \tabletypesize{\scriptsize}
  \tablecaption{Observations \label{tab:observations}}
  \tablewidth{0pt} \tablehead{ \colhead{SPT ID} & \colhead{$z$} &
    \colhead{UT Date} & \colhead{Instrument} &
    \colhead{Disperser/Filter} & \colhead{Masks} & \colhead{$N$} &
    \colhead{$t_{\mathrm{exp}}$ (h)} & \colhead{Resolution (\AA)}
  } \\
  \startdata % name  z  inst disp n_masks t resolving-power
SPT-CL J0000-5748  & $0.702$  & 2010 Sep 07 & GMOS-S & R150\_G5326 & 2 & 26 & 1.33 & 23.7 \\
SPT-CL J0014-4952  & $0.752$  & 2011 Aug 21 & FORS2 & GRIS\_300I/OG590 & 2 & 29 & 2.83  & 13.5 \\
SPT-CL J0037-5047  & $1.026$  & 2011 Aug 22 & FORS2 & GRIS\_300I/OG590 & 2 & 18 & 5.00  & 13.5 \\
SPT-CL J0040-4407  & $0.350$  & 2011 Sep 29 & GMOS-S & B600\_G5323 & 2 & 36 &  1.17 & 5.7 \\
SPT-CL J0118-5156  & $0.705$  & 2011 Sep 28 & GMOS-S & R400\_G5325, N\&S & 2 & 14 & 2.53  & 9.0 \\
SPT-CL J0205-5829  & $1.322$  & 2011 Sep 25 & IMACS & Gri-300-26.7/WB6300-950, f/2 & 1 & 9 & 11.00  & 5.2 \\
SPT-CL J0205-6432  & $0.744$  & 2011 Sep 30 & GMOS-S & R400\_G5325, N\&S & 2 & 15 & 2.67 & 9.0 \\
SPT-CL J0233-5819  & $0.664$  & 2011 Sep 29 & GMOS-S & R400\_G5325, N\&S & 1 & 10 & 1.33 & 9.0 \\
SPT-CL J0234-5831  & $0.415$  & 2010 Oct 08 & IMACS/GISMO & Gra-300-4.3/Z1-430-675, f/4 & 1 & 22 &  1.50 & 6.5 \\
SPT-CL J0240-5946  & $0.400$  & 2010 Oct 09 & IMACS/GISMO & Gra-300-4.3/Z1-430-675, f/4 & 1 & 25 & 1.00  & 6.4 \\
SPT-CL J0245-5302  & $0.300$  & 2011 Sep 29 & GMOS-S & B600\_G5323 & 2 & 29 & 0.83  & 7.0 \\
SPT-CL J0254-5857  & $0.437$  & 2010 Oct 08 & IMACS/GISMO & Gra-300-4.3/Z1-430-675, f/4 & 1 & 35 & 1.50  & 6.9 \\
SPT-CL J0257-5732  & $0.434$  & 2010 Oct 09 & IMACS/GISMO & Gra-300-4.3/Z1-430-675, f/4 & 1 & 22 & 1.50  & 6.6 \\
SPT-CL J0317-5935  & $0.469$  & 2010 Oct 09 & IMACS/GISMO & Gra-300-4.3/Z1-430-675, f/4 & 1 & 17 &  1.63 & 6.6 \\
SPT-CL J0433-5630  & $0.692$  & 2011 Jan 28 & IMACS/GISMO & Gri-300-17.5/Z2-520-775, f/2 & 1 & 22 & 1.00  & 5.7 \\
SPT-CL J0438-5419  & $0.422$  & 2011 Sep 28 & GMOS-S & R400\_G5325 & 1 & 18 & 0.75  & 9.0 \\
SPT-CL J0449-4901  & $0.790$  & 2011 Jan 28 & IMACS/GISMO & Gri-300-26.7/WB6300-950, f/2 & 1 & 20 & 1.63  & 5.6 \\
SPT-CL J0509-5342  & $0.462$  & 2009 Dec 12 & GMOS-S & R150\_G5326 & 2 & 18 & 1.00 & 23.7 \\
                                       &                  & 2012 Mar 23 & FORS2 & GRIS\_300V/GG435 & 1 & 4 & 2.37 & 13.7 \\
SPT-CL J0511-5154  & $0.645$  & 2011 Sep 30 & GMOS-S & R400\_G5325, N\&S & 2 & 15 & 2.67 & 9.0 \\
SPT-CL J0516-5430  & $0.294$  & 2010 Sep 17 & IMACS/GISMO & Gra-300-4.3/Z1-430-675, f/4 & 2 & 48 & 1.67 & 6.7 \\
SPT-CL J0528-5300  & $0.769$  & 2010 Jan 13 & GMOS-S  & R150\_G5326 & 2 & 20 &  3.00 & 23.7 \\
SPT-CL J0533-5005  & $0.881$  & 2008 Dec 05 & LDSS3 & VPH-Red & 1 & 4 & 0.63 & 5.4 \\
SPT-CL J0534-5937  & $0.576$  & 2008 Dec 05 & LDSS3 & VPH-Red & 1 & 3 & 0.45 & 5.5 \\
SPT-CL J0546-5345  & $1.066$  & 2010 Feb 11 & IMACS/GISMO & Gri-300-26.7/WB6300-950, f/2 & 1 & 21 & 3.00  & 5.7 \\
%                                         & GMOS-S & R150\_G5326 & 2 &  & 4.00  & 23.7 \\
SPT-CL J0551-5709  & $0.424$  & 2010 Sep 17 & IMACS/GISMO & Gra-300-4.3/Z1-430-675, f/4 & 2 & 34 & 1.42  & 6.8 \\
SPT-CL J0559-5249  & $0.609$  & 2009 Dec 07 & GMOS-S & R150\_G5326 & 2 & 37 & 1.33  & 23.7 \\
SPT-CL J2022-6323  & $0.383$  & 2010 Oct 09 & IMACS/GISMO & Gra-300-4.3/Z1-430-675, f/4  & 1 & 37 & 1.17  & 6.7 \\
SPT-CL J2032-5627  & $0.284$  & 2010 Oct 08 & IMACS/GISMO & Gra-300-4.3/Z1-430-675, f/4 & 1 & 31 & 1.17 & 6.8 \\
SPT-CL J2040-4451  & $1.478$  & 2012 Sep 15 & IMACS & Gri-300-26.7, f/2 & 2 & 14 & 11.30 & 9.3 \\
SPT-CL J2040-5725  & $0.930$  & 2010 Aug 13 & IMACS/GISMO & Gri-300-26.7/WB6300-950, f/2 & 1 & 5 & 3.00  & 5.0 \\
SPT-CL J2043-5035  & $0.723$  & 2011 Aug 27 & FORS2 & GRIS\_300I/OG590 & 2 & 21 & 4.00  & 13.5 \\
SPT-CL J2056-5459  & $0.719$  & 2010 Aug 14 & IMACS/GISMO & Gri-300-26.7/WB6300-950, f/2 & 1 & 12 & 2.00  & 5.3 \\
SPT-CL J2058-5608  & $0.607$  & 2011 Oct 01 & GMOS-S & R400\_G5325 & 2 & 9 & 1.67  & 9.0 \\
SPT-CL J2100-4548  & $0.712$  & 2011 Jul 23 & FORS2 & GRIS\_300I/OG590 & 2 & 19 & 1.50  & 13.5 \\
SPT-CL J2104-5224  & $0.799$  & 2011 Jul 21 & FORS2 & GRIS\_300I/OG590 & 2 & 23 & 2.83  & 13.5 \\
SPT-CL J2106-5844  & $1.131$  & 2010 Dec 08 & FORS2 & GRIS\_300I/OG590 & 1 & 15 & 3.00  & 13.5 \\
                                          & & 2010 Jun 07 & IMACS/GISMO & Gri-300-26.7/WB6300-950, f/2 & 1 & 4 & 8.00  & 4.5 \\
SPT-CL J2118-5055 & $0.625$  & 2011 May 26 & FORS2 & GRIS\_300I/OG590 & 2 & 22 & 1.33  & 13.5 \\
                                          &  & 2011 Sep 27 & GMOS-S & R400\_G5325, N\&S & 1 & 3 & 1.20 & 9.0 \\
SPT-CL J2124-6124 & $0.435$  & 2009 Sep 25 & IMACS/GISMO & Gra-300-4.3/Z1-430-675, f/4 & 1 & 24 & 1.50  & 7.0 \\
%SPT-CL J2130-6458 & $0.316$  & 2009 Sep 25, 2010 Sep 17 & IMACS/GISMO & Gra-300-4.3/Z1-430-675, f/4 & 2 &  & 2.00  & 7.1 \\
SPT-CL J2130-6458 & $0.316$  & 2010 Sep 17 & IMACS/GISMO & Gra-300-4.3/Z1-430-675, f/4 & 2 & 47 & 2.00  & 7.1 \\
SPT-CL J2135-5726 & $0.427$  & 2010 Sep 16 & IMACS/GISMO & Gra-300-4.3/Z1-430-675, f/4 & 1 & 33 & 1.00  & 6.8 \\
SPT-CL J2136-4704 & $0.425$  & 2011 Sep 29 & GMOS-S  & R400\_G5325 & 2 & 24 & 1.67  & 9.0 \\
SPT-CL J2136-6307 & $0.926$  & 2010 Aug 14 & IMACS/GISMO & Gri-300-26.7/WB6300-950, f/2 & 1 & 10 & 2.00  & 5.0 \\
SPT-CL J2138-6007 & $0.319$  & 2010 Sep 17 & IMACS/GISMO & Gra-300-4.3/Z1-430-675, f/4 & 1 & 34 & 1.50  & 6.8 \\
%SPT-CL J2145-5644 & $0.480$  & 2009 Sep 25, 2010 Sep 16 & IMACS/GISMO & Gra-300-4.3/Z1-430-675, f/4 & 2 &  & 2.92  & 7.4, 6.6 \\
SPT-CL J2145-5644 & $0.480$  & 2010 Sep 16 & IMACS/GISMO & Gra-300-4.3/Z1-430-675, f/4 & 2 & 37 & 2.92  & 7.4 \\
SPT-CL J2146-4633 & $0.931$  & 2011 Sep 25 & IMACS & Gri-300-26.7/WB6300-950, f/2 & 1 & 17 & 3.00  & 4.7 \\
SPT-CL J2146-4846 & $0.623$  & 2011 Oct 01 & GMOS-S & R400\_G5325 & 2 & 26 & 2.33  & 9.0 \\
SPT-CL J2148-6116 & $0.571$  & 2009 Sep 25 & IMACS/GISMO & Gra-300-4.3/Z1-430-675, f/4 & 1 & 30 & 1.50  & 7.1 \\
SPT-CL J2155-6048 & $0.539$  & 2011 Oct 01 & GMOS-S  & R400\_G5325 & 2 & 25 & 1.50  & 9.0 \\
SPT-CL J2248-4431 & $0.351$  & 2009 Jul 12 & IMACS/GISMO & Gra-300-4.3/Z1-430-675, f/4 & 1 & 15 & 1.33  & 10.9 \\
SPT-CL J2300-5331 & $0.262$  & 2010 Oct 08 & IMACS/GISMO & Gra-300-4.3/Z1-430-675, f/4 & 1 & 24 & 1.00  & 6.8 \\
SPT-CL J2301-5546 & $0.748$  & 2010 Aug 14 & IMACS/GISMO & Gri-300-26.7/WB6300-950, f/2 & 1 & 11 & 2.00  & 5.4 \\
SPT-CL J2325-4111 & $0.358$  & 2011 Sep 28 & GMOS-S & B600\_G5323 & 2 & 33 &  1.00 & 5.7 \\
SPT-CL J2331-5051 & $0.575$  & 2008 Dec 05 & LDSS3 & VPH-Red & 2 & 6 & 1.00 & 5.5 \\
                                         & & 2010 Sep 09 & GMOS-S & R150\_G5326 & 2 & 28 & 1.00  & 23.7  \\
                                         & & 2010 Oct 09 & IMACS/GISMO & Gra-300-4.3/Z2-520-775, f/4 & 2 & 62 &  3.50 & 6.7 \\
SPT-CL J2332-5358 & $0.403$  & 2009 Jul 12 & IMACS/GISMO & Gri-200-15.0/WB5694-9819, f/2 & 1 & 24 & 1.50  & 18.1 \\
                                      &                  & 2010 Sep 05 & FORS2 & GRIS\_300V & 2 & 29 & 4.38  & 13.7 \\
SPT-CL J2337-5942 & $0.776$  & 2010 Aug 14 & GMOS-S & R150\_G5326 & 2 & 19  & 3.00  & 23.7 \\
SPT-CL J2341-5119 & $1.003$  & 2010 Aug 14 & GMOS-S & R150\_G5326 & 2 & 15 & 6.00  & 23.7 \\
SPT-CL J2342-5411 & $1.075$  & 2010 Sep 09 & GMOS-S & R150\_G5326 & 1 & 11 & 3.00  & 23.7 \\
SPT-CL J2344-4243 & $0.595$ & 2011 Sep 30 & GMOS-S & R400\_G5325 & 2 & 32 & 2.33  & 9.0 \\
SPT-CL J2347-5158 & $0.869$ & 2010 Aug 13 & IMACS/GISMO & Gri-300-26.7/WB6300-950, f/2 & 1 & 12 & 2.50  & 5.0 \\
SPT-CL J2355-5056 & $0.320$ & 2010 Sep 17 & IMACS/GISMO & Gra-300-4.3/Z1-430-675, f/4 & 1 & 37 & 1.50  & 7.0 \\
SPT-CL J2359-5009 & $0.775$ & 2009 Nov 22 & GMOS-S & R150\_G5326 & 2 & 7 & 1.33  & 23.7 \\
                                       &  & 2010 Aug 14 & IMACS/GISMO & Gri-300-26.7/WB6300-950, f/2 & 1 & 22 & 2.00 & 5.4 \\
  \enddata
  \tablecomments{The instruments used for our observations are IMACS
    on Magellan Baade, LDSS3 on Magellan Clay, GMOS-S on Gemini South,
    and FORS2 on VLT Antu. The UT date of observation, details of the
    configuration and the number of observed multislit masks are
    given, as well as the number of member redshifts retrieved from
    the observation ($N \equiv N_{\mathrm{members}}$), and the total
    spectroscopic exposure time for all masks, $t_{\mathrm{exp}}$, in
    hours. The spectral resolution is the FWHM of sky lines in
    Angstroms, measured in the science exposures.}
\end{deluxetable*}

% \begin{deluxetable*}{lcccccc|}
%   \tabletypesize{\normalsize} \tablecaption{Observing
%   programs \label{tab:programs}} \tablewidth{0pt} \tablehead{
%   \colhead{\#} & \colhead{Telescope/Instrument} & \colhead{Semester}
%   & \colhead{P.I.} & \colhead{Program ID} &
%   \colhead{$N_{\mathrm{clusters}}$}
% } \\
%   \startdata
%   A & Magellan/IMACS & 2009A & Stubbs & Harvard-CfA & 2 \\
%   B & Gemini/GMOS-S & 2009B & Mohr & GS-2009B-Q-16 (NOAO) & 9 \\
%   C & Magellan/IMACS & 2009B & Stubbs & Harvard-CfA & 4 \\
%   D & Magellan/IMACS & 2010A & Brodwin & Harvard-CfA & 1 \\
%   E & Magellan/IMACS & 2010A & Foley & Harvard-CfA & 1 \\
%   F  & Magellan/IMACS & 2010B & Brodwin & Harvard-CfA & 6 \\
%   G & Magellan/IMACS & 2010B & Clocchiatti & CNTAC & 8 \\
%   H & Magellan/IMACS & 2010B & Stubbs & Harvard-CfA & 8 \\
%   I & VLT/FORS2 & 2010B & Carlstrom & 286.A-5021(DDT) & 1 \\
%   J & Magellan/IMACS & 2011A & Brodwin & Harvard-CfA & 2 \\
%   K & Magellan/IMACS & 2011B & Brodwin & Harvard-CfA & 2 \\
%   L & VLT/FORS2 & 2011A & Bazin & 087.A-0843 & 6 \\
%   M & Gemini/GMOS-S & 2011A & Stubbs & GS-2011A-C-3 (NOAO) & 9 \\
%   N & Gemini/GMOS-S & 2011B & Stubbs & GS-2011B-C-6 (NOAO) & 5 \\
%   \enddata
% \end{deluxetable*}

\subsubsection{Data Processing}\label{s:data}

We used the COSMOS reduction
package\footnote{http://code.obs.carnegiescience.edu/cosmos}
\citep{kelson03} for CCD reductions of IMACS and LDSS3 data, and
standard IRAF routines and
XIDL\footnote{http://www.ucolick.org/\~{}xavier/IDL/} routines for
GMOS and FORS2.
% The spectra are extracted using the optimal algorithm of
% \citet{horne86}.
Flux calibration and telluric line removal were performed using the
well-exposed continua of spectrophotometric standard stars
\citep{wade88, Foley03}.  Wavelength calibration is based on arc lamp
exposures, obtained at night in between science exposures in the case
of IMACS and LDSS3, and during daytime in the same configuration as
for science exposures for GMOS and FORS2.  In the case of daytime arc
frames, the wavelength calibration was refined using sky lines in the
science exposures.

The redshift determination was performed using cross-correlation with
the {\it fabtemp97} template in the RVSAO package for IRAF
\citep{kurtz98} or a proprietary template fitting method using the
SDSS DR2 templates, and validated by agreement with visually
identified absorption or emission features. A single method was used
for each cluster depending on the reduction workflow, and both perform
similarly. Comparison between the redshifts obtained from the
continuum and emission-line redshifts, when both are available from
the same spectrum, shows that the uncertainties on individual
redshifts \citep[twice the RVSAO uncertainty, see e.g.][]{quintana00}
correctly represent the statistical uncertainty of the fit.

\subsection{A few-$N_{\mathrm{members}}$ Spectroscopic
  Strategy}\label{ss:strategy}

Modern multi-object spectrographs use slit masks, so that the
investment in telescope time is quantized by how many masks are
allocated to each cluster. The optimization problem is, therefore, to
allocate the observation of $m$ masks across $n$ clusters so as to
minimize the uncertainty on the ensemble cluster mass normalization.

We pursue a strategy for spectroscopic observations informed by the
expectation \citep[from $N$-body simulations; see e.g.][]{kasun05,
  white10, saro12} that line-of-sight projection effects induce an
unavoidable intrinsic scatter of 12\% in log dispersion ($\ln \sigma$)
at fixed mass, implying a 35\% scatter in dynamical mass \citep[][see
Equation \ref{eq:saro_scatter} of the present paper]{saro12}.  As this
35\% intrinsic scatter needs to be added to the dynamical mass
uncertainty of any one cluster, for the purpose of mass calibration,
obtaining coarser dispersions on more clusters is more informative
than measuring higher-precision velocity dispersions on a few
clusters. Considering the results of those simulations and the
experience encapsulated in the velocity dispersion literature
\citep[e.g. ][]{girardi93}, we have adopted a target of
$N_{\mathrm{members}} \sim 20 - 30$, where $N_{\mathrm{members}}$ is
the number of spectroscopic member galaxies in a cluster.  This target
range of $N_{\mathrm{members}}$ can be obtained by observing two masks
per cluster on the spectrographs available to us.\footnote{Some of the
  observations presented here depart from this model and have only one
  slitmask with correspondingly fewer members. In some cases the
  second mask has yet to be observed, while in others observations
  were undertaken with different objectives (e.g. the identification
  and characterization of high-redshift clusters, the follow-up of
  bright sub-millimeter galaxies, and long slit observations from the
  early days of our follow-up program). Finally, some clusters of
  special interest were targeted with more than $2$ masks.} The use of
a red-sequence selection to target likely cluster members is a
necessary feature of this strategy, as a small number of multislit
masks only allows us to target a small fraction of the galaxies in the
region of the sky around the SZ center.

In discussions throughout this paper, we often use a
$N_{\mathrm{members}} \geq 15$ cut. We note that this number is chosen
somewhat arbitrarily for the conservative exclusion of systems with
very few members. As we will see in the resampling analysis of Section
\ref{s:resampling}, no special statistical transition happens at
$N_{\mathrm{members}} = 15$, and dispersions with fewer members could
potentially be used for reliable mass estimates.

Recent simulations and our data suggest that this choice of few-member
strategy may increase the scatter due to systematics in the measured
dispersions. This is discussed in Section \ref{s:estimators}.

%%%%%%%%%%%%%%%%%%%%
%% Results %%
%%%%%%%%%%%%%%%%%%%%

\section{Results}\label{s:results}

\subsection{Individual Galaxy Redshifts}
The full sample of redshifts for both member and non-member galaxies
is available in electronic format. In Table \ref{tab:redshifts}, we
present a subset composed of central galaxies, for the 50 clusters
where we have the central galaxy redshift. We have visually selected
the central galaxy for each cluster to be a large, bright, typically
cD-type galaxy that is close to the SZ center and that appears to be
central to the distribution of galaxies. For each galaxy, the table
lists the SPT ID of the associated cluster, a galaxy ID, right
ascension and declination, the redshift and redshift-measurement
method, and notable spectral features.

% Galaxy ID is J... ; see Brodwin et al.  Include method

\begin{deluxetable*}{lcrrcll}
  \tabletypesize{\scriptsize} \tablecaption{Galaxy redshifts
    \label{tab:redshifts}} \tablewidth{0pt} \tablehead{
    \colhead{Associated SPT ID} & \colhead{Galaxy ID} &
    \colhead{Galaxy R.A.} & \colhead{Galaxy Dec.} & \colhead{$z$} &
    % \colhead{$z_{\mathrm{emission}}$} &
    \colhead{$z$ method} & \colhead{Spectral features}
    \\
    \colhead{} & \colhead{} & \colhead{(J2000 deg.)} & \colhead{(J2000
      deg.)} & \colhead{} &
    % \colhead{} &
    \colhead{}
  } \\
  \startdata
  % SPT-CL J0245-5302 & J024515.79-530108.5 & 41.316 & -53.019 &
  % $0.3022 \pm 0.0001$ & $0.3022 \pm 0.0001$ & RVSAO & [\ion{O}{2}],
  % \ion{Ca}{2} H\&K \\
  SPT-CL J0000-5748 & J000059.99-574832.7 & $0.2500$ & $-57.8091$ & $0.7007 \pm 0.0002$ & template & [\ion{O}{2}], \ion{Ca}{2} H\&K \\
SPT-CL J0037-5047 & J003747.30-504718.9 & $9.4471$ & $-50.7886$ & $1.0302 \pm 0.0002$ & template & \ion{Ca}{2} H\&K \\
SPT-CL J0118-5156 & J011824.76-515628.6 & $19.6032$ & $-51.9413$ & $0.7021 \pm 0.0004$ & rvsao-xc & \ion{Ca}{2} H\&K \\
SPT-CL J0205-5829 & J020548.26-582848.4 & $31.4511$ & $-58.4801$ & $1.3218 \pm 0.0002$ & rvsao-xc & \ion{Ca}{2} H\&K \\
SPT-CL J0205-6432 & J020507.83-643226.8 & $31.2827$ & $-64.5408$ & $0.7430 \pm 0.0001$ & rvsao-xc & \ion{Ca}{2} H\&K \\
SPT-CL J0233-5819 & J023300.97-581937.0 & $38.2540$ & $-58.3270$ & $0.6600 \pm 0.0001$ & rvsao-xc & \ion{Ca}{2} H\&K \\
SPT-CL J0234-5831 & J023442.26-583124.7 & $38.6761$ & $-58.5235$ & $0.4146 \pm 0.0001$ & rvsao-xc & [\ion{O}{2}], \ion{Ca}{2} H\&K \\
SPT-CL J0240-5946 & J024038.38-594548.5 & $40.1599$ & $-59.7635$ & $0.4027 \pm 0.0002$ & rvsao-xc & \ion{Ca}{2} H\&K \\
SPT-CL J0245-5302 & J024524.82-530145.3 & $41.3534$ & $-53.0293$ & $0.3028 \pm 0.0001$ & rvsao-xc & \ion{Ca}{2} H\&K \\
SPT-CL J0254-5857 & J025415.47-585710.6 & $43.5645$ & $-58.9530$ & $0.4373 \pm 0.0001$ & rvsao-xc & \ion{Ca}{2} H\&K \\
SPT-CL J0257-5732 & J025720.95-573254.0 & $44.3373$ & $-57.5484$ & $0.4329 \pm 0.0001$ & rvsao-xc & \ion{Ca}{2} H\&K \\
SPT-CL J0317-5935 & J031715.84-593529.0 & $49.3160$ & $-59.5914$ & $0.4677 \pm 0.0001$ & rvsao-xc & \ion{Ca}{2} H\&K \\
SPT-CL J0433-5630 & J043301.03-563109.4 & $68.2543$ & $-56.5193$ & $0.6946 \pm 0.0002$ & rvsao-xc & \ion{Ca}{2} H\&K \\
SPT-CL J0438-5419 & J043817.62-541920.6 & $69.5734$ & $-54.3224$ & $0.4217 \pm 0.0002$ & rvsao-xc & \ion{Ca}{2} H\&K \\
SPT-CL J0449-4901 & J044904.03-490139.1 & $72.2668$ & $-49.0275$ & $0.7949 \pm 0.0002$ & rvsao-xc & \ion{Ca}{2} H\&K \\
SPT-CL J0509-5342 & J050921.37-534212.7 & $77.3390$ & $-53.7035$ & $0.4616 \pm 0.0002$ & template & [\ion{O}{2}], \ion{Ca}{2} H\&K \\
SPT-CL J0511-5154 & J051142.95-515436.6 & $77.9290$ & $-51.9102$ & $0.6488 \pm 0.0002$ & rvsao-xc & \ion{Ca}{2} H\&K \\
SPT-CL J0516-5430 & J051637.33-543001.5 & $79.1556$ & $-54.5004$ & $0.2970 \pm 0.0002$ & rvsao-xc & \ion{Ca}{2} H\&K \\
SPT-CL J0528-5300 & J052805.29-525953.1 & $82.0220$ & $-52.9981$ & $0.7670 \pm 0.0002$ & template & \ion{Ca}{2} H\&K \\
SPT-CL J0534-5937 & J053430.04-593653.8 & $83.6252$ & $-59.6150$ & $0.5757 \pm 0.0002$ & rvsao-xc & \ion{Ca}{2} H\&K \\
SPT-CL J0551-5709 & J055135.58-570828.6 & $87.8983$ & $-57.1413$ & $0.4243 \pm 0.0002$ & rvsao-xc & \ion{Ca}{2} H\&K \\
SPT-CL J0559-5249 & J055943.19-524926.2 & $89.9300$ & $-52.8240$ & $0.6104 \pm 0.0002$ & template & \ion{Ca}{2} H\&K \\
SPT-CL J2022-6323 & J202209.82-632349.3 & $305.5409$ & $-63.3970$ & $0.3736 \pm 0.0001$ & rvsao-em & [\ion{O}{2}] \\
SPT-CL J2032-5627 & J203214.04-562612.4 & $308.0585$ & $-56.4368$ & $0.2844 \pm 0.0002$ & rvsao-xc & \ion{Ca}{2} H\&K \\
SPT-CL J2043-5035 & J204317.52-503531.2 & $310.8230$ & $-50.5920$ & $0.7225 \pm 0.0005$ & template & \ion{Ca}{2} H\&K \\
SPT-CL J2056-5459 & J205653.57-545909.1 & $314.2232$ & $-54.9859$ & $0.7151 \pm 0.0002$ & rvsao-xc & \ion{Ca}{2} H\&K \\
SPT-CL J2058-5608 & J205822.28-560847.2 & $314.5928$ & $-56.1465$ & $0.6061 \pm 0.0002$ & rvsao-xc & [\ion{O}{2}], \ion{Ca}{2} H\&K \\
SPT-CL J2100-4548 & J210023.85-454834.6 & $315.0994$ & $-45.8096$ & $0.7148 \pm 0.0002$ & template & \ion{Ca}{2} H\&K \\
SPT-CL J2118-5055 & J211853.24-505559.5 & $319.7218$ & $-50.9332$ & $0.6253 \pm 0.0002$ & template & \ion{Ca}{2} H\&K \\
SPT-CL J2124-6124 & J212437.81-612427.7 & $321.1576$ & $-61.4077$ & $0.4375 \pm 0.0001$ & rvsao-xc & \ion{Ca}{2} H\&K \\
SPT-CL J2130-6458 & J213056.21-645840.4 & $322.7342$ & $-64.9779$ & $0.3161 \pm 0.0002$ & rvsao-xc & \ion{Ca}{2} H\&K \\
SPT-CL J2135-5726 & J213537.41-572630.7 & $323.9059$ & $-57.4419$ & $0.4305 \pm 0.0002$ & rvsao-xc & \ion{Ca}{2} H\&K \\
SPT-CL J2136-6307 & J213653.72-630651.5 & $324.2239$ & $-63.1143$ & $0.9224 \pm 0.0002$ & rvsao-xc & \ion{Ca}{2} H\&K \\
SPT-CL J2138-6007 & J213800.82-600753.8 & $324.5034$ & $-60.1316$ & $0.3212 \pm 0.0002$ & rvsao-xc & \ion{Ca}{2} H\&K \\
SPT-CL J2145-5644 & J214551.96-564453.5 & $326.4665$ & $-56.7482$ & $0.4813 \pm 0.0003$ & rvsao-xc & \ion{Ca}{2} H\&K \\
SPT-CL J2146-4633 & J214635.34-463301.7 & $326.6472$ & $-46.5505$ & $0.9282 \pm 0.0002$ & rvsao-xc & \ion{Ca}{2} H\&K \\
SPT-CL J2146-4846 & J214605.93-484653.3 & $326.5247$ & $-48.7815$ & $0.6177 \pm 0.0001$ & rvsao-xc & \ion{Ca}{2} H\&K \\
SPT-CL J2148-6116 & J214838.82-611555.9 & $327.1617$ & $-61.2655$ & $0.5649 \pm 0.0002$ & rvsao-xc & \ion{Ca}{2} H\&K \\
SPT-CL J2155-6048 & J215555.46-604902.8 & $328.9811$ & $-60.8175$ & $0.5419 \pm 0.0001$ & rvsao-xc & \ion{Ca}{2} H\&K \\
SPT-CL J2248-4431 & J224843.98-443150.8 & $342.1833$ & $-44.5308$ & $0.3482 \pm 0.0001$ & rvsao-xc & \ion{Ca}{2} H\&K \\
SPT-CL J2300-5331 & J230039.69-533111.4 & $345.1654$ & $-53.5198$ & $0.2630 \pm 0.0002$ & rvsao-xc & \ion{Ca}{2} H\&K \\
SPT-CL J2325-4111 & J232511.70-411213.7 & $351.2988$ & $-41.2038$ & $0.3624 \pm 0.0003$ & rvsao-xc & \ion{Ca}{2} H\&K \\
SPT-CL J2331-5051 & J233151.13-505154.1 & $352.9631$ & $-50.8650$ & $0.5786 \pm 0.0002$ & rvsao-xc & \ion{Ca}{2} H\&K \\
SPT-CL J2332-5358 & J233227.48-535828.2 & $353.1145$ & $-53.9745$ & $0.4041 \pm 0.0002$ & rvsao-xc & \ion{Ca}{2} H\&K \\
SPT-CL J2337-5942 & J233727.52-594204.8 & $354.3647$ & $-59.7014$ & $0.7788 \pm 0.0002$ & template & \ion{Ca}{2} H\&K \\
SPT-CL J2341-5119 & J234112.34-511944.9 & $355.3015$ & $-51.3291$ & $1.0050 \pm 0.0005$ & template & \ion{Ca}{2} H\&K \\
SPT-CL J2342-5411 & J234245.89-541106.1 & $355.6912$ & $-54.1850$ & $1.0808 \pm 0.0003$ & template & \ion{Ca}{2} H\&K \\
SPT-CL J2344-4243 & J234443.90-424312.1 & $356.1829$ & $-42.7200$ & $0.5981 \pm 0.0008$ & rvsao-em & [\ion{O}{2}] \\
SPT-CL J2355-5056 & J235547.48-505540.5 & $358.9479$ & $-50.9279$ & $0.3184 \pm 0.0002$ & rvsao-xc & \ion{Ca}{2} H\&K \\
SPT-CL J2359-5009 & J235942.81-501001.7 & $359.9284$ & $-50.1671$ & $0.7709 \pm 0.0003$ & rvsao-xc & \ion{Ca}{2} H\&K \\

  \enddata
  % \tablenotetext{a}{Redshift errors are twice those given by RVSAO.}
  \tablecomments{ Redshifts of individual galaxies. This is a partial
    listing, and the full table is available electronically. The
    entries listed here are the central galaxies, a subset of our
    observations.
    % the central galaxy selection is explained in Section
    % \ref{ss:bcg}.
    For each galaxy, the table lists the SPT ID of the associated
    cluster, a galaxy ID, right ascension and declination, the
    redshift $z$ and associated uncertainty, redshift measurement
    method, and notable spectral features. The labels of the ``$z$
    method'' column are ``rvsao-xc'' and ``rvsao-em'', respectively,
    for the RVSAO cross-correlation to absorption features and fit to
    emission lines, and ``template'' for an in-house template-fitting
    method using the SDSS DR2 templates.  }
\end{deluxetable*}

\subsection{Cluster Redshifts and Velocity Dispersions}\label{ss:meta}

Table \ref{tab:meta} lists the cluster redshifts and velocity
dispersions measured from the galaxy redshifts.

The cluster redshift $z$ is the biweight average \citep{beers90} of
member galaxy redshifts (see below) with an uncertainty given by the
standard error, as explained in Section \ref{ss:ci}.
% with an uncertainty derived from the statistical jackknife
% \citep[see e.g.][]{mosteller77}.
Once the cluster redshift is computed, the galaxy proper velocities
$v_i$ are obtained from their redshifts $z_i$ by $v_i =
c(z_i-z)/(1+z)$ \citep{danese80}.  The velocity dispersion
$\sigma_{\mathrm{BI}}$ is the square root of the biweight sample
variance of proper velocities, the uncertainty of which we found to be
well described by $0.92 \sigma_{\mathrm{BI}} /
\sqrt{N_{\mathrm{members}}-1}$ when including the effect of membership
selection (Section \ref{ss:ci}; see Section \ref{s:estimators} for the
formula of the biweight sample variance). We also report the
dispersion $\sigma_G$ determined from the gapper estimator, which is a
preferred measurement, according to \citet{beers90}, for those
clusters with fewer than $15$ member redshifts.

The cluster redshifts and velocity dispersions are calculated using
only galaxies identified as members, where membership is established
using iterative $3\sigma$ clipping on the velocities \citep{yahil77,
  mamon10, saro12}. The center at each iteration of $3\sigma$-clipping
is the biweight average, and $\sigma$ is calculated from the biweight
variance, or the gapper estimator in the case where there are fewer
than $15$ members. We do not make a hard velocity cut; the initial
estimate of $\sigma$ used in the iterative clipping is determined from
the galaxies located within $4000$ km\,s$^{-1}$ of the center, in the
rest frame.

\begin{figure*}
  \begin{center}
    \vspace*{-0.1in} \hspace*{-0.9in} \epsscale{1.4} \plotone{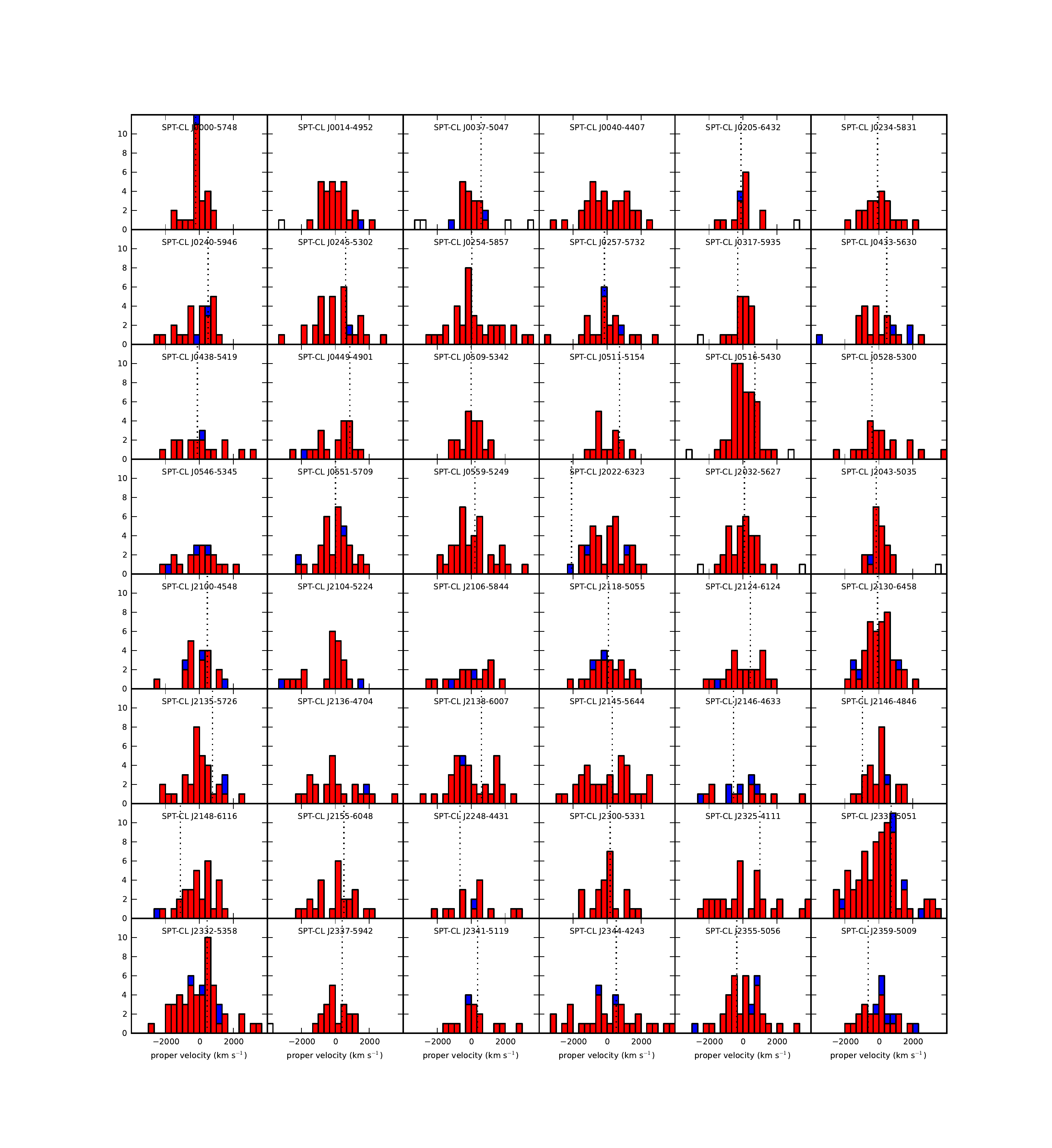}
    \vspace*{-0.8in}
\caption{Histograms showing the proper velocities of galaxies selected for each cluster, where colors correspond to: (red) passive galaxies, (blue) emission-line galaxies, and (white) non-members.  The central galaxy proper velocity is marked with a dotted line, though we note that this was not measured for six clusters, mostly at high redshift ($z \gtrsim 0.8$).
  \label{fig:vel_hist}}
\end{center}
\end{figure*}

Figure \ref{fig:vel_hist} shows the velocity histogram for each
cluster with 15 members or more, as well as an indication of
emission-line objects and our determination of member and non-member
galaxies.

Some entries in Table \ref{tab:meta} have a star-shaped flag $\star$
in the SPT ID column, which highlights possibly less reliable
dispersion measurements. These include $8$ clusters that have fewer
than $15$ measured member redshifts\footnote{Once again,
  $N_{\mathrm{members}} = 15$ is a somewhat arbitrary cutoff. See note
  at the end of Section \ref{ss:strategy}.}, as well as
SPT-CL~J0205-6432 with $N_{\mathrm{members}} = 15$, for which the
gapper and biweight dispersions differ by more than one sigma. Since
these are not independent measurements but rather two estimates of the
same quantity from the same data, we consider a one-sigma discrepancy
to be large and an indication that the sampling is inadequate.
% Finally, the value of the velocity dispersion of SPT-CL J2325-4111
% with $N_{\mathrm{members}} = 33$ is found to be unstable under
% resampling in Section \ref{ss:resampling}. A stable value could be
% extracted from the same data, but would require a modification to
% the membership selection algorithm that needs to be investigated in
% future work.

% \LongTables \tabletypesize{\scriptsize}
\begin{deluxetable*}{lrccrrr}
  \tabletypesize{\scriptsize} \tablecaption{Cluster redshifts and
    velocity dispersions
    \label{tab:meta}} \tablewidth{0pt} \tablehead{
    \colhead{SPT ID (and flag)} & \colhead{$N$}  & \colhead{$a$} & \colhead{$z$} & \colhead{$\sigma_{\mathrm{SPT}} $}  & \colhead{$\sigma_{\mathrm{G}} $} & \colhead{$\sigma_{\mathrm{BI}} $}  \\
    \colhead{} & \colhead{} & \colhead{($R_{200c, \mathrm{SPT}}$)}&
    \colhead{} & \colhead{(km\,s$^{-1}$)} & \colhead{(km\,s$^{-1}$)} &
    \colhead{(km\,s$^{-1}$)}
  } \\
  \startdata SPT-CL J0000-5748 & $26$ & $1.0$ & $0.7019(06)$ & $935$ & $598 \pm 109$ & $563 \pm 104$ \\
SPT-CL J0014-4952 & $29$ & $1.3$ & $0.7520(09)$ & $1004$ & $812 \pm 140$ & $811 \pm 141$ \\
SPT-CL J0037-5047 & $18$ & $1.6$ & $1.0262(09)$ & $945$ & $550 \pm 121$ & $555 \pm 124$ \\
SPT-CL J0040-4407 & $36$ & $0.4$ & $0.3498(10)$ & $1171$ & $1275 \pm 196$ & $1277 \pm 199$ \\
SPT-CL J0118-5156$\star$ & $14$ & $0.9$ & $0.7050(14)$ & $865$ & $948 \pm 239$ & $986 \pm 252$ \\
SPT-CL J0205-5829\tablenotemark{a} & $9$ & $1.3$ & $1.3219(16)$ & $1101$ & - & - \\
SPT-CL J0205-6432$\star$ & $15$ & $1.1$ & $0.7436(10)$ & $862$ & $687 \pm 167$ & $340 \pm 84$ \\
SPT-CL J0233-5819$\star$ & $10$ & $0.9$ & $0.6635(14)$ & $884$ & $783 \pm 238$ & $800 \pm 245$ \\
SPT-CL J0234-5831 & $22$ & $0.3$ & $0.4149(09)$ & $1076$ & $929 \pm 185$ & $926 \pm 186$ \\
SPT-CL J0240-5946 & $25$ & $0.4$ & $0.4004(09)$ & $948$ & $999 \pm 186$ & $1014 \pm 190$ \\
SPT-CL J0245-5302 & $30$ & - & $0.3001(10)$ & - & $1245 \pm 210$ & $1235 \pm 211$ \\
SPT-CL J0254-5857 & $35$ & $0.4$ & $0.4371(12)$ & $1071$ & $1431 \pm 223$ & $1483 \pm 234$ \\
SPT-CL J0257-5732 & $23$ & $0.6$ & $0.4337(12)$ & $800$ & $1220 \pm 237$ & $1157 \pm 227$ \\
SPT-CL J0317-5935 & $17$ & $0.5$ & $0.4691(06)$ & $832$ & $473 \pm 108$ & $473 \pm 109$ \\
SPT-CL J0433-5630 & $23$ & $0.7$ & $0.6919(15)$ & $817$ & $1260 \pm 244$ & $1232 \pm 242$ \\
SPT-CL J0438-5419 & $18$ & $0.5$ & $0.4223(16)$ & $1211$ & $1428 \pm 315$ & $1422 \pm 317$ \\
SPT-CL J0449-4901 & $20$ & $0.6$ & $0.7898(15)$ & $972$ & $1067 \pm 223$ & $1090 \pm 230$ \\
SPT-CL J0509-5342 & $21$ & $0.8$ & $0.4616(07)$ & $963$ & $670 \pm 136$ & $678 \pm 139$ \\
SPT-CL J0511-5154 & $15$ & $0.9$ & $0.6447(11)$ & $873$ & $778 \pm 189$ & $791 \pm 194$ \\
SPT-CL J0516-5430 & $48$ & $0.4$ & $0.2940(05)$ & $995$ & $721 \pm 96$ & $724 \pm 97$ \\
SPT-CL J0528-5300 & $21$ & $1.2$ & $0.7694(17)$ & $857$ & $1397 \pm 284$ & $1318 \pm 271$ \\
SPT-CL J0533-5005 & $4$ & $0.4$ & $0.8813(04)$ & $826$ & - & - \\
SPT-CL J0534-5937 & $3$ & $0.4$ & $0.5757(04)$ & $782$ & - & - \\
SPT-CL J0546-5345\tablenotemark{b} & $21$ & $0.8$ & $1.0661(18)$ & $1080$ & $1162 \pm 236$ & $1191 \pm 245$ \\
SPT-CL J0551-5709 & $34$ & $0.7$ & $0.4243(08)$ & $848$ & $962 \pm 152$ & $966 \pm 155$ \\
SPT-CL J0559-5249 & $37$ & $0.8$ & $0.6092(10)$ & $1072$ & $1135 \pm 172$ & $1146 \pm 176$ \\
SPT-CL J2022-6323 & $37$ & $0.4$ & $0.3832(08)$ & $847$ & $1076 \pm 163$ & $1080 \pm 166$ \\
SPT-CL J2032-5627 & $31$ & $0.3$ & $0.2841(06)$ & $898$ & $771 \pm 128$ & $777 \pm 131$ \\
SPT-CL J2040-4451\tablenotemark{c}\hspace{-0.3mm}$\star$ & $14$ & $1.5$ & $1.4780(25)$ & $989$ & $1111 \pm 280$ & $676 \pm 173$ \\
SPT-CL J2040-5725 & $5$ & $0.9$ & $0.9295(36)$ & $890$ & - & - \\
SPT-CL J2043-5035 & $21$ & $1.1$ & $0.7234(07)$ & $969$ & $509 \pm 104$ & $524 \pm 108$ \\
SPT-CL J2056-5459$\star$ & $12$ & $0.7$ & $0.7185(12)$ & $891$ & $704 \pm 193$ & $642 \pm 178$ \\
SPT-CL J2058-5608 & $9$ & $0.9$ & $0.6065(18)$ & $780$ & - & - \\
SPT-CL J2100-4548 & $20$ & $1.4$ & $0.7121(11)$ & $803$ & $874 \pm 183$ & $854 \pm 180$ \\
SPT-CL J2104-5224 & $23$ & $1.5$ & $0.7990(14)$ & $849$ & $1176 \pm 228$ & $1153 \pm 226$ \\
SPT-CL J2106-5844\tablenotemark{d} & $18$ & $1.0$ & $1.1312(21)$ & $1287$ & $1216 \pm 268$ & $1228 \pm 274$ \\
SPT-CL J2118-5055 & $25$ & $1.2$ & $0.6249(11)$ & $855$ & $981 \pm 182$ & $982 \pm 184$ \\
SPT-CL J2124-6124 & $24$ & $0.6$ & $0.4354(11)$ & $916$ & $1151 \pm 218$ & $1153 \pm 221$ \\
SPT-CL J2130-6458 & $47$ & $0.5$ & $0.3164(06)$ & $882$ & $897 \pm 120$ & $903 \pm 122$ \\
SPT-CL J2135-5726 & $33$ & $0.4$ & $0.4269(09)$ & $976$ & $1020 \pm 164$ & $1029 \pm 167$ \\
SPT-CL J2136-4704 & $24$ & $0.6$ & $0.4247(14)$ & $870$ & $1461 \pm 277$ & $1461 \pm 280$ \\
SPT-CL J2136-6307$\star$ & $10$ & $0.8$ & $0.9258(25)$ & $883$ & $1244 \pm 377$ & $1269 \pm 389$ \\
SPT-CL J2138-6007 & $34$ & $0.3$ & $0.3185(10)$ & $1014$ & $1269 \pm 201$ & $1303 \pm 209$ \\
SPT-CL J2145-5644 & $37$ & $0.5$ & $0.4798(13)$ & $1025$ & $1634 \pm 248$ & $1638 \pm 251$ \\
SPT-CL J2146-4633 & $18$ & $1.0$ & $0.9318(28)$ & $1057$ & $1840 \pm 406$ & $1817 \pm 405$ \\
SPT-CL J2146-4846 & $26$ & $0.9$ & $0.6230(08)$ & $872$ & $772 \pm 140$ & $784 \pm 144$ \\
SPT-CL J2148-6116 & $30$ & $0.6$ & $0.5707(09)$ & $894$ & $969 \pm 164$ & $966 \pm 165$ \\
SPT-CL J2155-6048 & $25$ & $0.9$ & $0.5393(12)$ & $787$ & $1157 \pm 215$ & $1162 \pm 218$ \\
SPT-CL J2248-4431 & $15$ & $0.2$ & $0.3512(15)$ & $1417$ & $1304 \pm 317$ & $1301 \pm 320$ \\
SPT-CL J2300-5331 & $24$ & $0.3$ & $0.2623(08)$ & $816$ & $887 \pm 168$ & $920 \pm 177$ \\
SPT-CL J2301-5546$\star$ & $11$ & $0.7$ & $0.7479(22)$ & $847$ & $1242 \pm 357$ & $1261 \pm 367$ \\
SPT-CL J2325-4111 & $33$ & $0.6$ & $0.3579(15)$ & $1048$ & $1926 \pm 310$ & $1921 \pm 312$ \\
SPT-CL J2331-5051 & $78$ & $0.9$ & $0.5748(08)$ & $970$ & $1363 \pm 141$ & $1382 \pm 145$ \\
SPT-CL J2332-5358 & $53$ & $0.6$ & $0.4020(08)$ & $1016$ & $1253 \pm 158$ & $1240 \pm 158$ \\
SPT-CL J2337-5942 & $19$ & $0.9$ & $0.7764(10)$ & $1181$ & $700 \pm 150$ & $707 \pm 153$ \\
SPT-CL J2341-5119 & $15$ & $1.1$ & $1.0025(17)$ & $1091$ & $1111 \pm 270$ & $959 \pm 236$ \\
SPT-CL J2342-5411$\star$ & $11$ & $1.5$ & $1.0746(27)$ & $893$ & $1278 \pm 368$ & $1268 \pm 369$ \\
SPT-CL J2344-4243\tablenotemark{e} & $32$ & $0.7$ & $0.5952(18)$ & $1317$ & $1824 \pm 298$ & $1878 \pm 310$ \\
SPT-CL J2347-5158$\star$ & $12$ & - & $0.8693(11)$ & - & $630 \pm 173$ & $635 \pm 176$ \\
SPT-CL J2355-5056 & $37$ & $0.5$ & $0.3200(08)$ & $856$ & $1124 \pm 170$ & $1104 \pm 169$ \\
SPT-CL J2359-5009 & $26$ & $0.9$ & $0.7747(11)$ & $889$ & $951 \pm 173$ & $950 \pm 175$ 

  \enddata
  % \tablenotetext{a}{Redshift errors are twice those given by RVSAO.}
  \tablecomments{This table shows the number $N$ ($\equiv
    N_{\mathrm{members}}$) of spectroscopic members as determined by
    iterative $3\sigma$ clipping, the aperture radius $a$ within which
    they were sampled in units of $R_{200c, \mathrm{SPT}}$, the robust
    biweight average redshift $z$ with the uncertainty in the last two
    digits in parentheses, the ``equivalent dispersion'' calculated
    from the SZ-based SPT mass $\sigma_{\mathrm{SPT}}$ (see Section
    \ref{ss:stack}), and the measured gapper scale
    $\sigma_{\mathrm{G}}$ and biweight dispersion
    $\sigma_{\mathrm{BI}}$. The star flag $\star$ in the SPT ID column
    indicates potentially less reliable dispersion measurements (see
    Section \ref{ss:meta}).  } \tablenotetext{a}{for
    SPT-CL~J0205-5829, see also \citet{stalder12}}
  \tablenotetext{b}{for SPT-CL~J0546-5345, see also \citet{brodwin10}}
  \tablenotetext{c}{for SPT-CL~J2040-4451, see also \citet{bayliss13}}
  \tablenotetext{d}{for SPT-CL~J2106-5844, see also \citet{foley11}}
  \tablenotetext{e}{for SPT-CL~J2344-4243, see also
    \citet{mcdonald12}}
\end{deluxetable*}

\subsubsection{The Stacked Cluster} \label{ss:stack} To examine the
ensemble phase-space galaxy selection, we produce a stacked cluster
from our observations; this stacked cluster will also be useful for
evaluating our confidence intervals via resampling (see Section
\ref{ss:ci}). We generate it in a way that is independent of cluster
membership determination.  As the calculation of the velocity
dispersion and membership selection are unavoidably intertwined, we
use the SPT mass --- the other uniform mass measurement that we have
for all clusters --- to normalize the velocities before stacking.

\begin{figure*}[htbp]
  % \epsscale{1.20}
  \plotone{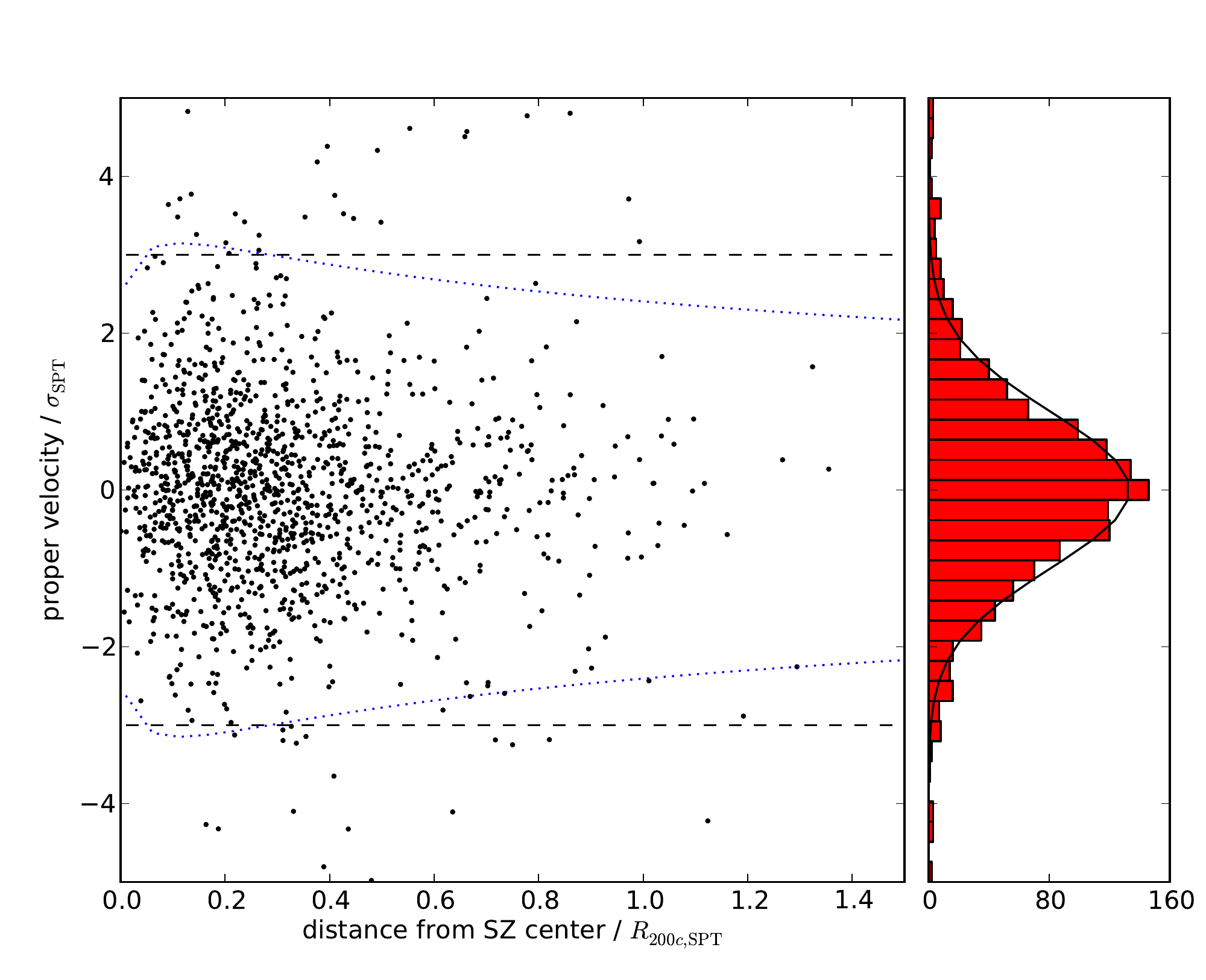}
  % \vskip-0.8in
\caption{ Stacked cluster, constructed using the dispersion equivalent to the SPT mass. {\it Left panel:} phase-space diagram of velocities. The black dashed line is a 3-sigma cut, and the blue dotted line is a radially-dependent $2.7\sigma(R)$ cut; see the text, Section \ref{ss:stack} for more details. These cuts would be applied iteratively in membership selection. {\it Right panel:} histogram of proper velocities, and Gaussian distribution with a mean of zero, a standard deviation of one, and an area equal to that of the histogram.
  \label{fig:stack_sz}}
\end{figure*}

We make a stacked proper-velocity distribution independent of any
measurement of the velocity dispersion by calculating the ``equivalent
dispersion'' from the SPT mass. We convert the $M_{500c,\mathrm{SPT}}$
to $M_{200c,\mathrm{SPT}}$ assuming an NFW profile and the
\cite{duffy08} concentration, and then convert the
$M_{200c,\mathrm{SPT}}$ to a $\sigma_{\mathrm{SPT}}$ (in km\,s$^{-1}$)
using the \cite{saro12} scaling relation. This $\sigma_{\mathrm{SPT}}$
is listed in Table \ref{tab:meta} for reference.  We also normalize
the distance to the SZ center by $R_{200c,\mathrm{SPT}}$. The
resulting phase-space diagram of the normalized proper velocities
$v_i/\sigma_{\mathrm{SPT}}$ vs $r_i/R_{200c,\mathrm{SPT}}$ is shown in
Figure \ref{fig:stack_sz}. For reference, different velocity cuts are
plotted. The black dashed line is a $3\sigma$ cut.
% The red dot-dashed line is the ``phase-space method''
% \citep{denhartog96, biviano06, white10} velocity cut, where the mass
% as a function of radius is calculated from an NFW profile of typical
% concentration instead of the dynamical mass; it is close to a
% $2\sigma$ cut.
The blue dotted line is a radially-dependent $2.7\sigma(R)$ cut, where
again the $\sigma(R)$ is from an NFW profile; this velocity cut is
found to be optimal for rejecting interlopers by \cite{mamon10}
(although when considering systems without red-sequence selection).
% All of these cuts would be applied iteratively in membership
% selection.
While $3\sigma$ clipping was a natural choice of membership selection
algorithm (given our sometimes small sample size for individual
clusters), these different cuts demonstrate that we were generally
successful at selecting member galaxies.  The histogram of proper
velocities is shown in the right panel, together with a Gaussian of
mean zero and standard deviation of one. The agreement between the
distributions is difficult to quantify due to the expected presence of
non-members in the histogram. We will see in Section \ref{s:comp} that
we measure a systematic bias in normalization.

\subsection{Data in the Literature: Summary and
  Comparison}\label{ss:comp_act}

Table \ref{tab:lit} contains spectroscopic redshifts and velocity
dispersions from the literature for clusters detected by the SPT. 
Notably, 14 of these clusters are from \cite{sifon12}, which presents 
spectrosopic follow-up of galaxy clusters that were detected by the 
ACT. Because SPT and ACT are both SZ surveys based in the southern 
hemisphere, there is some overlap between the galaxy clusters detected 
with the two telescopes. We independently obtained data for five
of the clusters that appear in \cite{sifon12}, and there is some overlap 
between the cluster members for which we have measured redshifts. These 
clusters are of some interest for evaluating our follow-up strategy, because 
the typical number of SPT-reported member galaxiesper dispersion is 25 
(for $N_{\mathrm{members}} \ge 15$), while for the overlapping 
\cite{sifon12} sample it is $55$. All of the overlapping cluster redshifts 
and dispersions are consistent between our work and \cite{sifon12} at 
the one-sigma level, except for the velocity dispersion of 
SPT-CL~J0528-5300. Its velocity histogram shows extended structure 
(Figure \ref{fig:vel_hist}). The galaxies responsible for this extended 
structure are kept by the membership selection algorithm in this paper, 
but they were either not observed or not classified as cluster members 
by \cite{sifon12}. It is not possible to determine from the data in 
hand whether this discrepancy is statistical or systematic in origin.

We note that ACT-CL~J0616-5227, also studied in \cite{sifon12}, is
seen in SPT maps but is excluded from the survey because of its
proximity to a point source.

% We have not had an opportunity to evaluate the extent of overlap
% between the selected galaxies, which would make the errors
% correlated, as the data from \cite{sifon12} are not available to us
% at the time of writing We do know, from that article, that their
% $48$ members for SPT-CL~J0546-5345 include our 21 redshifts that
% were published in \cite{brodwin10}, and that the redshifts measured
% in common agree within two sigma.

\section{Statistical Methodology in the Few-$N_{\mathrm{members}}$ Regime}\label{s:resampling}
In this section, we explore the statistical issues surrounding our
obtaining reliable estimates of velocity dispersions and associated
confidence intervals.
% Our results are not new knowledge as the science of statistics is
% concerned, but we sort through a few issues that are unclear or
% inconsistent in the velocity dispersion literature and can become
% important in the few-$N_{\mathrm{member}}$ regime, where we need to
% be attentive to the fact that $N$ and $(N-1)$ differ by enough to
% perturb our results.
A key element in our approach is ``resampling'', in which we extract
and analyze subsets of the data, either on a cluster-by-cluster basis,
or from the stacked cluster that we constructed from the entire
catalog. This allows us to generate large numbers of
``pseudo-observations'' to address statistical questions where we have
too few observations to directly answer.

% We resample our data with the goals of
% \begin{itemize}
% \item selecting the estimators of the average and variance (from
%   which to calculate the redshift and dispersion, respectively) and
%   of the confidence intervals on the estimates;
% \item exploring the behavior of the estimates as a function of
%   $N_{\mathrm{members}}$, as part of this validation;
% \item exploring the effects of membership selection.
% \end{itemize}

\subsection{Unbiased Estimators}\label{s:estimators}
Estimators and confidence intervals for velocity dispersions are
discussed in \cite{beers90}, which the reader is encouraged to
review. They present estimators such as the biweight, that are {\it
  resistant} and {\it robust}\footnote{Resistance means that the
  estimate does not change much when a number of data points are
  replaced by other values; the median is a well-known example of a
  resistant estimator. Robustness means that the estimate does not
  change much when the distribution from which the data points are
  drawn is varied.}.  As we are exploring the properties of the
few-$N_{\mathrm{members}}$ regime, we would also like our estimators
to be {\it unbiased}, meaning that the mean estimate should be
independent of the number of points that are sampled.
% In other words, small number of members per cluster should not
% introduce biases, beyond yielding larger statistical uncertainties.

The first point that we would like to make on the subject is that the
biweight dispersion (or more correctly, the associated variance) as
presented in \cite{beers90}, is biased for samples, in the same way
that the population variance, $\sum_i (v_i - \bar v)^2 / n$, is biased
and the sample variance, $\sum_i (v_i - \bar v)^2 / (n-1)$, is
not\footnote{The fact that this estimator is biased is often
  acknowledged by researchers who use an unbiased version of the
  biweight, yet cite \cite{beers90}. Also, the implementation of the
  Fortran code companion to \citet{beers90} contains a partial
  correction of this bias, in a factor of $\sqrt{n/(n-1)}$ multiplying
  the dispersion. See rostat.f, version 1.2, February 1991. Retrieved
  April 2012 from
  http://www.pa.msu.edu/ftp/pub/beers/posts/rostat/rostat.f.}.

We use the biweight sample variance, which does not suffer from this
bias \citep[see, e.g.,][]{mosteller77}: \be \sigma_{\mathrm{BI}}^2 =
N_{\mathrm{members}} \frac{\sum_{|u_i| < 1} (1-u_i^2)^4 (v_i-\bar v)^2
}{D(D-1)} \ee where $v_i$ are the proper velocities, $\bar v$ their
average, \be D = \sum_{|u_i| < 1} (1-u_i^2)(1-5u_i^2), \ee and $u_i$
is the usual biweight weighting \be u_i = \frac{v_i-\bar v}{9
  \mathrm{MAD}(v_i)}, \ee where $\mathrm{MAD}(v_i)$ is the median
absolute deviation of the velocities.
% We have verified, by applying it to our pseudo-samples, that this
% estimator is less biased at small $N_{\mathrm{members}}$ than the
% version of \cite{beers90}, and also less biased than the better
% version in its associated Fortran code.

We calculate cluster redshifts using the same biweight average
estimator that is presented in \cite{beers90}. Unlike the more subtle
case of the variance, the biweight average is unbiased for all
$N_{\mathrm{members}}$.

\begin{deluxetable*}{lrrr|lrllc}
  \tabletypesize{\scriptsize} \tablecaption{Cluster redshifts and
    velocity dispersions from the literature
    \label{tab:lit}} \tablewidth{0pt} \tablehead{
    \multicolumn{4}{c}{{\bf SPT~/~this work}} &
    \multicolumn{5}{c}{{\bf Literature}} \\
    \colhead{SPT ID} & \colhead{$N$} & \colhead{$z$} &
    \colhead{$\sigma_{\mathrm{BI}} $} &
    \colhead{Lit. ID} & \colhead{$N$} & \colhead{$z$} & \colhead{$\sigma_{\mathrm{BI}} $}  & \colhead{Ref.} \\
    \colhead{} & \colhead{} & \colhead{} & \colhead{(km\,s$^{-1}$)} &
    \colhead{} & \colhead{} & \colhead{} & \colhead{(km\,s$^{-1}$)} &
    \colhead{}
  } \\
  \startdata $z < 0.3$ \\
SPT-CL J0235-5121 & - & - & - $\phantom{222}$ & ACT-CL J0235-5121 & $82$ & $0.2777 \pm 0.0005$ & $1063 \pm 101$ & 1 \\
SPT-CL J0328-5541 & - & - & - $\phantom{222}$ & A3126 & $38$ & $0.0844$ & $1041$ & 3 \\
SPT-CL J0431-6126 & - & - & - $\phantom{222}$ & A3266 & $132$ & $0.0594 \pm 0.0003$ & $1182^{+100}_{-85}$ & 2 \\
SPT-CL J0658-5556 & - & - & - $\phantom{222}$ & 1E0657-56 & $71$ & $0.2958 \pm 0.0003$ & $1249^{+109}_{-100}$ & 4 \\
SPT-CL J2012-5649 & - & - & - $\phantom{222}$ & A3667 & $123$ & $0.0550 \pm 0.0004$ & $1208^{+95}_{-84}$ & 2 \\
SPT-CL J2201-5956 & - & - & - $\phantom{222}$ & A3827 & $22$ & $0.0983 \pm 0.0010$ & $1103^{+252}_{-138}$ & 5 \\
\hline 
 \\
$z \ge 0.3$ \\
SPT-CL J0102-4915 & - & - & - $\phantom{222}$ & ACT-CL J0102-4915 & $89$ & $0.8701 \pm 0.0009$ & $1321 \pm 106$ & 1 \\
SPT-CL J0232-5257 & - & - & - $\phantom{222}$ & ACT-CL J0232-5257 & $64$ & $0.5559 \pm 0.0007$ & $884 \pm 110$ & 1 \\
SPT-CL J0236-4938 & - & - & - $\phantom{222}$ & ACT-CL J0237-4939 & $65$ & $0.3344 \pm 0.0007$ & $1280 \pm 89$ & 1 \\
SPT-CL J0304-4921 & - & - & - $\phantom{222}$ & ACT-CL J0304-4921 & $71$ & $0.3922 \pm 0.0007$ & $1109 \pm 89$ & 1 \\
SPT-CL J0330-5228 & - & - & - $\phantom{222}$ & ACT-CL J0330-5227 & $71$ & $0.4417 \pm 0.0008$ & $1238 \pm 98$ & 1 \\
SPT-CL J0346-5439 & - & - & - $\phantom{222}$ & ACT-CL J0346-5438 & $88$ & $0.5297 \pm 0.0007$ & $1075 \pm 74$ & 1 \\
SPT-CL J0438-5419 & $18$ & $0.4223 \pm 0.0016$ & $1422 \pm 317$ & ACT-CL J0438-5419 & $65$ & $0.4214 \pm 0.0009$ & $1324 \pm 105$ & 1 \\
SPT-CL J0509-5342 & $21$ & $0.4616 \pm 0.0007$ & $678 \pm 139$ & ACT-CL J0509-5341 & $76$ & $0.4607 \pm 0.0005$ & $846 \pm 111$ & 1 \\
SPT-CL J0521-5104 & - & - & - $\phantom{222}$ & ACT-CL J0521-5104 & $24$ & $0.6755 \pm 0.0016$ & $1150 \pm 163$ & 1 \\
SPT-CL J0528-5300 & $21$ & $0.7694 \pm 0.0017$ & $1318 \pm 271$ & ACT-CL J0528-5259 & $55$ & $0.7678 \pm 0.0007$ & $928 \pm 111$ & 1 \\
SPT-CL J0546-5345 & $21$ & $1.0661 \pm 0.0018$ & $1191 \pm 245$ & ACT-CL J0546-5345 & $48$ & $1.0663 \pm 0.0014$ & $1082 \pm 187$ & 1 \\
SPT-CL J0559-5249 & $37$ & $0.6092 \pm 0.0010$ & $1146 \pm 176$ & ACT-CL J0559-5249 & $31$ & $0.6091 \pm 0.0014$ & $1219 \pm 118$ & 1 \\
SPT-CL J2351-5452 & - & - & - $\phantom{222}$ & SCSOJ235138-545253 & $30$ & $0.3838 \pm 0.0008$ & $855^{+108}_{-96}$ & 6 \\
  \enddata
  \tablecomments{ Number of member-galaxy redshifts $N$ ($\equiv
    N_{\mathrm{members}}$), cluster redshift and velocity dispersion
    for clusters found in the literature that are also SPT
    detections. In the cases where we are presenting our own
    spectroscopic observations, some of the information from Table
    \ref{tab:meta} is repeated in the left half of the present table
    for reference.  } \tablerefs{(1) \cite{sifon12}; (2)
    \cite{girardi96}; (3) \citet[][this paper does not contain
    confidence intervals]{struble99}; (4) \cite{barrena2002}; (5)
    \cite{katgert98}; (6) \cite{buckleygeer11}.}
\end{deluxetable*}

% Because it comes at no cost in computation or complexity, we argue
% that this biweight sample variance is the biweight estimator that
% should always be used.

% But statistics are subtle and bias can come in different ways.  This
% estimator (like the sample variance) is an unbiased estimator of
% variance but not of the dispersion. In other words, when sampling a
% Gaussian distribution of variance $\sigma^2$, the following
% expectations values hold: \be \mathbb{E}(\sigma_{\mathrm{BI}}^2) =
% \sigma^2 \text{ but } \mathbb{E}(\sigma_{\mathrm{BI}}) \neq \sigma.
% \ee This statement is $N_{\mathrm{members}}$-dependent, and for any
% reasonable $N_{\mathrm{members}}$ (such as more than 15),
% effectively $\mathbb{E}(\sigma_{\mathrm{BI}}) = \sigma$.

A second issue related to bias at few $N_{\mathrm{members}}$ is the
possibility that the presence of velocity substructure would bias the
estimation of the dispersion. We tested whether that was the case by
extracting smaller pseudo-observations from the $18$ individual
clusters for which we obtained $30$ or more member velocities. We did
not use the stacked cluster here as substructure would be lost in the
averaging. For each cluster, we randomly drew $1000$
pseudo-observations with $10 \le N_{\mathrm{members}} \le 25$. The
cluster redshift and dispersion from those smaller, random samples was
computed and compared to the value that was measured with the full
data set.

Figure \ref{fig:bi_var_stack} shows the results of this resampling
analysis as a function of $N_{\mathrm{members}}$; the black solid line
is the average relative error $ \left< (\sigma_{\mathrm{BI}} -
  \sigma_{\mathrm{pseudo-obs}}) / \sigma_{\mathrm{BI}} \right>$ of the
sample velocity dispersion of all samples across all clusters, while
the colored solid lines depict the average relative error for the
individual clusters. The average relative error departs from zero at
the percent level. From this, we conclude that the observation of only
a small number of velocities per cluster does not introduce
significant bias in the measurement of the velocity dispersion for an
ensemble of clusters.

However, we see that for some individual clusters that have many
measured galaxy velocities, the distribution of velocities is such
that measuring fewer members in a pseudo-observation yields, on
average, a velocity dispersion that can have several to many percent
difference with the one obtained with more members.  This is a way in
which observing few member galaxies will increase the scatter of
observed velocity dispersions at fixed mass. The size of our sample
does not allow us to pursue this effect thoroughly, but Figure
\ref{fig:bi_var_stack} shows that this systematic increase in the
scatter is of order 5\%, relative to dispersions computed with more
than 30 members. \cite{saro12} isolate the scatter that is not due to
statistical effects and also find that the scatter due to systematics
increases at few-$N_{\mathrm{members}}$ and that this effect is most
significant when $N_{\mathrm{members}}$ is less than $\sim 30$.
% Looked at the what it looks like with only normal disribution.  py
% ruel12.py jul10.h5 -s q_n -c no_vel_gap q_n_gaussian

\subsection{Confidence intervals}\label{ss:ci}

We now turn to the calculation of the statistical uncertainty on our
measured redshifts and velocity dispersions. \cite{beers90} describe a
number of different ways in which the confidence intervals on biweight
estimators can be calculated.  They conclude that the statistical
jackknife and the statistical bootstrap both yield satisfactory
confidence intervals.  Broadly speaking, both of these methods
estimate the confidence intervals by looking at the internal
variability of a sample.  The statistical
jackknife %\citep[see, e.g.][]{mosteller77}
constructs a confidence interval for an estimate from how much it
varies when data points are removed.  The bootstrap generates a
probability distribution function for the estimate from resampling the
observed values with replacement a large number of times, often 1000
or more. The confidence intervals can then be found from the
percentiles of this distribution.  Many publications after
\cite{beers90} have chosen the bootstrap; different practices seen in
its use, with papers quoting asymmetric confidence intervals and
others symmetric ones, have promtped us to inspect our uncertainties
carefully.
% When applied to the dispersion, the median of the statistical
% bootstrap distribution is biased with respect to the value of the
% estimate, which often leads to strongly asymmetric confidence
% intervals, even with second-order (``BCa'') bias correction
% \citep[this bias is a well-known statistical phenomenon; see,
% e.g.,][]{efron87}.  For instance, \cite{fadda96, girardi96} quote
% asymmetric intervals, as in $1053^{+164}_{-108}$ km\,s$^{-1}$, while
% others like \cite{zhang11} and \cite{sifon12} quote symmetric
% intervals from the bootstrap, as in $(1053\,\pm\,139)$ km\,s$^{-1}$.

The reason for using the statistical bootstrap or jackknife is the
absence of an analytic expression for the distribution of the errors,
given that the source distribution of velocities is unknown, as is the
distribution of measured biweight dispersions.  We use the stacked
cluster as the best model of a cluster with our selection of potential
member galaxies.
% , of which some are in fact non-members and interlopers, non-members
% that occupy the same projected phase-space as members.
As explained in Section \ref{ss:stack}, the availability of SPT masses
for all clusters allows us to construct this stacked cluster
independently of cluster membership determination or dispersion
measurements.  We draw a large number of pseudo-observations with
replacement from the stacked cluster, perform member selection, and
calculate the cluster's redshift and velocity dispersion from each
pseudo-observation. Thus, we generate a probability distribution
function for those quantities.

We find that the distribution of the measured cluster redshift is
close to a normal distribution whose standard deviation is well
described by: \be \Delta z = \frac{1}{c}
\frac{\sigma_{\mathrm{BI}}(1+z)}{\sqrt{N_{\mathrm{members}}}}.  \ee
This is the ``usual'' standard error; the $1/c$ factor converts
between velocity and redshift, and the $1+z$ factor is needed because
$\sigma_{\mathrm{BI}}$ is defined in the rest frame.
% ; it is the same factor encountered in the calculation of the proper
% velocities in Section \ref{ss:meta}.
At any given $N_{\mathrm{members}}$, the average bootstrap and
jackknife uncertainties also reproduce this standard error.
% with an uncertainty derived from the statistical jackknife
% \citep[see e.g.][]{mosteller77}.

% Drawing from this is not the same as drawing independent samples
% from a parent distribution, however we believe that it more closely
% matches the reality of observing clusters of galaxies, where a
% finite number of galaxies exist above a given luminosity cut. This
% galaxy velocity dispersion is linked to the dark matter dispersion.

In the case of the velocity dispersion, the bootstrap and jackknife
give confidence intervals that are too narrow. Simply put, those
estimators use a sample's internal variability to infer likely
properties of the population from which it was drawn.  However, the
variability is reduced by the membership selection, and the effect of
that step is not included in the confidence interval.

The distribution of biweight sample dispersions measured in
pseudo-observations after $3\sigma$-clipping membership selection is
also observed to be close to a normal distribution in our resampling
analysis. We set out to model the standard deviation of this
distribution, which is the uncertainty that we are looking for.

If we draw observations from a normal distribution of variance
$\sigma^2$ and calculate the velocity dispersion as the ``usual''
(non-biweight) sample standard deviation from $n$ members, without a
membership selection step, then the distribution of the measured
standard deviation is related to a chi distribution with $n-1$ degrees
of freedom. Indeed, the sample standard deviation $s$ is \be s =
\sqrt{\frac{1}{n-1} \sum_{i=1}^n (v_i - \bar v_i)^2 } \ee

This implies that \bea
\frac{s}{\sigma} &=&  \sqrt{\frac{1}{n-1} \sum_{i=1}^n \frac{(v_i - \bar v_i)^2}{\sigma^2} } \\
\Rightarrow \quad \sqrt{n-1}\frac{s}{\sigma} &=& \sqrt{ \sum_{i=1}^n
  \frac{(v_i - \bar v_i)^2}{\sigma^2} } \sim \chi_{n-1} \eea which is
the definition of a chi distribution.

The variance of the chi distribution varies very little between $k=10$
and $k=100$ degrees of freedom: \bea
\Var \, \chi_{k} &=& k - 2\left( \frac{\Gamma((k+1)/2)}{\Gamma(k/2)}  \right)^2 \\
&=& 0.49 \quad (k=10) \\
&=& 0.50 \quad (k=100). \eea
% from math import gamma n - 2*( gamma((n+1.)/2.) / gamma(n/2.) )**2

Therefore, taking the square root on each side to find the standard
deviation $\Delta$ of the dispersion estimate $s$: \bea
\Delta (\sqrt{n-1}\frac{s}{\sigma}) &=& \sqrt{0.5} = 0.7 \\
\Rightarrow \quad \Delta s &=& 0.7 \sigma / \sqrt{n-1} \eea

Following the above, we model the uncertainty as \bea
\label{eq:param}
\Delta \sigma_{\mathrm{BI}} &=& C_{\mathrm{BI}} \sigma_{\mathrm{BI}} /
\sqrt{N_{\mathrm{members}}-1} \eea where $C_{\mathrm{BI}}$ is a
constant.  We also parameterize the uncertainty on the gapper
measurement in the same way, with a constant $C_{\mathrm{G}}$.

Figure \ref{fig:error_disp} shows the relative error in
$\sigma_{\mathrm{BI}}$ measured from the resampling analysis, as a
function of $N_{\mathrm{members}}$, for $10 \le N_{\mathrm{members}}
\le 60$.  The solid black line shows the average error, and the blue
dashed line is the asymmetric root-mean-square error.

We find the numerical value of $C_{\mathrm{BI}}$ as the mean ratio of
the RMS error and $1 / \sqrt{N_{\mathrm{members}}-1}$.
% (this is the relative error; the absolute error would multiply this
% by $\sigma_{\mathrm{BI}}$).
We find that $C_{\mathrm{BI}} = 0.92$. The green dotted line of Figure
\ref{fig:error_disp} shows the uncertainty given by our model, $\pm
0.92 / \sqrt{N_{\mathrm{members}}-1}$.  Similarly for the gapper
scale, we find that $C_{\mathrm{G}} = 0.91$.

Therefore, we find that the 3-$\sigma$ membership selection combined
with the biweight estimation of the dispersion gives an uncertainty
increased by 30\% compared with random sampling from a normal
distribution, and also compared to the bootstrap and jackknife
estimates.  The larger errors are caused by non-Gaussianity in the
velocity distribution and by the cluster membership selection, which
can both include non-members and reject true members, generically
leading to increased scatter in the measured dispersion.

\begin{figure*}[htbp]
  \epsscale{1.0} \plotone{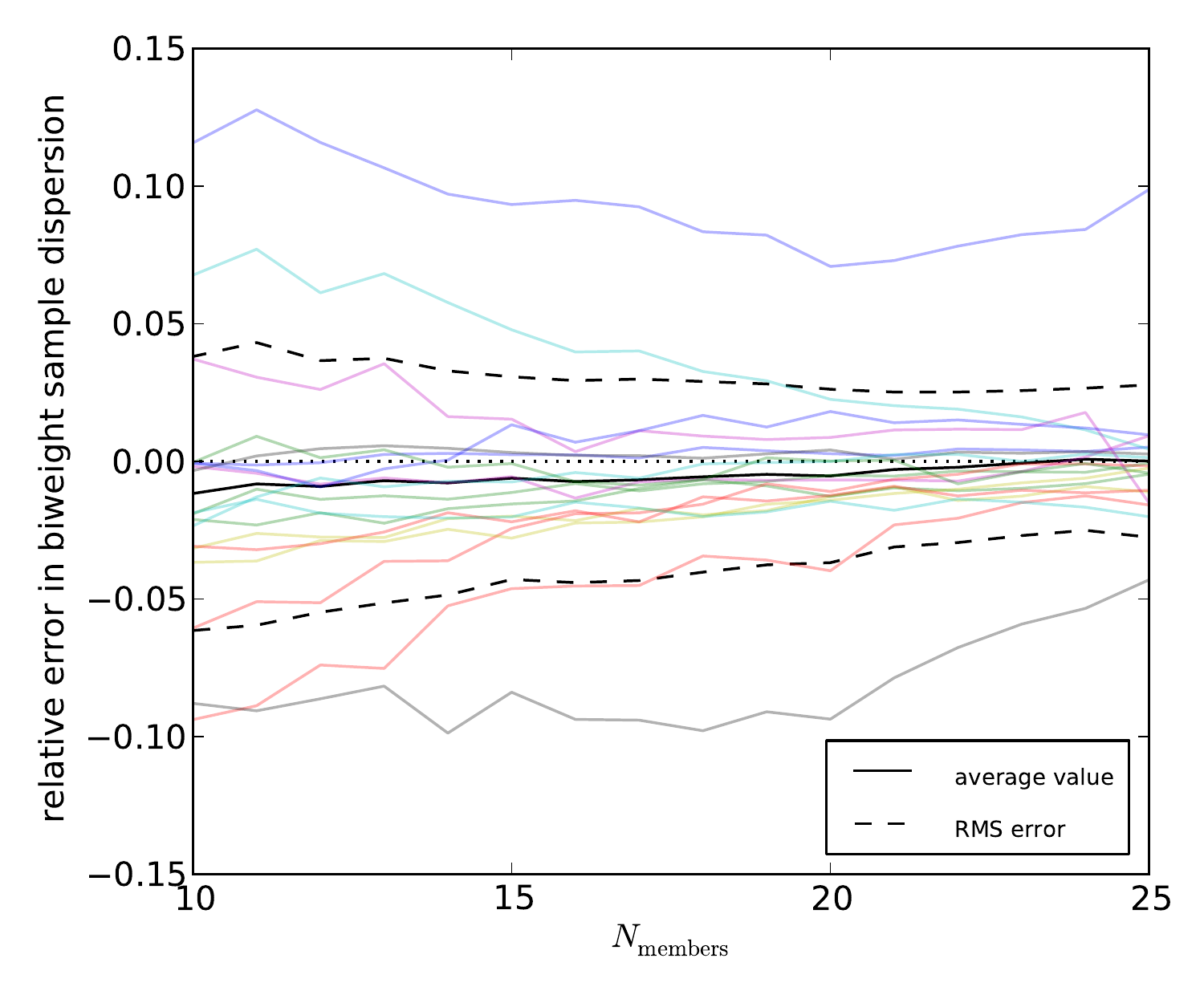}
\caption{ Average relative error $ \left< (\sigma_{\mathrm{BI}} -
  \sigma_{\mathrm{pseudo-obs}}) / \sigma_{\mathrm{BI}} \right>$ in the measured dispersion for pseudo-observations sampled from each individual cluster with more than 30 members, as described in Section \ref{s:estimators}. The colored solid lines show the average across pseudo-observations for each individual cluster. The black solid line is the average across clusters, and the dashed lines show the one-sigma range in the distribution of colored traces. The dotted line is identically zero. 1000 pseudo-observations were drawn with replacement at each $N_{\mathrm{members}}$, and three-sigma clipping membership selection was applied to each pseudo-observation before computation of the dispersion. The average relative error departs from zero at the percent level; from this we conclude that the observation of only a small number of velocities per cluster does not introduce bias in the measurement of the velocity dispersion, in an ensemble sense. However, it presents an additional source of error for individual clusters, which will increase the measurement scatter.
\label{fig:bi_var_stack}}
\end{figure*}

\begin{figure*}[htbp]
  \epsscale{1.0} \plotone{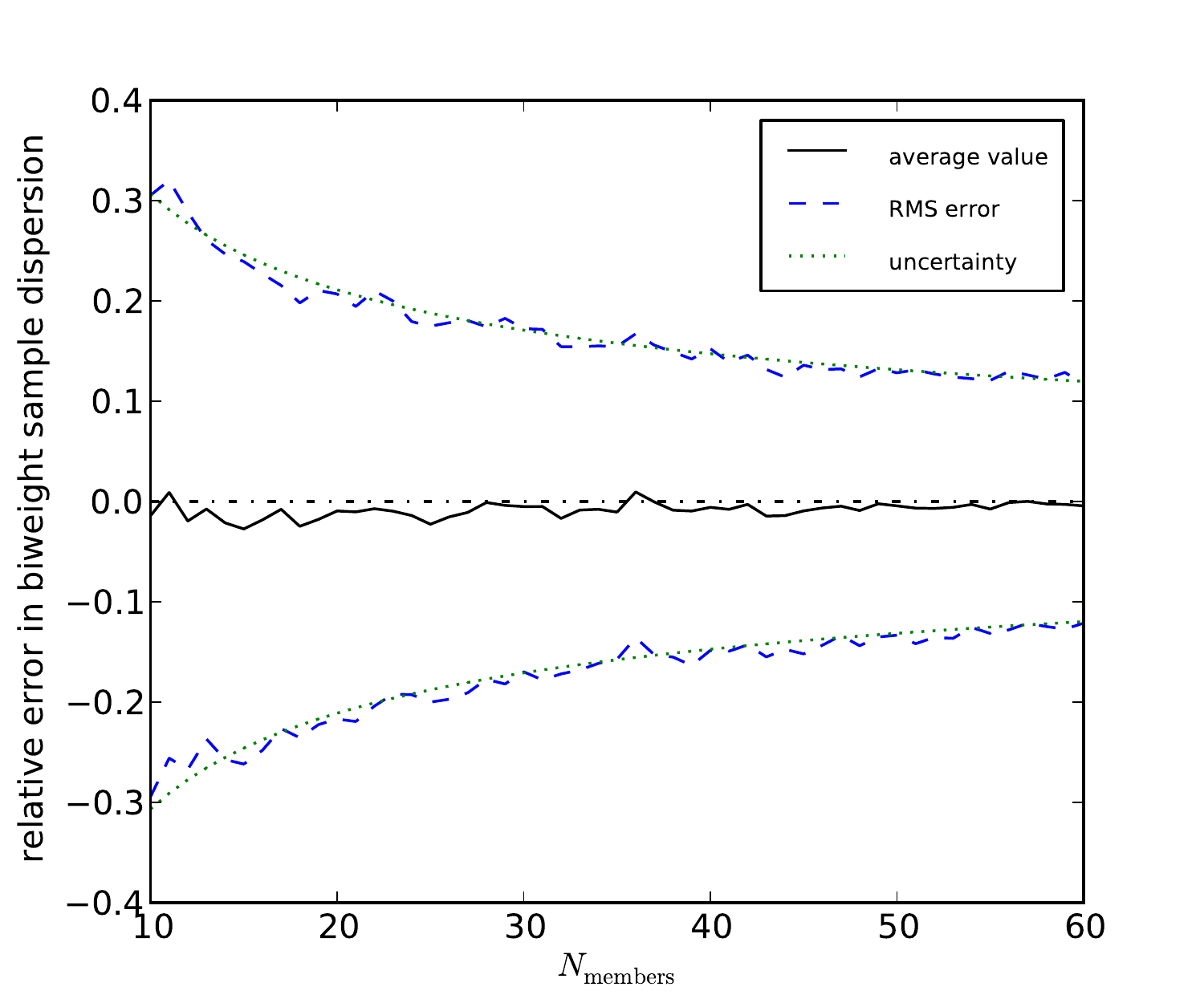}
  % \vskip-0.8in
\caption{Statistical uncertainty and relative error in the measured dispersion for pseudo-observations sampled from the stacked cluster, as described in Section \ref{ss:ci}. The black solid line shows the average error, the blue dash-dotted line is the asymmetric root-mean-square error, and the green dotted line is the (relative) uncertainty, $\pm 0.92 / \sqrt{N_{\mathrm{members}}-1}$. The parameter $C_{\mathrm{BI}} = 0.92$ from Equation \ref{eq:param} was fit using this root-mean-square error, hence the agreement of the lines. The dot-dashed line is identically zero. 1000 pseudo-observations were drawn with replacement at each $N_{\mathrm{members}}$, and three-sigma clipping membership selection was applied to each pseudo-observation before computation of the dispersion.
  \label{fig:error_disp}}
\end{figure*}

We note that this effect is different than the systematic scatter
shown in Figure \ref{fig:bi_var_stack}, where the measured dispersion
changed significantly for some individual clusters when resampling
with fewer galaxy members.  This latter effect likely has both
physical (e.g., velocity sub-structure in the cluster) and measurement
(e.g., member selection, interlopers) origins.  However, both effects
will be present at some level in any dispersion measurement, and the
results here are important benchmarks for simulations to compare to
and reproduce.

\section{Comparison of Velocity Dispersions with Other Observables}
\label{s:comp}
In this section, we compare our cluster velocity dispersion
measurements with gas-based observables and estimates of the cluster
mass.  In particular, we measure the normalization and the scatter of
scaling relations between the two observables, and compare these to
our expectations from simulations. We neglect effects related to the
SZ cluster selection, variation of the cosmology, or potentially
correlated intrinsic scatter between observables, and leave the
accounting of these effects to future work.  However, this comparison
is still useful in understanding how our velocity dispersion mass
estimates compare to those using other methods, and can also help
identify systematics.

\subsection{Comparison with SPT Masses}\label{ss:comp_sz}
% Sources of scatter: D_SZ = 0.24, and B_SZ = 1.33, so the Mass-Mass
% scatter is 0.24 /Ê1.33 = 0.18, _*.343 = 0.0619 in dispersion Mean,
% median relative error on SZ mass 0.208908500962 0.204235974259 *.343
% = That includes the intrinsic scatter => 7% in dispersion

% Mean, median error on X-ray mass 0.143853069056 0.118554609271 The
% intrinsic scatter in mass given a Yx is supposed to be only 7%

% Fixed M500, actually expects larger scatter than fixed M200

\begin{figure*}[htbp]
  \epsscale{1.20} \plotone{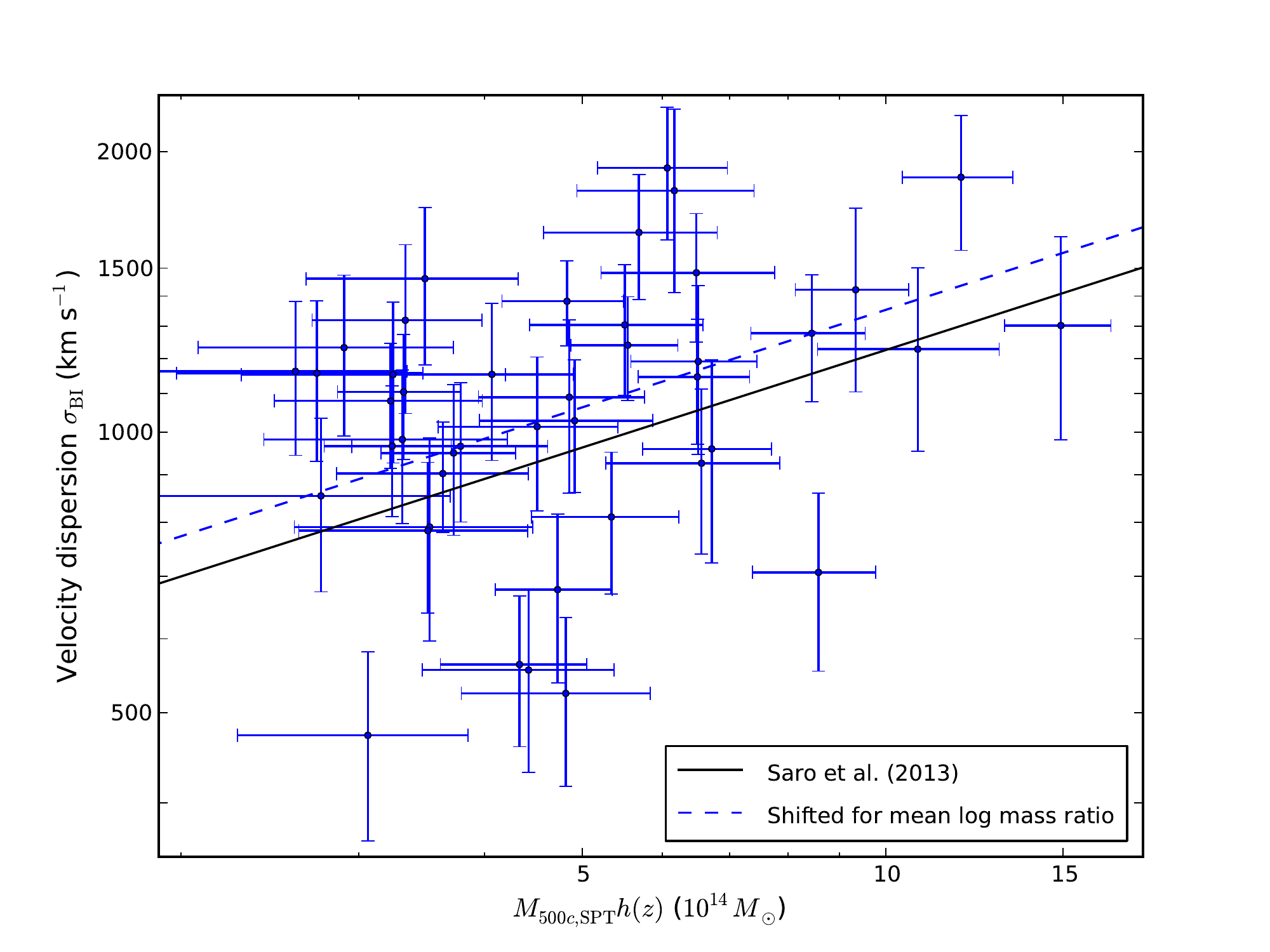}
  % \vskip-0.8in
  \caption{ Cluster biweight velocity dispersions from Table
    \ref{tab:meta} as a function of SZ-based SPT masses (Table
    \ref{tab:spt}, Section \ref{ss:obsspt}) for clusters with
    $N_{\mathrm{members}} \ge 15$ and $z \ge 0.3$. The figure also
    shows the scaling relationship between velocity dispersion and
    mass expected from dark-matter simulations as a solid line
    \citep{saro12}. The dashed line is this same scaling relationship,
    shifted to show the average scaling relationship implied by the
    mean log mass ratio of the data points. \label{fig:scaling}}
\end{figure*}

Figure \ref{fig:scaling} shows 43 cluster biweight velocity
dispersions from Table \ref{tab:meta} plotted against the masses
estimated from their SPT SZ signal (combined with X-ray observations
where applicable; Table \ref{tab:spt}, Section \ref{ss:obsspt}).  The
clusters that are included are those with $N_{\mathrm{members}} \ge
15$ and $z \ge 0.3$, except for SPT-CL~J0205-6432, which was flagged
as having a potentially less reliable dispersion measurement in
Section \ref{ss:meta}.  We also plot, as a solid line, the predicted
scaling between dispersion and mass from \cite{saro12}
\begin{equation}
  \label{eq:saro_scaling} 
  M_{200c, \,\mathrm{dyn}} = \left( \frac{\sigma_{\mathrm{DM}}}{A \times h_{70}(z)^C} \right)^B 10^{15}
  M_\odot
\end{equation}
where $A = 939$, $B = 2.91$ and $C = 0.33$, with negligible
statistical uncertainty compared to the systematic uncertainty, whose
floor is evaluated to be at 5\% in dispersion
\citep{evrard08}. $\sigma_{\mathrm{DM}}$ is the dispersion computed
from dark-matter subhalos, which are identified as galaxies in
simulations. This has a different functional form but is consistent
with the \cite{evrard08} scaling relation.

Our measurements appear to have a systematic offset relative to the
model prediction. To quantify this offset, we compute the mean of the
log mass ratio, $\ln(M_{200c, \,\mathrm{dyn}}/M_{200c,
  \,\mathrm{SPT}})$.  For each cluster $i$, we compute this log mass
ratio, and its associated uncertainty $\sigma_{\ln M_\mathrm{ratio},
  i}$.  The uncertainty in the ratio is estimated from the quadrature
sum of the fractional uncertainty in the SZ and dynamical mass
estimates. To the latter, we add the expected intrinsic scatter in
dynamical to true-mass as estimated by \cite{saro12}:
\begin{equation}
  \label{eq:saro_scatter} 
  \Delta_{\ln(M_{200c, \,\mathrm{dyn}}/M_{200c})} = 0.3 + 0.075z.
\end{equation}
The uncertainty on the SZ-based SPT mass already includes the effect
of intrinsic scatter.

The weighted average log mass ratio is
\begin{equation}
  \left< \ln(M_{200c, \,\mathrm{dyn}}/M_{200c, \,\mathrm{SPT}}) \right> = 0.33 \pm 0.10,
\end{equation}
where the weights for each data point are given by $1/\sigma_{\ln
  M_\mathrm{ratio}, i}^2$, and the uncertainty on the average is given
by $1/\sigma^2 = \sum_i 1/\sigma_{\ln M_\mathrm{ratio}, i}^2$.
% If the two mass estimates were equal, the average log mass ratio
% would be zero; it does not suffer from Eddington bias.  0.28 \pm
% 0.14 when adding uncertainty uniformly to take the scatter mismatch
% into account This is similar to the mass comparison in
% \cite{high12}.
This average log ratio means that the dynamical mass is $\exp(0.33) =
1.39$ times the SPT mass estimate.

Figure \ref{fig:scaling} shows as a dashed blue line how the N-body
scaling relation is shifted if the log mass ratio is shifted by $0.33$
to make the mass estimates coincide. Because the slope is $1/B =
1/2.91$, the offset in log dispersion is $(0.33 \pm 0.10)/2.91 = 0.11
\pm 0.03$.  In other words, the measured velocity dispersions are on
average $\exp(0.11) = 1.12$ times their expected value given the
N-body simulation work and the current normalization of the SPT mass
estimate. The size of this normalization offset is consistent with the
expected size of systematic biases, as discussed in Section
\ref{ss:systematics}.

We quantify the level of Gaussianity in the dispersion estimates
around the best-fit dispersion-mass relation by performing the
Anderson-Darling test on the residuals.  We find that the residuals
are non-Gaussian at the 95\% confidence level.  If we remove the two
clusters with the lowest dispersions (SPT-CL~J0317-5935 and
SPT-CL~J2043-5035), we find per the Anderson-Darling test that the
residuals are consistent with a normal distribution.  This suggests
that the scatter in $\ln \sigma$ is normal -- i.e., that the
dispersion distribution is log-normal -- with a tail towards low
dispersion, as might be suspected from the distribution of data points
in Figure 5.

If the statistical uncertainty on dispersion measurements of
individual clusters has been correctly estimated and is much larger
than any systematic uncertainty, then the fractional scatter in $\ln
\sigma$ at fixed mass should roughly equal the average fractional
uncertainty in the individual measurements.  The mean uncertainty in
log dispersion at fixed mass is 0.24, including the intrinsic scatter
of the scaling relation and the uncertainty on the SPT mass. Analysis
of mock observations from simulated clusters indicate that the
combination of intrinsic, statistical, and systematic effects would
lead to a log-normal scatter of 0.26 in dispersion at fixed mass
\citep[][]{saro12}. Both numbers are smaller but in general agreement
with the measured scatter in $\ln \sigma$ at fixed mass, $(0.31 \pm
0.03)$. Systematic effects can increase the scatter, as discussed in
Section \ref{ss:systematics}.

\subsection{Comparison with X-ray Observations}\label{ss:xray}

In this section, we compare the velocity dispersion measurements to
X-ray observables and mass estimates, and contrast these results with
predictions from simulations.  We also compare our results to those
when using a separate low-redshift sample of comparable-mass clusters
with similar velocity dispersion and X-ray observables.

For the clusters in this work, we primarily use X-ray measurements
from a \emph{Chandra X-ray Visionary Project} to observe the 80 most
significantly detected clusters by the SPT at $z > 0.4$ (PI:
B. Benson).  This cluster sample has been observed and analyzed in a
uniform fashion to derive cluster mass-observables
\citep[][]{benson13} and cluster cooling properties
\citep{mcdonald13}.
% Slight deviations from previously-published values for some clusters
% \citep{andersson10} in the characteristic ICM temperature, $T_X$,
% and $Y_X$-derived mass, $M_{500c,Y_X}$, are due to improvements in
% the X-ray reduction and analysis pipeline (e.g., centroiding,
% substructure masking, Galactic/extragalactic background modeling,
% updated calibration, etc).
In Table \ref{tab:xray_lit}, we give the X-ray measured ICM
temperature, $T_X$, and the $Y_X$-derived cluster mass,
$M_{500c,\mathrm{Y_X}}$, for the 28 clusters that overlap with the
sample from \citet[][]{benson13}.

We also plot our results alongside velocity dispersion and X-ray
measurements of comparable-mass low-redshift clusters taken from the
literature.  For the X-ray measurements, we use measurements of $T_X$
and $M_{500c,\mathrm{Y_X}}$ from the low-$z$ sample of
\cite{vikhlinin09b}, which were produced following an analysis
identical to that used in \citet[][]{benson13}. The velocity
dispersions for many of those galaxy clusters were calculated in a
uniform way in \cite{girardi96}. These velocity dispersion
measurements were made with a different galaxy selection and more
cluster members, and so will carry different systematics from our
own. They nonetheless provide an interesting baseline for comparison.
We will see that the scatter of those data points is smaller that that
of our sample.  Taking instrinsic scatter and mass uncertainties into
account, the measured scatter of the literature sample at fixed mass
is consistent both with our analysis from Section \ref{ss:ci} and with
the \cite{girardi96} uncertainties, and therefore is due to the lower
statistical uncertainty.

\begin{figure*}[htbp]
  \epsscale{1.2} \plotone{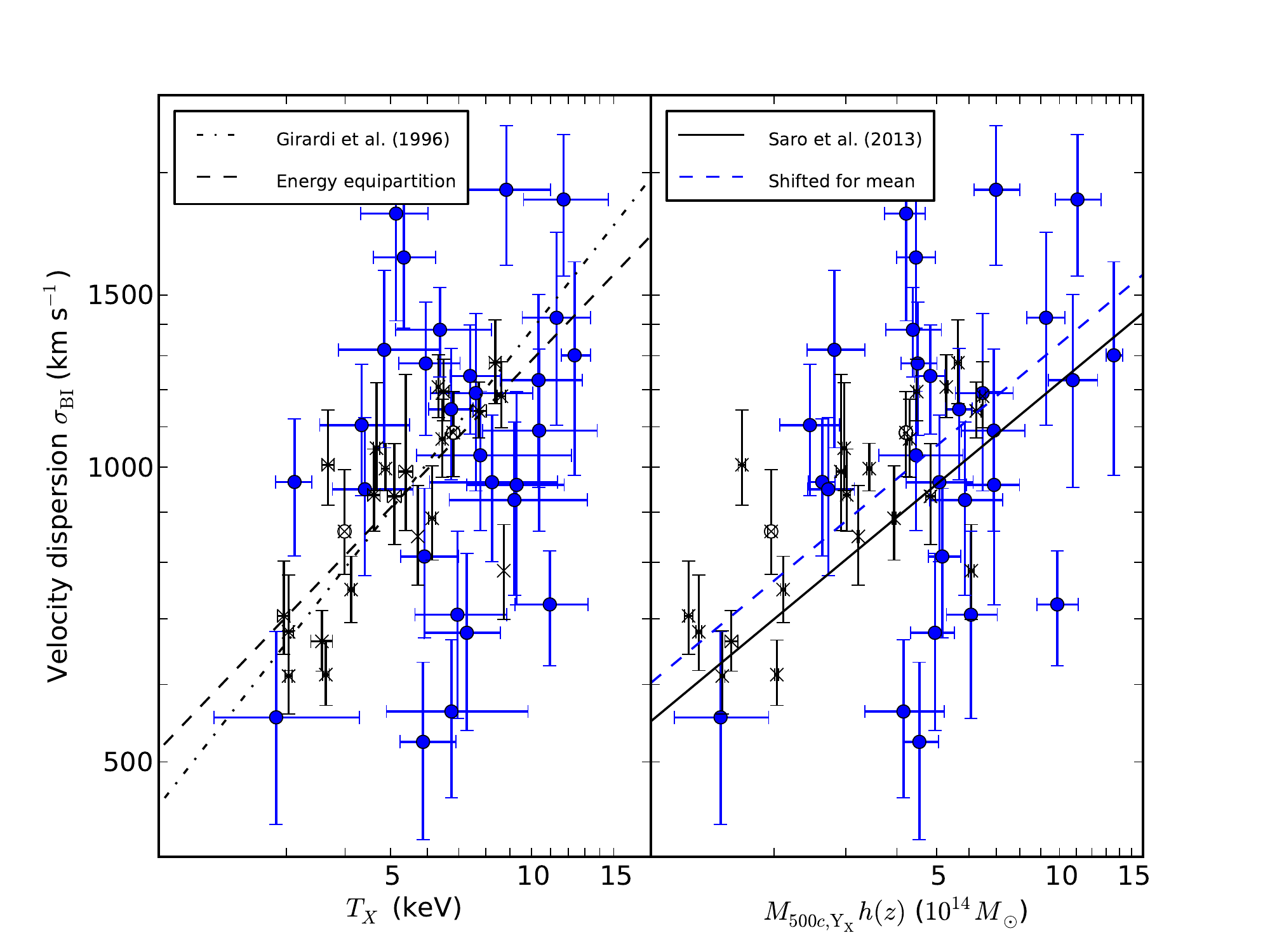}
  % \vskip-0.8in
\caption{Velocity dispersion compared to X-ray properties. The blue points are our sample, and the black crosses are the data from the literature, with X-ray data from \cite{vikhlinin09b} and dispersions from \cite{girardi96}; two of them are also low-redshift SPT detections and are circled.
{\it Left panel:} velocity dispersion vs. X-ray temperature. The dot-dashed line is the best-fit scaling relation from \cite{girardi96}. The dashed line shows the scaling expected if galaxies and gas were both in equilibrium with the gravitational potential.
{\it Right panel:} velocity dispersion vs. $M_{500c,\mathrm{Y_X}}$. The solid line is the scaling relation from \citep{saro12}, and the dashed line is this same scaling relationship, shifted to show the average scaling relationship implied by the mean log mass ratio of the data points.
\label{fig:xray}}
\end{figure*}

\begin{deluxetable}{lrrrrr}
  \tabletypesize{\scriptsize} \tablecaption{X-ray and velocity
    dispersion data
    \label{tab:xray_lit}} \tablewidth{0pt} \tablehead{
    \colhead{Cluster ID} & \colhead{$z$}  & \colhead{$N$} & \colhead{$\sigma_{\mathrm{BI}} $} & \colhead{$T_X$} & \colhead{$M_{500c,\mathrm{Y_X}}$} \\
    \colhead{} & \colhead{} & \colhead{} & \colhead{(km\,s$^{-1}$)} &
    \colhead{(keV)} & \colhead{($10^{14} M_{\odot}$)}
  } \\
  \startdata {\it SPT-CL} \\
J0000-5748 & $0.702$ & $26$ & $563 \pm 104$ & $6.75^{+3.09}_{-1.85}$ & $4.11^{+1.06}_{-0.80}$ \\
J0014-4952 & $0.752$ & $29$ & $811 \pm 141$ & $5.91^{+1.09}_{-0.65}$ & $4.97^{+0.54}_{-0.38}$ \\
J0037-5047 & $1.026$ & $18$ & $555 \pm 124$ & $2.85^{+1.44}_{-0.75}$ & $1.22^{+0.38}_{-0.28}$ \\
J0040-4407 & $0.350$ & $36$ & $1277 \pm 199$ & $5.95^{+1.09}_{-0.74}$ & $5.42^{+0.62}_{-0.49}$ \\
J0234-5831 & $0.415$ & $22$ & $926 \pm 186$ & $9.20^{+3.98}_{-2.52}$ & $6.83^{+1.62}_{-1.24}$ \\
J0438-5419 & $0.422$ & $18$ & $1422 \pm 317$ & $11.32^{+2.07}_{-1.76}$ & $10.74^{+1.19}_{-1.11}$ \\
J0449-4901 & $0.790$ & $20$ & $1090 \pm 230$ & $10.39^{+3.43}_{-2.53}$ & $6.50^{+1.24}_{-1.08}$ \\
J0509-5342 & $0.462$ & $21$ & $678 \pm 139$ & $7.28^{+1.30}_{-1.38}$ & $5.62^{+0.65}_{-0.72}$ \\
J0516-5430 & $0.294$ & $48$ & $724 \pm 97$ & $10.95^{+2.27}_{-1.73}$ & $12.25^{+1.52}_{-1.31}$ \\
J0528-5300 & $0.769$ & $21$ & $1318 \pm 271$ & $4.85^{+1.56}_{-0.98}$ & $2.68^{+0.50}_{-0.38}$ \\
J0546-5345 & $1.066$ & $21$ & $1191 \pm 245$ & $7.61^{+2.45}_{-1.52}$ & $5.22^{+0.98}_{-0.74}$ \\
J0551-5709 & $0.424$ & $34$ & $966 \pm 155$ & $3.12^{+0.28}_{-0.28}$ & $3.04^{+0.23}_{-0.23}$ \\
J0559-5249 & $0.609$ & $37$ & $1146 \pm 176$ & $6.74^{+0.76}_{-0.71}$ & $5.93^{+0.45}_{-0.45}$ \\
J2043-5035 & $0.723$ & $21$ & $524 \pm 108$ & $5.87^{+1.03}_{-0.63}$ & $4.44^{+0.50}_{-0.38}$ \\
J2106-5844 & $1.131$ & $18$ & $1228 \pm 274$ & $10.36^{+2.49}_{-1.79}$ & $8.37^{+1.26}_{-1.07}$ \\
J2135-5726 & $0.427$ & $33$ & $1029 \pm 167$ & $7.78^{+4.41}_{-2.10}$ & $5.15^{+1.56}_{-0.97}$ \\
J2145-5644 & $0.480$ & $37$ & $1638 \pm 251$ & $5.34^{+0.90}_{-0.74}$ & $5.00^{+0.57}_{-0.52}$ \\
J2148-6116 & $0.571$ & $30$ & $966 \pm 165$ & $8.24^{+3.14}_{-2.18}$ & $5.42^{+1.14}_{-0.93}$ \\
J2248-4431 & $0.351$ & $15$ & $1301 \pm 320$ & $12.37^{+1.01}_{-0.77}$ & $16.35^{+0.84}_{-0.70}$ \\
J2325-4111 & $0.358$ & $33$ & $1921 \pm 312$ & $8.84^{+2.16}_{-1.55}$ & $8.39^{+1.19}_{-0.98}$ \\
J2331-5051 & $0.575$ & $78$ & $1382 \pm 145$ & $6.38^{+1.84}_{-1.25}$ & $4.66^{+0.81}_{-0.66}$ \\
J2332-5358\tablenotemark{a} & $0.402$ & $53$ & $1240 \pm 158$ & $7.40^{+1.20}_{-0.70}$ & $5.66^{+0.48}_{-0.48}$ \\
J2337-5942 & $0.776$ & $19$ & $707 \pm 153$ & $6.95^{+1.91}_{-1.31}$ & $5.76^{+0.92}_{-0.74}$ \\
J2341-5119 & $1.002$ & $15$ & $959 \pm 236$ & $9.30^{+2.45}_{-2.02}$ & $5.77^{+0.89}_{-0.83}$ \\
J2344-4243 & $0.595$ & $32$ & $1878 \pm 310$ & $11.72^{+2.88}_{-2.10}$ & $11.64^{+1.64}_{-1.36}$ \\
J2359-5009 & $0.775$ & $26$ & $950 \pm 175$ & $4.41^{+1.18}_{-0.65}$ & $2.58^{+0.41}_{-0.28}$ \\
\\
 \hline 
 \\
{\it Literature} \\
A3571 & $0.039$ & $70$ & $1085^{+110}_{-107}$ & $6.81 \pm 0.10$ & $5.90 \pm 0.06$ \\
A2199 & $0.030$ & $51$ & $860^{+134}_{-83}$ & $3.99 \pm 0.10$ & $2.77 \pm 0.05$ \\
A496 & $0.033$ & $151$ & $750^{+61}_{-56}$ & $4.12 \pm 0.07$ & $2.96 \pm 0.04$ \\
A3667 & $0.056$ & $123$ & $1208^{+95}_{-84}$ & $6.33 \pm 0.06$ & $7.35 \pm 0.07$ \\
A754 & $0.054$ & $83$ & $784^{+90}_{-85}$ & $8.73 \pm 0.00$ & $8.47 \pm 0.13$ \\
A85 & $0.056$ & $131$ & $1069^{+105}_{-92}$ & $6.45 \pm 0.10$ & $5.98 \pm 0.07$ \\
A1795 & $0.062$ & $87$ & $887^{+116}_{-83}$ & $6.14 \pm 0.10$ & $5.46 \pm 0.06$ \\
A3558 & $0.047$ & $206$ & $997^{+61}_{-51}$ & $4.88 \pm 0.10$ & $4.78 \pm 0.07$ \\
A2256 & $0.058$ & $47$ & $1279^{+136}_{-117}$ & $8.37 \pm 0.24$ & $7.84 \pm 0.15$ \\
A3266 & $0.060$ & $132$ & $1182^{+100}_{-85}$ & $8.63 \pm 0.18$ & $9.00 \pm 0.13$ \\
A401 & $0.074$ & $123$ & $1142^{+80}_{-70}$ & $7.72 \pm 0.30$ & $8.63 \pm 0.24$ \\
A2052 & $0.035$ & $62$ & $679^{+97}_{-59}$ & $3.03 \pm 0.07$ & $1.84 \pm 0.03$ \\
Hydra-A & $0.055$ & $82$ & $614^{+52}_{-43}$ & $3.64 \pm 0.06$ & $2.83 \pm 0.03$ \\
A119 & $0.044$ & $80$ & $850^{+108}_{-92}$ & $5.72 \pm 0.00$ & $4.50 \pm 0.03$ \\
A2063 & $0.034$ & $91$ & $664^{+50}_{-45}$ & $3.57 \pm 0.19$ & $2.21 \pm 0.08$ \\
A1644 & $0.048$ & $92$ & $937^{+107}_{-77}$ & $4.61 \pm 0.14$ & $4.21 \pm 0.09$ \\
A3158 & $0.058$ & $35$ & $1046^{+174}_{-99}$ & $4.67 \pm 0.07$ & $4.13 \pm 0.05$ \\
MKW3s & $0.045$ & $30$ & $612^{+69}_{-52}$ & $3.03 \pm 0.05$ & $2.09 \pm 0.03$ \\
A3395 & $0.051$ & $107$ & $934^{+123}_{-100}$ & $5.10 \pm 0.17$ & $6.74 \pm 0.18$ \\
A399 & $0.071$ & $92$ & $1195^{+94}_{-79}$ & $6.49 \pm 0.17$ & $6.18 \pm 0.11$ \\
A576 & $0.040$ & $48$ & $1006^{+138}_{-91}$ & $3.68 \pm 0.11$ & $2.34 \pm 0.05$ \\
A2634 & $0.030$ & $69$ & $705^{+97}_{-61}$ & $2.96 \pm 0.09$ & $1.74 \pm 0.04$ \\
A3391 & $0.055$ & $55$ & $990^{+254}_{-128}$ & $5.39 \pm 0.19$ & $4.06 \pm 0.10$ \\

  \enddata
  \tablenotetext{a}{XMM X-ray data from \cite{andersson10}}
  \tablecomments{SPT data and data from the literature used in Figure
    \ref{fig:xray}. For the SPT data, the redshift, number of
    member-galaxy redshifts $N$ ($\equiv N_{\mathrm{members}}$) and
    velocity dispersion from Table \ref{tab:meta} are repeated for
    reference, and the X-ray temperature and $M_{500c,\mathrm{Y_X}}$
    are from the same {\it Chandra XVP} program, except for one case
    that are marked. The literature clusters draw their velocity
    dispersion from \cite{girardi96} and X-ray properties from
    \cite{vikhlinin09b}.  }
\end{deluxetable}

Figure \ref{fig:xray} shows the velocity dispersion versus X-ray
temperature and versus $M_{500c,\mathrm{Y_X}}$. The blue points are
our data, and the black crosses are the data from the literature;
these literature data are listed for reference in Table
\ref{tab:xray_lit}.

The left panel of Figure \ref{fig:xray} shows dispersion versus
$T_X$. The empirical best-fit scaling relation from \cite{girardi96},
where $\sigma \propto T_X^{0.61}$, is plotted as a solid line; this
scaling relation is consistent with the \cite{vikhlinin09b}
temperatures used here, although it was fit using X-ray temperatures
from a different source, \cite{david93}. The comparison to the
temperature is especially interesting in that there is, to first
order, a simple correspondence between temperature and velocity
dispersion. Assuming that the galaxies and gas are both in equilibrium
with the potential \citep[see, e.g.,][]{voit04}, then $\sigma^2 =
k_\mathrm{B} T_X / (\mu m_p)$, where $m_p$ is the proton mass, and
$\mu$ the mean molecular weight \citep[we take $\mu = 0.58$;
see][]{girardi96}. This energy equipartition line is plotted as a
dashed line in the left panel of Figure \ref{fig:xray}. Real clusters
show a deviation from this simple model, but it offers an interesting
theoretical baseline, one independent of data or simulations. This
relation implies that the temperature and velocity dispersion have a
similar redshift evolution, which is why the quantities in this plot
are uncorrected for redshift.

The X-ray $Y_X$ observable, while not independent from $T_X$, is
expected to be significantly less sensitive to cluster mergers than
$T_X$, with simulations predicting $Y_X$ to have both a lower scatter
and to be a less biased mass indicator \citep[see,
e.g.,][]{kravtsov06a, fabjan11}. For this reason, we also plot the
velocity dispersion against $M_{500c,\mathrm{Y_X}}$ (times a
redshift-evolution factor), in the right panel of Figure
\ref{fig:xray}. The dot-dashed line is the scaling relation predicted
from the simulation analysis of \cite{saro12}.

Computing the average log ratio of the dynamical and $Y_X$-based
masses gives $0.26 \pm 0.12$, corresponding to a bias of $0.09 \pm
0.04$ in log dispersion. This was computed, in the previous section,
to be $0.33 \pm 0.10$ in the case of dynamical and SPT masses,
corresponding to $0.11 \pm 0.03$ in log dispersion.  The residuals of
the dispersion-$M_{500c,\mathrm{Y_X}}$ relation have a measured
scatter in dispersion of $0.31 \pm 0.03$, which is the same as the
measurement made using the SZ-based SPT mass.  The Anderson-Darling
test gives similar results to the residuals of the previous section,
suggesting a normal scatter in $\ln \sigma$ with a tail towards low
dispersion.

While there is very good agreement between the scaling relations
comparing the dispersion to the SPT and X-ray mass estimates, we note
that the results are not independent. Nine of the clusters included in
this work from \cite{reichardt12} quoted joint SZ and X-ray mass
estimates, which we have included in our sample of SPT mass estimates.
In addition, the SPT significance-mass relation used in the SZ mass
estimates was in part calibrated from a sub-sample of SPT clusters
with X-ray mass estimates, which have effectively calibrated the SPT
cluster mass normalization. Regardless, the majority of clusters in
this work have SPT mass estimates derived only from the SPT SZ
measurements, which have very different noise properties from the
X-ray measurements, therefore the agreement in the measured scatters
is not entirely trivial.

\subsection{Systematics}\label{ss:systematics}

There are two different, although related, systematics that affect the
interpretation of velocity dispersion measurements: systematics that
can affect the measurement of the velocity dispersion of galaxies, and
a possible velocity bias between the galaxies and the underlying dark
matter halo. The velocity bias cannot be empirically measured in our
data.  However, both effects have been quantified in recent cluster
simulation studies \citep{saro12, gifford13b, munari13, wu13}.  In
principle, the velocity bias could explain part of the offset between
our measurements and the predicted relation from N-body simulations,
described in Sections \ref{ss:comp_sz} and \ref{ss:xray}.  The
velocity bias has been estimated to be on order of $\sim$5\% by
\cite{evrard08}. More recent studies have found a spread in the
velocity bias of $\sim$10\% when comparing different tracers and
algorithms for predicting the galaxy population \citep{gifford13b} and
comparing dark-matter with hydrodynamic simulations \citep{munari13,
  wu13}.

The measured velocity dispersion can be biased from the true value by
the galaxy selection of the measurements, in principle being affected
by systematics relating to the luminosity, color, and offset from the
cluster center of the galaxies. The observations will also have some
amount of imperfect membership determination, due to the presence of
interlopers.

Of those effects, the luminosity of the selected galaxies has the
potential to create the largest bias, according to recent simulation
work showing that brighter galaxies have a smaller velocity dispersion
\citep[][]{saro12, old13, wu13, gifford13b}. Observing only the 25
brightest galaxies of the halo leads to the velocity dispersion being
biased low by as much as $\sim$5-10\%. These results are difficult to
directly compare to our measurements, because the simulated
observations use the $N$ brightest galaxies, while real ones target a
more varied population. Nonetheless, it is true that brighter galaxies
are targeted in priority in our observations.  As far as our data is
concerned, we took the $18$ clusters for which we obtained $30$ or
more member velocities, and compared the dispersion of the 15
brightest galaxies (among those observed spectroscopically) with our
best value. The bright galaxies have a dispersion that is $(5 \pm
4)$\% lower than the measured dispersion.

The effect of the radius at which the galaxies are sampled is
discussed in \cite{sifon12}. They conclude that there are too many
uncertainties to accurately correct for a potential bias. Regardless,
they estimate the systematic bias compared to sampling all the way to
the virial radius by using mock observations of a simulated
cluster. They find an average correction of 0.91 to the velocity
dispersions and 0.79 to the dynamical masses; in other words, the
measured velocity dispersions are biased high by 10\%. That a small
aperture radius should bias the velocity dispersions high is in line
with the results of \cite{saro12} and \cite{gifford13b}.

We performed a related test of the radial dependence of the dispersion
using our best-sampled clusters. For the $18$ clusters with $30$ or
more member velocities, we compared the dispersion of the half of the
galaxies that are the most central with the half that are further away
from the center. There is no statistically significant difference
between the most central galaxies and the most distant, their
dispersions differing by $(-2 \pm 6)$\%. Our data sample 
cluster member galaxies out to a projected radius that is typically 
$\sim$0.5 Mpc h$_{70}^{-1}$, which is generally less that the virial radius. 
As a result, our data are not always directly comparable to the 
numbers quoted from the literature.

\cite{gifford13b} also explores the effect of measuring velocity dispersion 
from galaxies that are a mix of red (passive) and blue 
(in-falling) cluster members.  including blue galaxies
alongside red galaxies in the spectroscopic sample, and find that
including a few blue galaxies only has a small differential effect on
the measured dispersion.

In addition to causing a bias in the measurement, systematics can also
increase the scatter. The resampling of Section \ref{s:estimators}
implies that there is an increase in the scatter at
few-$N_{\mathrm{members}}$ due to the different shapes of the velocity
distribution of individual clusters. \cite{saro12} find that the
scatter due to systematics is most significant when
$N_{\mathrm{members}} \lesssim 30$.

More feedback between statistical studies of much larger spectroscopic
samples than the present one and simulation work will be needed to
understand precisely how those effects affect the measured velocity
dispersion. One could imagine using the color, magnitude, position and
number of the galaxies with a spectroscopic redshift to compute a
correction factor to the dispersion, or relative weights for the
proper velocities, that would eliminate the systematic bias and
scatter from the sources discussed above.

%%%%%%%%%%%%%%%%%%%%
%% Conclusion %%
%%%%%%%%%%%%%%%%%%%%

\section{Conclusions}\label{s:conclusion}
We have reported the first results of our systematic campaign of
spectroscopic follow-up of galaxy clusters detected in the SPT-SZ
survey. We have measured cluster redshifts and velocity dispersions
from this data and conducted several tests to investigate the
robustness of these measurements and the correlation between the
velocity dispersions and other measures of cluster mass. The main
findings from these tests are:

\begin{itemize}
\item We find our strategy of obtaining redshift and velocity
  dispersion estimates from a small number of galaxies per cluster
  (typically, $N_{\mathrm{members}} \lesssim 30$) to be valid. By
  performing resampling tests that extract subsamples from a larger
  parent distribution, we observe no bias as a function of
  $N_{\mathrm{members}}$ in the redshift and percent-level bias in the
  dispersion measurements. We find, however, that the scatter is
  increased at few-$N_{\mathrm{members}}$; this systematic increase is
  due to the shapes of the velocity distribution of individual
  clusters (Section \ref{s:estimators}).

\item We fit an expression for the statistical confidence interval of
  the biweight dispersion after membership selection. It is given by
  Equation \ref{eq:param}, and we find that $C_{\mathrm{BI}} =
  0.92$. This interval is $\sim$30\% larger than the intervals
  commonly obtained in the velocity dispersion literature by using the
  statistical bootstrap. The larger width is due to the membership
  selection step and the shape of the observed velocity distribution.

\item We compare the velocity dispersions to the SZ-based SPT mass
  $M_{500c,\mathrm{SPT}}$, as well as to X-ray temperature
  measurements and $M_{500c,\mathrm{Y_X}}$.  In both comparisons with
  a mass, the measured velocity dispersions are larger by $\sim 10$\%
  on average than expected given the dispersion-mass scaling relation
  from dark-matter simulations and their SZ-based SPT or X-ray mass
  estimates. This offset is consistent with the size of several potential 
  systematic biases in the measurement of dispersions.  However, a 
  more complete understanding of its origin should include additional 
  measures of total mass (e.g., weak lensing), and a self-consistent 
  analysis that includes marginalization over uncertainties in cosmology 
  and the observableÕs scaling relation with mass. We present such an 
  analysis in \citet{bocquet2014}. 
  %Bocquet et al.~(2014). 
  The $\sim 30$\% measured log-normal 
  scatter in the dispersion measurements at fixed mass is slightly larger than, 
  but generally consistent with, the expectation from simulations.
\end{itemize}

A more complete understanding of the dispersion-mass relation, which 
more closely coupled observationally strategies across a range of simulations, 
would help to reduce systematic uncertainties. 
Observed velocity dispersions could depend in a systematic way on the
color, magnitude, and spatial selection of cluster galaxies targeted
for spectroscopic measurement. Work with simulations has improved our 
understanding of the magnitude of systematic sources of uncertainty 
in velocity dispersion mass estimates, but there is has not yet been 
a convergence of results among different simulations. A better 
quantification of systematic errors will require a combination of 
detailed, large-volume simulations and samples of clusters with many 
spectroscopic members. The ultimate goal should be a formula 
that maps a catalog of data -- individual galaxy positions, magnitudes, 
colors, and recession velocities -- into a cluster mass estimate that 
incorporates the various biases and uncertainties that result from 
the properties of the galaxy population that are used to estimate 
that cluster mass estimate. Such a formula will ultimately 
allow for better cosmological constraints from cluster surveys, which 
are currently limited by systematic uncertainties in the cluster mass 
calibration.

\acknowledgments

This paper includes spectroscopic data gathered with the 6.5-meter
Magellan Telescopes located at Las Campanas Observatory, Chile. Time
was allocated through Harvard-CfA (PIs Bayliss, Brodwin, Foley, and
Stubbs) and the Chilean National TAC (PI Clocchiatti).  Gemini South
access was obtained through NOAO. (PI Mohr, GS-2009B-Q-16, and PI
Stubbs, GS-2011A-C-3 and GS-2011B-C-6).  The VLT programs were granted
through DDT (PI Carlstrom, 286.A-5021) and ESO (PI Bazin, 087.A-0843,
and PI Chapman, 285.A-5034 and 088.A-0902).

Optical imaging data from the Blanco 4~m at Cerro Tololo Interamerican
Observatories (programs 2005B-0043, 2009B-0400, 2010A-0441,
2010B-0598) is included in this work. Additional imaging data were
obtained with the 6.5~m Magellan Telescopes and the Swope telescope,
which are located at the Las Campanas Observatory in Chile.  This work
is based in part on observations made with the Spitzer Space Telescope
(PIDs 60099, 70053), which is operated by the Jet Propulsion
Laboratory, California Institute of Technology under a contract with
NASA. Support for this work was provided by NASA through an award
issued by JPL/Caltech.

The South Pole Telescope program is supported by the National Science
Foundation through grant ANT-0638937.  Partial support is also
provided by the NSF Physics Frontier Center grant PHY-0114422 to the
Kavli Institute of Cosmological Physics at the University of Chicago,
the Kavli Foundation, and the Gordon and Betty Moore Foundation.
Galaxy cluster research at Harvard is supported by NSF grant
AST-1009012.  Galaxy cluster research at SAO is supported in part by
NSF grants AST-1009649 and MRI-0723073. Support for X-ray analysis was
provided by NASA through Chandra Award Nos. 12800071, 12800088, and
13800883 issued by the Chandra X-Ray Observatory Center, which is
operated by the Smithsonian Astrophysical Observatory for and on
behalf of NASA. The McGill group acknowledges funding from the
National Sciences and Engineering Research Council of Canada, Canada
Research Chairs program, and the Canadian Institute for Advanced
Research.  X-ray research at the CfA is supported through NASA
Contract NAS 8-03060. The Munich group was supported by The Cluster of
Excellence ``Origin and Structure of the Universe'', funded by the
Excellence Initiative of the Federal Government of Germany, EXC
project number 153.  R.J.F.\ is supported by a Clay Fellowship.
B.A.B\ is supported by a KICP Fellowship, M.Bautz and M.M. acknowledge
support from contract 2834-MIT-SAO-4018 from the Pennsylvania State
University to the Massachusetts Institute of Technology. M.D.\
acknowledges support from an Alfred P.\ Sloan Research Fellowship,
W.F.\ and C.J.\ acknowledge support from the Smithsonian
Institution. B.S. acknowledges support from the Brinson Foundation. 
A.C. received support from PFB-06 CATA, Chile. % RRESPONSE
This research used resources of the National Energy Research
Scientific Computing Center, which is supported by the Office of
Science of the U.S. Department of Energy under Contract
No. DE-AC02-05CH11231.

{\it Facilities:} \facility{Blanco (MOSAIC~II)}, \facility{Gemini-S
  (GMOS)}, \facility{Magellan:Baade (IMACS)}, \facility{Magellan:Clay
  (LDSS3)}, \facility{South Pole Telescope}, \facility{{\it
    Spitzer}/IRAC}, \facility{Swope}, \facility{VLT:Antu (FORS2)}.

\bibliographystyle{fapj} 
\bibliography{oir.bib}

\begin{thebibliography}{59}
\expandafter\ifx\csname natexlab\endcsname\relax\def\natexlab#1{#1}\fi

\bibitem[{{Allington-Smith} {et~al.}(1994){Allington-Smith}, {Breare}, {Ellis},
  {Gellatly}, {Glazebrook}, {Jorden}, {Maclean}, {Oates}, {Shaw}, {Tanvir},
  {Taylor}, {Taylor}, {Webster}, \& {Worswick}}]{Allington-Smith94}
{Allington-Smith}, J., {et~al.} 1994, \pasp, 106, 983

\bibitem[{{Andersson} {et~al.}(2011){Andersson}, {Benson}, {Ade}, {Aird},
  {Armstrong}, {Bautz}, {Bleem}, {Brodwin}, {Carlstrom}, {Chang}, {Crawford},
  {Crites}, {de Haan}, {Desai}, {Dobbs}, {Dudley}, {Foley}, {Forman},
  {Garmire}, {George}, {Gladders}, {Halverson}, {High}, {Holder}, {Holzapfel},
  {Hrubes}, {Jones}, {Joy}, {Keisler}, {Knox}, {Lee}, {Leitch}, {Lueker},
  {Marrone}, {McMahon}, {Mehl}, {Meyer}, {Mohr}, {Montroy}, {Murray}, {Padin},
  {Plagge}, {Pryke}, {Reichardt}, {Rest}, {Ruel}, {Ruhl}, {Schaffer}, {Shaw},
  {Shirokoff}, {Song}, {Spieler}, {Stalder}, {Staniszewski}, {Stark}, {Stubbs},
  {Vanderlinde}, {Vieira}, {Vikhlinin}, {Williamson}, {Yang}, {Zahn}, \&
  {Zenteno}}]{andersson10}
{Andersson}, K., {et~al.} 2011, \apj, 738, 48

\bibitem[{{Appenzeller} {et~al.}(1998){Appenzeller}, {Fricke}, {F{\"u}rtig},
  {G{\"a}ssler}, {H{\"a}fner}, {Harke}, {Hess}, {Hummel}, {J{\"u}rgens},
  {Kudritzki}, {Mantel}, {Meisl}, {Muschielok}, {Nicklas}, {Rupprecht},
  {Seifert}, {Stahl}, {Szeifert}, \& {Tarantik}}]{appenzeller98}
{Appenzeller}, I., {et~al.} 1998, The Messenger, 94, 1

\bibitem[{{Barrena} {et~al.}(2002){Barrena}, {Biviano}, {Ramella}, {Falco}, \&
  {Seitz}}]{barrena2002}
{Barrena}, R., {Biviano}, A., {Ramella}, M., {Falco}, E.~E., \& {Seitz}, S.
  2002, \aap, 386, 816

\bibitem[{{Bayliss} {et~al.}(2013){Bayliss}, {Ashby}, {Ruel}, {Brodwin},
  {Aird}, {Bautz}, {Benson}, {Bleem}, {Bocquet}, {Carlstrom}, {Chang}, {Cho},
  {Clocchiatti}, {Crawford}, {Crites}, {Desai}, {Dobbs}, {Dudley}, {Foley},
  {Forman}, {George}, {Gettings}, {Gladders}, {Gonzalez}, {de Haan},
  {Halverson}, {High}, {Holder}, {Holzapfel}, {Hoover}, {Hrubes}, {Jones},
  {Joy}, {Keisler}, {Knox}, {Lee}, {Leitch}, {Liu}, {Lueker}, {Luong-Van},
  {Mantz}, {Marrone}, {Mawatari}, {McDonald}, {McMahon}, {Mehl}, {Meyer},
  {Miller}, {Mocanu}, {Mohr}, {Montroy}, {Murray}, {Padin}, {Plagge}, {Pryke},
  {Reichardt}, {Rest}, {Ruhl}, {Saliwanchik}, {Saro}, {Sayre}, {Schaffer},
  {Shirokoff}, {Song}, {Stalder}, {Suhada}, {Spieler}, {Stanford},
  {Staniszewski}, {Stark}, {Story}, {Stubbs}, {van Engelen}, {Vanderlinde},
  {Vieira}, {Vikhlinin}, {Williamson}, {Zahn}, \& {Zenteno}}]{bayliss13}
{Bayliss}, M.~B., {et~al.} 2013, ArXiv e-prints, 1307.2903

\bibitem[{{Beers} {et~al.}(1990){Beers}, {Flynn}, \& {Gebhardt}}]{beers90}
{Beers}, T.~C., {Flynn}, K., \& {Gebhardt}, K. 1990, \aj, 100, 32

\bibitem[{{Benson} {et~al.}(2013){Benson}, {de Haan}, {Dudley}, {Reichardt},
  {Aird}, {Andersson}, {Armstrong}, {Ashby}, {Bautz}, {Bayliss}, {Bazin},
  {Bleem}, {Brodwin}, {Carlstrom}, {Chang}, {Cho}, {Clocchiatti}, {Crawford},
  {Crites}, {Desai}, {Dobbs}, {Foley}, {Forman}, {George}, {Gladders},
  {Gonzalez}, {Halverson}, {Harrington}, {High}, {Holder}, {Holzapfel},
  {Hoover}, {Hrubes}, {Jones}, {Joy}, {Keisler}, {Knox}, {Lee}, {Leitch},
  {Liu}, {Lueker}, {Luong-Van}, {Mantz}, {Marrone}, {McDonald}, {McMahon},
  {Mehl}, {Meyer}, {Mocanu}, {Mohr}, {Montroy}, {Murray}, {Natoli}, {Padin},
  {Plagge}, {Pryke}, {Rest}, {Ruel}, {Ruhl}, {Saliwanchik}, {Saro}, {Sayre},
  {Schaffer}, {Shaw}, {Shirokoff}, {Song}, {Spieler}, {Stalder},
  {Staniszewski}, {Stark}, {Story}, {Stubbs}, {Suhada}, {van Engelen},
  {Vanderlinde}, {Vieira}, {Vikhlinin}, {Williamson}, {Zahn}, \&
  {Zenteno}}]{benson11}
{Benson}, B.~A., {et~al.} 2013, \apj, 763, 147

\bibitem[{{Benson} {et~al.}(2014){Benson}, {de Haan}, {Dudley}, {Reichardt},
  {Aird}, {Andersson}, {Armstrong}, {Ashby}, {Bautz}, {Bayliss}, {Bazin},
  {Bleem}, {Brodwin}, {Carlstrom}, {Chang}, {Cho}, {Clocchiatti}, {Crawford},
  {Crites}, {Desai}, {Dobbs}, {Foley}, {Forman}, {George}, {Gladders},
  {Gonzalez}, {Halverson}, {Harrington}, {High}, {Holder}, {Holzapfel},
  {Hoover}, {Hrubes}, {Jones}, {Joy}, {Keisler}, {Knox}, {Lee}, {Leitch},
  {Liu}, {Lueker}, {Luong-Van}, {Mantz}, {Marrone}, {McDonald}, {McMahon},
  {Mehl}, {Meyer}, {Mocanu}, {Mohr}, {Montroy}, {Murray}, {Natoli}, {Padin},
  {Plagge}, {Pryke}, {Rest}, {Ruel}, {Ruhl}, {Saliwanchik}, {Saro}, {Sayre},
  {Schaffer}, {Shaw}, {Shirokoff}, {Song}, {Spieler}, {Stalder},
  {Staniszewski}, {Stark}, {Story}, {Stubbs}, {Suhada}, {van Engelen},
  {Vanderlinde}, {Vieira}, {Vikhlinin}, {Williamson}, {Zahn}, \&
  {Zenteno}}]{benson13}
------. 2014, in prep

\bibitem[{{Bocquet} {et~al.}(2014){Bocquet}, {Saro}, {Mohr}, {Aird}, {Ashby},
  {Bautz}, {Bayliss}, {Bazin}, {Benson}, {Bleem}, {Brodwin}, {Carlstrom},
  {Chang}, {Chiu}, {Cho}, {Clocchiatti}, {Crawford}, {Crites}, {Desai}, {de
  Haan}, {Dietrich}, {Dobbs}, {Foley}, {Forman}, {Gangkofner}, {George},
  {Gladders}, {Gonzalez}, {Halverson}, {Hennig}, {Hlavacek-Larrondo}, {Holder},
  {Holzapfel}, {Hrubes}, {Jones}, {Keisler}, {Knox}, {Lee}, {Leitch}, {Liu},
  {Lueker}, {Luong-Van}, {Marrone}, {McDonald}, {McMahon}, {Meyer}, {Mocanu},
  {Murray}, {Padin}, {Pryke}, {Reichardt}, {Rest}, {Ruel}, {Ruhl},
  {Saliwanchik}, {Sayre}, {Schaffer}, {Shirokoff}, {Spieler}, {Stalder},
  {Stanford}, {Staniszewski}, {Stark}, {Story}, {Stubbs}, {Vanderlinde},
  {Vieira}, {Vikhlinin}, {Williamson}, {Zahn}, \& {Zenteno}}]{bocquet2014}
{Bocquet}, S., {et~al.} 2014, ArXiv e-prints, 1407.2942

\bibitem[{{Brodwin} {et~al.}(2010){Brodwin}, {Ruel}, {Ade}, {Aird},
  {Andersson}, {Ashby}, {Bautz}, {Bazin}, {Benson}, {Bleem}, {Carlstrom},
  {Chang}, {Crawford}, {Crites}, {de Haan}, {Desai}, {Dobbs}, {Dudley},
  {Fazio}, {Foley}, {Forman}, {Garmire}, {George}, {Gladders}, {Gonzalez},
  {Halverson}, {High}, {Holder}, {Holzapfel}, {Hrubes}, {Jones}, {Joy},
  {Keisler}, {Knox}, {Lee}, {Leitch}, {Lueker}, {Marrone}, {McMahon}, {Mehl},
  {Meyer}, {Mohr}, {Montroy}, {Murray}, {Padin}, {Plagge}, {Pryke},
  {Reichardt}, {Rest}, {Ruhl}, {Schaffer}, {Shaw}, {Shirokoff}, {Song},
  {Spieler}, {Stalder}, {Stanford}, {Staniszewski}, {Stark}, {Stubbs},
  {Vanderlinde}, {Vieira}, {Vikhlinin}, {Williamson}, {Yang}, {Zahn}, \&
  {Zenteno}}]{brodwin10}
{Brodwin}, M., {et~al.} 2010, \apj, 721, 90

\bibitem[{{Buckley-Geer} {et~al.}(2011){Buckley-Geer}, {Lin}, {Drabek},
  {Allam}, {Tucker}, {Armstrong}, {Barkhouse}, {Bertin}, {Brodwin}, {Desai},
  {Frieman}, {Hansen}, {High}, {Mohr}, {Lin}, {Ngeow}, {Rest}, {Smith}, {Song},
  \& {Zenteno}}]{buckleygeer11}
{Buckley-Geer}, E.~J., {et~al.} 2011, \apj, 742, 48

\bibitem[{{Danese} {et~al.}(1980){Danese}, {de Zotti}, \& {di
  Tullio}}]{danese80}
{Danese}, L., {de Zotti}, G., \& {di Tullio}, G. 1980, \aap, 82, 322

\bibitem[{{David} {et~al.}(1993){David}, {Slyz}, {Jones}, {Forman}, {Vrtilek},
  \& {Arnaud}}]{david93}
{David}, L.~P., {Slyz}, A., {Jones}, C., {Forman}, W., {Vrtilek}, S.~D., \&
  {Arnaud}, K.~A. 1993, \apj, 412, 479

\bibitem[{{Desai} {et~al.}(2012){Desai}, {Armstrong}, {Mohr}, {Semler}, {Liu},
  {Bertin}, {Allam}, {Barkhouse}, {Bazin}, {Buckley-Geer}, {Cooper}, {Hansen},
  {High}, {Lin}, {Lin}, {Ngeow}, {Rest}, {Song}, {Tucker}, \&
  {Zenteno}}]{desai12}
{Desai}, S., {et~al.} 2012, \apj, 757, 83

\bibitem[{{Dressler} {et~al.}(2006){Dressler}, {Hare}, {Bigelow}, \&
  {Osip}}]{dressler06}
{Dressler}, A., {Hare}, T., {Bigelow}, B.~C., \& {Osip}, D.~J. 2006, in Society
  of Photo-Optical Instrumentation Engineers (SPIE) Conference Series, Vol.
  6269, Society of Photo-Optical Instrumentation Engineers (SPIE) Conference
  Series

\bibitem[{{Duffy} {et~al.}(2008){Duffy}, {Schaye}, {Kay}, \& {Dalla
  Vecchia}}]{duffy08}
{Duffy}, A.~R., {Schaye}, J., {Kay}, S.~T., \& {Dalla Vecchia}, C. 2008,
  \mnras, 390, L64

\bibitem[{{Evrard} {et~al.}(2008){Evrard}, {Bialek}, {Busha}, {White}, {Habib},
  {Heitmann}, {Warren}, {Rasia}, {Tormen}, {Moscardini}, {Power}, {Jenkins},
  {Gao}, {Frenk}, {Springel}, {White}, \& {Diemand}}]{evrard08}
{Evrard}, A.~E., {et~al.} 2008, \apj, 672, 122

\bibitem[{{Fabjan} {et~al.}(2011){Fabjan}, {Borgani}, {Rasia}, {Bonafede},
  {Dolag}, {Murante}, \& {Tornatore}}]{fabjan11}
{Fabjan}, D., {Borgani}, S., {Rasia}, E., {Bonafede}, A., {Dolag}, K.,
  {Murante}, G., \& {Tornatore}, L. 2011, \mnras, 416, 801

\bibitem[{{Foley} {et~al.}(2011){Foley}, {Andersson}, {Bazin}, {de Haan},
  {Ruel}, {Ade}, {Aird}, {Armstrong}, {Ashby}, {Bautz}, {Benson}, {Bleem},
  {Bonamente}, {Brodwin}, {Carlstrom}, {Chang}, {Clocchiatti}, {Crawford},
  {Crites}, {Desai}, {Dobbs}, {Dudley}, {Fazio}, {Forman}, {Garmire}, {George},
  {Gladders}, {Gonzalez}, {Halverson}, {High}, {Holder}, {Holzapfel}, {Hoover},
  {Hrubes}, {Jones}, {Joy}, {Keisler}, {Knox}, {Lee}, {Leitch}, {Lueker},
  {Luong-Van}, {Marrone}, {McMahon}, {Mehl}, {Meyer}, {Mohr}, {Montroy},
  {Murray}, {Padin}, {Plagge}, {Pryke}, {Reichardt}, {Rest}, {Ruhl},
  {Saliwanchik}, {Saro}, {Schaffer}, {Shaw}, {Shirokoff}, {Song}, {Spieler},
  {Stalder}, {Stanford}, {Staniszewski}, {Stark}, {Story}, {Stubbs},
  {Vanderlinde}, {Vieira}, {Vikhlinin}, {Williamson}, \& {Zenteno}}]{foley11}
{Foley}, R.~J., {et~al.} 2011, \apj, 731, 86

\bibitem[{{Foley} {et~al.}(2003){Foley}, {Papenkova}, {Swift}, {Filippenko},
  {Li}, {Mazzali}, {Chornock}, {Leonard}, \& {Van Dyk}}]{Foley03}
------. 2003, \pasp, 115, 1220

\bibitem[{{Gifford} {et~al.}(2013){Gifford}, {Miller}, \& {Kern}}]{gifford13b}
{Gifford}, D., {Miller}, C., \& {Kern}, N. 2013, \apj, 773, 116

\bibitem[{{Girardi} {et~al.}(1993){Girardi}, {Biviano}, {Giuricin},
  {Mardirossian}, \& {Mezzetti}}]{girardi93}
{Girardi}, M., {Biviano}, A., {Giuricin}, G., {Mardirossian}, F., \&
  {Mezzetti}, M. 1993, \apj, 404, 38

\bibitem[{{Girardi} {et~al.}(1996){Girardi}, {Fadda}, {Giuricin},
  {Mardirossian}, {Mezzetti}, \& {Biviano}}]{girardi96}
{Girardi}, M., {Fadda}, D., {Giuricin}, G., {Mardirossian}, F., {Mezzetti}, M.,
  \& {Biviano}, A. 1996, \apj, 457, 61

\bibitem[{{Hasselfield} {et~al.}(2013){Hasselfield}, {Hilton}, {Marriage},
  {Addison}, {Barrientos}, {Battaglia}, {Battistelli}, {Bond}, {Crichton},
  {Das}, {Devlin}, {Dicker}, {Dunkley}, {D{\"u}nner}, {Fowler}, {Gralla},
  {Hajian}, {Halpern}, {Hincks}, {Hlozek}, {Hughes}, {Infante}, {Irwin},
  {Kosowsky}, {Marsden}, {Menanteau}, {Moodley}, {Niemack}, {Nolta}, {Page},
  {Partridge}, {Reese}, {Schmitt}, {Sehgal}, {Sherwin}, {Sievers}, {Sif{\'o}n},
  {Spergel}, {Staggs}, {Swetz}, {Switzer}, {Thornton}, {Trac}, \&
  {Wollack}}]{hasselfield13}
{Hasselfield}, M., {et~al.} 2013, \jcap, 7, 8

\bibitem[{{High} {et~al.}(2012){High}, {Hoekstra}, {Leethochawalit}, {de Haan},
  {Abramson}, {Aird}, {Armstrong}, {Ashby}, {Bautz}, {Bayliss}, {Bazin},
  {Benson}, {Bleem}, {Brodwin}, {Carlstrom}, {Chang}, {Cho}, {Clocchiatti},
  {Conroy}, {Crawford}, {Crites}, {Desai}, {Dobbs}, {Dudley}, {Foley},
  {Forman}, {George}, {Gladders}, {Gonzalez}, {Halverson}, {Harrington},
  {Holder}, {Holzapfel}, {Hoover}, {Hrubes}, {Jones}, {Joy}, {Keisler}, {Knox},
  {Lee}, {Leitch}, {Liu}, {Lueker}, {Luong-Van}, {Mantz}, {Marrone},
  {McDonald}, {McMahon}, {Mehl}, {Meyer}, {Mocanu}, {Mohr}, {Montroy},
  {Murray}, {Natoli}, {Nurgaliev}, {Padin}, {Plagge}, {Pryke}, {Reichardt},
  {Rest}, {Ruel}, {Ruhl}, {Saliwanchik}, {Saro}, {Sayre}, {Schaffer}, {Shaw},
  {Schrabback}, {Shirokoff}, {Song}, {Spieler}, {Stalder}, {Staniszewski},
  {Stark}, {Story}, {Stubbs}, {{\v S}uhada}, {Tokarz}, {van Engelen},
  {Vanderlinde}, {Vieira}, {Vikhlinin}, {Williamson}, {Zahn}, \&
  {Zenteno}}]{high12}
{High}, F.~W., {et~al.} 2012, \apj, 758, 68

\bibitem[{{High} {et~al.}(2010){High}, {Stalder}, {Song}, {Ade}, {Aird},
  {Allam}, {Armstrong}, {Barkhouse}, {Benson}, {Bertin}, {Bhattacharya},
  {Bleem}, {Brodwin}, {Buckley-Geer}, {Carlstrom}, {Challis}, {Chang},
  {Crawford}, {Crites}, {de Haan}, {Desai}, {Dobbs}, {Dudley}, {Foley},
  {George}, {Gladders}, {Halverson}, {Hamuy}, {Hansen}, {Holder}, {Holzapfel},
  {Hrubes}, {Joy}, {Keisler}, {Lee}, {Leitch}, {Lin}, {Lin}, {Loehr}, {Lueker},
  {Marrone}, {McMahon}, {Mehl}, {Meyer}, {Mohr}, {Montroy}, {Morell}, {Ngeow},
  {Padin}, {Plagge}, {Pryke}, {Reichardt}, {Rest}, {Ruel}, {Ruhl}, {Schaffer},
  {Shaw}, {Shirokoff}, {Smith}, {Spieler}, {Staniszewski}, {Stark}, {Stubbs},
  {Tucker}, {Vanderlinde}, {Vieira}, {Williamson}, {Wood-Vasey}, {Yang},
  {Zahn}, \& {Zenteno}}]{high10}
------. 2010, \apj, 723, 1736

\bibitem[{{Hook} {et~al.}(2004){Hook}, {J{\o}rgensen}, {Allington-Smith},
  {Davies}, {Metcalfe}, {Murowinski}, \& {Crampton}}]{hook04}
{Hook}, I.~M., {J{\o}rgensen}, I., {Allington-Smith}, J.~R., {Davies}, R.~L.,
  {Metcalfe}, N., {Murowinski}, R.~G., \& {Crampton}, D. 2004, \pasp, 116, 425

\bibitem[{{Kasun} \& {Evrard}(2005)}]{kasun05}
{Kasun}, S.~F., \& {Evrard}, A.~E. 2005, \apj, 629, 781

\bibitem[{{Katgert} {et~al.}(1998){Katgert}, {Mazure}, {den Hartog}, {Adami},
  {Biviano}, \& {Perea}}]{katgert98}
{Katgert}, P., {Mazure}, A., {den Hartog}, R., {Adami}, C., {Biviano}, A., \&
  {Perea}, J. 1998, \aaps, 129, 399

\bibitem[{{Kelson}(2003)}]{kelson03}
{Kelson}, D.~D. 2003, \pasp, 115, 688

\bibitem[{{Komatsu} {et~al.}(2011){Komatsu}, {Smith}, {Dunkley}, {Bennett},
  {Gold}, {Hinshaw}, {Jarosik}, {Larson}, {Nolta}, {Page}, {Spergel},
  {Halpern}, {Hill}, {Kogut}, {Limon}, {Meyer}, {Odegard}, {Tucker}, {Weiland},
  {Wollack}, \& {Wright}}]{komatsu11}
{Komatsu}, E., {et~al.} 2011, \apjs, 192, 18

\bibitem[{{Kravtsov} {et~al.}(2006){Kravtsov}, {Vikhlinin}, \&
  {Nagai}}]{kravtsov06a}
{Kravtsov}, A.~V., {Vikhlinin}, A., \& {Nagai}, D. 2006, \apj, 650, 128

\bibitem[{{Kurtz} \& {Mink}(1998)}]{kurtz98}
{Kurtz}, M.~J., \& {Mink}, D.~J. 1998, \pasp, 110, 934

\bibitem[{{Mamon} {et~al.}(2010){Mamon}, {Biviano}, \& {Murante}}]{mamon10}
{Mamon}, G.~A., {Biviano}, A., \& {Murante}, G. 2010, \aap, 520, A30

\bibitem[{{Marriage} {et~al.}(2011){Marriage}, {Acquaviva}, {Ade}, {Aguirre},
  {Amiri}, {Appel}, {Barrientos}, {Battistelli}, {Bond}, {Brown}, {Burger},
  {Chervenak}, {Das}, {Devlin}, {Dicker}, {Bertrand Doriese}, {Dunkley},
  {D{\"u}nner}, {Essinger-Hileman}, {Fisher}, {Fowler}, {Hajian}, {Halpern},
  {Hasselfield}, {Hern{\'a}ndez-Monteagudo}, {Hilton}, {Hilton}, {Hincks},
  {Hlozek}, {Huffenberger}, {Handel Hughes}, {Hughes}, {Infante}, {Irwin},
  {Baptiste Juin}, {Kaul}, {Klein}, {Kosowsky}, {Lau}, {Limon}, {Lin},
  {Lupton}, {Marsden}, {Martocci}, {Mauskopf}, {Menanteau}, {Moodley},
  {Moseley}, {Netterfield}, {Niemack}, {Nolta}, {Page}, {Parker}, {Partridge},
  {Quintana}, {Reese}, {Reid}, {Sehgal}, {Sherwin}, {Sievers}, {Spergel},
  {Staggs}, {Swetz}, {Switzer}, {Thornton}, {Trac}, {Tucker}, {Warne},
  {Wilson}, {Wollack}, \& {Zhao}}]{marriage11b}
{Marriage}, T.~A., {et~al.} 2011, \apj, 737, 61

\bibitem[{{McDonald} {et~al.}(2012){McDonald}, {Bayliss}, {Benson}, {Foley},
  {Ruel}, {Sullivan}, {Veilleux}, {Aird}, {Ashby}, {Bautz}, {Bazin}, {Bleem},
  {Brodwin}, {Carlstrom}, {Chang}, {Cho}, {Clocchiatti}, {Crawford}, {Crites},
  {de Haan}, {Desai}, {Dobbs}, {Dudley}, {Egami}, {Forman}, {Garmire},
  {George}, {Gladders}, {Gonzalez}, {Halverson}, {Harrington}, {High},
  {Holder}, {Holzapfel}, {Hoover}, {Hrubes}, {Jones}, {Joy}, {Keisler}, {Knox},
  {Lee}, {Leitch}, {Lieu}, {Lueker}, {Luong-Van}, {Mantz}, {Marrone},
  {McMahon}, {Mehl}, {Meyer}, {Miller}, {Mocanu}, {Mohr}, {Montroy}, {Murray},
  {Natoli}, {Padin}, {Plagge}, {Pryke}, {Rawle}, {Reichardt}, {Rest}, {Rex},
  {Ruhl}, {Saliwanchik}, {Saro}, {Sayre}, {Schaffer}, {Shaw}, {Shirokoff},
  {Simcoe}, {Song}, {Spieler}, {Stalder}, {Staniszewski}, {Stark}, {Story},
  {Stubbs}, {Suhada}, {van Engelen}, {Vanderlinde}, {Vieira}, {Vikhlinin},
  {Williamson}, {Zahn}, \& {Zenteno}}]{mcdonald12}
{McDonald}, M., {et~al.} 2012, \nat, 488, 349

\bibitem[{{McDonald} {et~al.}(2013){McDonald}, {Benson}, {Vikhlinin},
  {Stalder}, {Bleem}, {de Haan}, {Lin}, {Aird}, {Ashby}, {Bautz}, {Bayliss},
  {Bocquet}, {Brodwin}, {Carlstrom}, {Chang}, {Cho}, {Clocchiatti}, {Crawford},
  {Crites}, {Desai}, {Dobbs}, {Dudley}, {Foley}, {Forman}, {George},
  {Gettings}, {Gladders}, {Gonzalez}, {Halverson}, {High}, {Holder},
  {Holzapfel}, {Hoover}, {Hrubes}, {Jones}, {Joy}, {Keisler}, {Knox}, {Lee},
  {Leitch}, {Liu}, {Lueker}, {Luong-Van}, {Mantz}, {Marrone}, {McMahon},
  {Mehl}, {Meyer}, {Miller}, {Mocanu}, {Mohr}, {Montroy}, {Murray},
  {Nurgaliev}, {Padin}, {Plagge}, {Pryke}, {Reichardt}, {Rest}, {Ruel}, {Ruhl},
  {Saliwanchik}, {Saro}, {Sayre}, {Schaffer}, {Shirokoff}, {Song}, {{\v
  S}uhada}, {Spieler}, {Stanford}, {Staniszewski}, {Stark}, {Story}, {van
  Engelen}, {Vanderlinde}, {Vieira}, {Williamson}, {Zahn}, \&
  {Zenteno}}]{mcdonald13}
------. 2013, \apj, 774, 23

\bibitem[{{Mosteller} \& {Tukey}(1977)}]{mosteller77}
{Mosteller}, F., \& {Tukey}, J.~W. 1977, {Data analysis and regression. A
  second course in statistics} (Addison-Wesley)

\bibitem[{{Munari} {et~al.}(2013){Munari}, {Biviano}, {Borgani}, {Murante}, \&
  {Fabjan}}]{munari13}
{Munari}, E., {Biviano}, A., {Borgani}, S., {Murante}, G., \& {Fabjan}, D.
  2013, \mnras, 430, 2638

\bibitem[{{Old} {et~al.}(2013){Old}, {Gray}, \& {Pearce}}]{old13}
{Old}, L., {Gray}, M.~E., \& {Pearce}, F.~R. 2013, \mnras, 434, 2606

\bibitem[{{Planck Collaboration} {et~al.}(2013){Planck Collaboration}, {Ade},
  {Aghanim}, {Armitage-Caplan}, {Arnaud}, {Ashdown}, {Atrio-Barandela},
  {Aumont}, {Aussel}, {Baccigalupi}, \& et~al.}]{planck13}
{Planck Collaboration}, {et~al.} 2013, ArXiv e-prints, 1303.5089

\bibitem[{{Planck Collaboration} {et~al.}(2011){Planck Collaboration}, {Ade},
  {Aghanim}, {Arnaud}, {Ashdown}, {Aumont}, {Baccigalupi}, {Balbi}, {Banday},
  {Barreiro}, \& et~al.}]{planck11-5.1a_arxiv}
------. 2011, \aap, 536, A8

\bibitem[{{Quintana} {et~al.}(2000){Quintana}, {Carrasco}, \&
  {Reisenegger}}]{quintana00}
{Quintana}, H., {Carrasco}, E.~R., \& {Reisenegger}, A. 2000, \aj, 120, 511

\bibitem[{{Reichardt} {et~al.}(2013){Reichardt}, {Stalder}, {Bleem}, {Montroy},
  {Aird}, {Andersson}, {Armstrong}, {Ashby}, {Bautz}, {Bayliss}, {Bazin},
  {Benson}, {Brodwin}, {Carlstrom}, {Chang}, {Cho}, {Clocchiatti}, {Crawford},
  {Crites}, {de Haan}, {Desai}, {Dobbs}, {Dudley}, {Foley}, {Forman}, {George},
  {Gladders}, {Gonzalez}, {Halverson}, {Harrington}, {High}, {Holder},
  {Holzapfel}, {Hoover}, {Hrubes}, {Jones}, {Joy}, {Keisler}, {Knox}, {Lee},
  {Leitch}, {Liu}, {Lueker}, {Luong-Van}, {Mantz}, {Marrone}, {McDonald},
  {McMahon}, {Mehl}, {Meyer}, {Mocanu}, {Mohr}, {Murray}, {Natoli}, {Padin},
  {Plagge}, {Pryke}, {Rest}, {Ruel}, {Ruhl}, {Saliwanchik}, {Saro}, {Sayre},
  {Schaffer}, {Shaw}, {Shirokoff}, {Song}, {Spieler}, {Staniszewski}, {Stark},
  {Story}, {Stubbs}, {{\v S}uhada}, {van Engelen}, {Vanderlinde}, {Vieira},
  {Vikhlinin}, {Williamson}, {Zahn}, \& {Zenteno}}]{reichardt12}
{Reichardt}, C.~L., {et~al.} 2013, \apj, 763, 127

\bibitem[{{Saro} {et~al.}(2013){Saro}, {Mohr}, {Bazin}, \& {Dolag}}]{saro12}
{Saro}, A., {Mohr}, J.~J., {Bazin}, G., \& {Dolag}, K. 2013, \apj, 772, 47

\bibitem[{{Sif{\'o}n} {et~al.}(2013){Sif{\'o}n}, {Menanteau}, {Hasselfield},
  {Marriage}, {Hughes}, {Barrientos}, {Gonz{\'a}lez}, {Infante}, {Addison},
  {Baker}, {Battaglia}, {Bond}, {Crichton}, {Das}, {Devlin}, {Dunkley},
  {D{\"u}nner}, {Gralla}, {Hajian}, {Hilton}, {Hincks}, {Kosowsky}, {Marsden},
  {Moodley}, {Niemack}, {Nolta}, {Page}, {Partridge}, {Reese}, {Sehgal},
  {Sievers}, {Spergel}, {Staggs}, {Thornton}, {Trac}, \& {Wollack}}]{sifon12}
{Sif{\'o}n}, C., {et~al.} 2013, \apj, 772, 25

\bibitem[{{Song} {et~al.}(2012){Song}, {Zenteno}, {Stalder}, {Desai}, {Bleem},
  {Aird}, {Armstrong}, {Ashby}, {Bayliss}, {Bazin}, {Benson}, {Bertin},
  {Brodwin}, {Carlstrom}, {Chang}, {Cho}, {Clocchiatti}, {Crawford}, {Crites},
  {de Haan}, {Dobbs}, {Dudley}, {Foley}, {George}, {Gettings}, {Gladders},
  {Gonzalez}, {Halverson}, {Harrington}, {High}, {Holder}, {Holzapfel},
  {Hoover}, {Hrubes}, {Joy}, {Keisler}, {Knox}, {Lee}, {Leitch}, {Liu},
  {Lueker}, {Luong-Van}, {Marrone}, {McDonald}, {McMahon}, {Mehl}, {Meyer},
  {Mocanu}, {Mohr}, {Montroy}, {Natoli}, {Nurgaliev}, {Padin}, {Plagge},
  {Pryke}, {Reichardt}, {Rest}, {Ruel}, {Ruhl}, {Saliwanchik}, {Saro}, {Sayre},
  {Schaffer}, {Shaw}, {Shirokoff}, {{\v S}uhada}, {Spieler}, {Stanford},
  {Staniszewski}, {Stark}, {Story}, {Stubbs}, {van Engelen}, {Vanderlinde},
  {Vieira}, {Williamson}, \& {Zahn}}]{song12}
{Song}, J., {et~al.} 2012, \apj, 761, 22

\bibitem[{{Stalder} {et~al.}(2013){Stalder}, {Ruel}, {{\v S}uhada}, {Brodwin},
  {Aird}, {Andersson}, {Armstrong}, {Ashby}, {Bautz}, {Bayliss}, {Bazin},
  {Benson}, {Bleem}, {Carlstrom}, {Chang}, {Cho}, {Clocchiatti}, {Crawford},
  {Crites}, {de Haan}, {Desai}, {Dobbs}, {Dudley}, {Foley}, {Forman}, {George},
  {Gettings}, {Gladders}, {Gonzalez}, {Halverson}, {Harrington}, {High},
  {Holder}, {Holzapfel}, {Hoover}, {Hrubes}, {Jones}, {Joy}, {Keisler}, {Knox},
  {Lee}, {Leitch}, {Liu}, {Lueker}, {Luong-Van}, {Mantz}, {Marrone},
  {McDonald}, {McMahon}, {Mehl}, {Meyer}, {Mocanu}, {Mohr}, {Montroy},
  {Murray}, {Natoli}, {Nurgaliev}, {Padin}, {Plagge}, {Pryke}, {Reichardt},
  {Rest}, {Ruhl}, {Saliwanchik}, {Saro}, {Sayre}, {Schaffer}, {Shaw},
  {Shirokoff}, {Song}, {Spieler}, {Stanford}, {Staniszewski}, {Stark}, {Story},
  {Stubbs}, {van Engelen}, {Vanderlinde}, {Vieira}, {Vikhlinin}, {Williamson},
  {Zahn}, \& {Zenteno}}]{stalder12}
{Stalder}, B., {et~al.} 2013, \apj, 763, 93

\bibitem[{{Staniszewski} {et~al.}(2009){Staniszewski}, {Ade}, {Aird}, {Benson},
  {Bleem}, {Carlstrom}, {Chang}, {Cho}, {Crawford}, {Crites}, {de Haan},
  {Dobbs}, {Halverson}, {Holder}, {Holzapfel}, {Hrubes}, {Joy}, {Keisler},
  {Lanting}, {Lee}, {Leitch}, {Loehr}, {Lueker}, {McMahon}, {Mehl}, {Meyer},
  {Mohr}, {Montroy}, {Ngeow}, {Padin}, {Plagge}, {Pryke}, {Reichardt}, {Ruhl},
  {Schaffer}, {Shaw}, {Shirokoff}, {Spieler}, {Stalder}, {Stark},
  {Vanderlinde}, {Vieira}, {Zahn}, \& {Zenteno}}]{staniszewski09}
{Staniszewski}, Z., {et~al.} 2009, \apj, 701, 32

\bibitem[{{Struble} \& {Rood}(1999)}]{struble99}
{Struble}, M.~F., \& {Rood}, H.~J. 1999, \apjs, 125, 35

\bibitem[{{Sunyaev} \& {Zel'dovich}(1972)}]{sunyaev72}
{Sunyaev}, R.~A., \& {Zel'dovich}, Y.~B. 1972, Comments on Astrophysics and
  Space Physics, 4, 173

\bibitem[{{Vanderlinde} {et~al.}(2010){Vanderlinde}, {Crawford}, {de Haan},
  {Dudley}, {Shaw}, {Ade}, {Aird}, {Benson}, {Bleem}, {Brodwin}, {Carlstrom},
  {Chang}, {Crites}, {Desai}, {Dobbs}, {Foley}, {George}, {Gladders}, {Hall},
  {Halverson}, {High}, {Holder}, {Holzapfel}, {Hrubes}, {Joy}, {Keisler},
  {Knox}, {Lee}, {Leitch}, {Loehr}, {Lueker}, {Marrone}, {McMahon}, {Mehl},
  {Meyer}, {Mohr}, {Montroy}, {Ngeow}, {Padin}, {Plagge}, {Pryke}, {Reichardt},
  {Rest}, {Ruel}, {Ruhl}, {Schaffer}, {Shirokoff}, {Song}, {Spieler},
  {Stalder}, {Staniszewski}, {Stark}, {Stubbs}, {van Engelen}, {Vieira},
  {Williamson}, {Yang}, {Zahn}, \& {Zenteno}}]{vanderlinde10}
{Vanderlinde}, K., {et~al.} 2010, \apj, 722, 1180

\bibitem[{{Vikhlinin} {et~al.}(2009){Vikhlinin}, {Burenin}, {Ebeling},
  {Forman}, {Hornstrup}, {Jones}, {Kravtsov}, {Murray}, {Nagai}, {Quintana}, \&
  {Voevodkin}}]{vikhlinin09b}
{Vikhlinin}, A., {et~al.} 2009, \apj, 692, 1033

\bibitem[{{Voit}(2005)}]{voit04}
{Voit}, G.~M. 2005, Reviews of Modern Physics, 77, 207

\bibitem[{{Wade} \& {Horne}(1988)}]{wade88}
{Wade}, R.~A., \& {Horne}, K. 1988, \apj, 324, 411

\bibitem[{{White} {et~al.}(2010){White}, {Cohn}, \& {Smit}}]{white10}
{White}, M., {Cohn}, J.~D., \& {Smit}, R. 2010, \mnras, 408, 1818

\bibitem[{{Williamson} {et~al.}(2011){Williamson}, {Benson}, {High},
  {Vanderlinde}, {Ade}, {Aird}, {Andersson}, {Armstrong}, {Ashby}, {Bautz},
  {Bazin}, {Bertin}, {Bleem}, {Bonamente}, {Brodwin}, {Carlstrom}, {Chang},
  {Chapman}, {Clocchiatti}, {Crawford}, {Crites}, {de Haan}, {Desai}, {Dobbs},
  {Dudley}, {Fazio}, {Foley}, {Forman}, {Garmire}, {George}, {Gladders},
  {Gonzalez}, {Halverson}, {Holder}, {Holzapfel}, {Hoover}, {Hrubes}, {Jones},
  {Joy}, {Keisler}, {Knox}, {Lee}, {Leitch}, {Lueker}, {Luong-Van}, {Marrone},
  {McMahon}, {Mehl}, {Meyer}, {Mohr}, {Montroy}, {Murray}, {Padin}, {Plagge},
  {Pryke}, {Reichardt}, {Rest}, {Ruel}, {Ruhl}, {Saliwanchik}, {Saro},
  {Schaffer}, {Shaw}, {Shirokoff}, {Song}, {Spieler}, {Stalder}, {Stanford},
  {Staniszewski}, {Stark}, {Story}, {Stubbs}, {Vieira}, {Vikhlinin}, \&
  {Zenteno}}]{williamson11}
{Williamson}, R., {et~al.} 2011, \apj, 738, 139

\bibitem[{{Wu} {et~al.}(2013){Wu}, {Hahn}, {Evrard}, {Wechsler}, \&
  {Dolag}}]{wu13}
{Wu}, H.-Y., {Hahn}, O., {Evrard}, A.~E., {Wechsler}, R.~H., \& {Dolag}, K.
  2013, ArXiv e-prints, 1307.0011

\bibitem[{{Yahil} \& {Vidal}(1977)}]{yahil77}
{Yahil}, A., \& {Vidal}, N.~V. 1977, \apj, 214, 347

\end{thebibliography}
\end{document}